\documentclass[aps,prb,reprint,twocolumn,superscriptaddress]{revtex4-1}
\usepackage{mhchem}
\usepackage[hyperindex,breaklinks]{hyperref}
\usepackage{xcolor}
\hypersetup{
    colorlinks,
    linkcolor={red!50!black},
    citecolor={blue!50!black},
    urlcolor={blue!80!black}
}
\usepackage{graphicx} % Required for inserting images
\usepackage{amsmath}
\usepackage{appendix}
\usepackage{mathtools}
\usepackage{wrapfig}
\usepackage{comment}
\usepackage{float}
\usepackage{soul}
\usepackage[normalem]{ulem}

\allowdisplaybreaks
\newcommand*{\AB}{A\textsubscript{p}B\textsubscript{q}}
\newcommand{\hd}{\Delta H^\text{d}_{\text{f}}}
\newcommand{\hp}{\Delta H^\text{p}_{\text{f}}}

\newcommand{\yt}{y^\text{1}_{\text{vac}}}
\newcommand{\yu}{y^\text{2}_{\text{vac}}}
\newcommand{\yA}{y^\text{1}_{\text{A}}}
\newcommand{\yB}{y^\text{1}_{\text{B}}}
\newcommand{\yi}{y^\text{3}_{\text{A}^{\text{i}}}}
\newcommand{\ye}{y^\text{3}_{\text{e\textsubscript{-}}}}
\newcommand{\yh}{y^\text{3}_{\text{h\textsubscript{+}}}}
\newcommand{\yvac}{y^\text{3}_{\text{vac}}}
\newcommand{\yii}{y_{\text{i}}}
\newcommand{\yjj}{y_{\text{j}}}
\newcommand{\yaddvac}{y^\text{4}_{\text{vac}}}

\newcommand{\edit}[1]{\textcolor{black}{#1}}

\begin{document}
\title{Defect energy formalism for CALPHAD thermodynamics of dilute point defects: Theory}
\author{Amir M. Orvati Movaffagh}
\affiliation{Civil, Materials, and Environmental Engineering, University of Illinois Chicago, Chicago, IL, USA}
\author{Adetoye Adekoya}
\affiliation{Materials Science and Engineering, Northwestern University, Evanston, IL, USA}
%\author{G. Jeffrey Snyder}
%\affiliation{Materials Science and Engineering, Northwestern University, Evanston, IL, USA}
\author{Sara Kadkhodaei}
\email{sarakad@uic.edu}
\affiliation{Civil, Materials, and Environmental Engineering, University of Illinois Chicago, Chicago, IL, USA}

\date{\today}

\begin{abstract}
The thermodynamics of point defects is crucial for determining the functional properties of various materials. Typically, defect stability is assessed using grand-canonical defect formation energy, which requires deducing the equilibrium chemical potential or Fermi level that governs atom and electron exchange with the environment. This process is complicated by the interplay of chemical potential and Fermi level and their dependence on composition and temperature. Typically, the grand-canonical formation energy is incorporated as an additional term to the bulk Gibbs energy, creating a defect-centric framework where each new defect state necessitates a distinct Gibbs energy formulation. The calculation of phase diagrams (CALPHAD) method offers a more flexible alternative by integrating defect energies into the total Gibbs energy model, allowing for easier extrapolation to more complex compositions. Additionally, CALPHAD unifies the analysis of chemically and electronically driven defects using chemical composition as the primary variable. However, the Compound Energy Formalism (CEF) used in CALPHAD has limitations, including a lack of clear connections between defect formation energies and Gibbs energy parameters and an exponential increase in complexity with added chemical or charge variations. We present the theoretical derivation of the Defect Energy Formalism (DEF), which we have recently proposed. DEF overcomes the limitations of CEF by establishing explicit relationships between the absolute defect energies—independent of chemical potential or Fermi level—and the Gibbs energy parameters of defective compounds. This results in a first-principles model for dilute defects, eliminating the need for model fitting to experimental or simulation data. Additionally, DEF reduces the inherent complexity of CEF by applying the superposition of absolute defect energies, making it feasible for modeling multi-component and chemically complex compounds. This paper presents a formal, general derivation of DEF and offers guidelines for its application, promising more accurate and efficient thermodynamic modeling of defective materials. %The DEF derivation is based on the principles of linear mapping and superposition of defect formation energies. %Usually, the grand-canonical formation energy is expressed as an additional term to the bulk Gibbs energy, resulting in a defect-centric and system-specific framework. 
\end{abstract}
\maketitle

\section{Introduction}\label{sec:intro}
Thermodynamics of point defects determines the functional properties of many materials in several applications, from semiconductors (thermoelectrics \cite{1,2,3,4,5,6}, photovoltaics \cite{7,8,9,10,11,12,13,14,15}) to insulators (optical materials \cite{16,17,18,19,20,21,22,23}, ion-conducting materials \cite{22,24,25,26,27,28}, non-stoichiometric materials \cite{28,29,30,31,32}). The thermodynamics of isolated defects in solids is well-established through formation energies, typically defined as the excess grand-potential energy in a grand-canonical ensemble, offering a defect-centric perspective \cite{Rogal2014,57,58}. First principles methods like density functional theory (DFT) have significantly advanced, enabling precise calculations of formation energies for various point defects (vacancies, interstitials, antisites) across different charge states \cite{Freysoldt2014,52,53,54,55,56}. In this defect-centric approach, the defect formation free energy is considered an additional contribution to the total free energy, expressed as ($G=G(\text{bulk})+\Delta G_{\text{defect}}$). However, this method is inherently system-specific, requiring the determination of a new $\Delta G_{\text{defect}}$ term for each new composition or higher-order multi-component system. The CALPHAD approach offers an adaptable alternative by incorporating the influence of defects directly into the comprehensive description of total free energy rather than treating them as isolated excess terms. In this framework, defects are introduced as distinct entities that interact with other components, such as chemical elements. Their impact on the Gibbs energy is integrated into interaction parameters within the CALPHAD model. CALPHAD’s hierarchical formulation enables straightforward extrapolation to higher-order multi-component systems. For instance, interaction parameters initially determined for a binary compound can be leveraged to predict free energy across other compositions, including ternary compounds. 
\begin{figure}\label{fig:DEFconcept}
    \centering
        \includegraphics[width=\columnwidth]{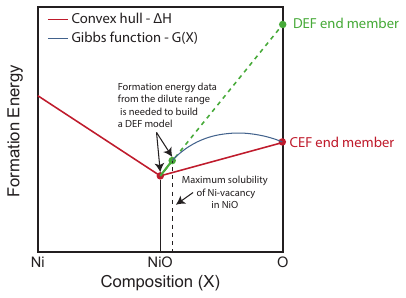} % Replace with your image file
        \caption{\textbf{Linear Mapping Principle underlying DEF}. Comparing CEF and DEF for thermodynamic modeling of dilute Ni vacancies in \ce{Ni_{1-$\delta$}O}. The schematic Gibbs energy is given by the blue curve, approximated in CALPHAD  using interpolation of end-member parameters. Traditional CEF uses the fully defective end-member, which in this case is \ce{O}, and effectively approximates the blue curve with the red line, which greatly underestimates the $G$ of \ce{NiO} with dilute Ni vacancies. The DEF method, in contrast, approximates the blue curve with the green line derived from the slope as it approaches the pristine compound (`dilute limit'). The DEF provides a much more accurate estimate of $G$ in the composition region pertinent to dilute defect concentrations.}
\end{figure}

The CALPHAD method for collecting, reporting, and computing thermodynamic data has been hugely successful both in industry and academia, leading to the rapid development of metal alloys for numerous applications \cite{101,102,103,104,105,106,107,108,109,110,111,112}. For ordered ionic compounds and semiconductors, the compound energy formalism (CEF) has often been utilized \cite{hillert1970regular,Harvig1971AnEV,SUNDMAN1981297,andersson1986compound,HILLERT2001161,FRISK2001177}, which divides the lattice into different sublattices. Each sublattice can host various constituents (e.g., chemical element or vacancy). An ``end member" represents a specific combination of these constituents on the sublattices, where each constituent exclusively occupies one sublattice. CEF formulates the Gibbs energy using these end members as the primary first-order parameters, while higher-order parameters characterize interactions among constituents within individual sublattices \cite{HILLERT2001161}. Defective compounds (e.g., non-stoichiometric oxides) can be recognized in CEF by defining point defects as constituents on the relevant sublattice. For example, non-stoichiometry of \ce{Ni_{1-$\delta$}O} can be defined as \ce{(Ni,Vac)(O)} (see Figure \ref{fig:DEFconcept}). %For example, oxygen non-stoichiometry in ceria \ce{CeO_{2-$\delta$}} can be defined as \ce{(Ce)(O,Vac)_2} %For example, Ni-doped \ce{(Ca,Sr)(Co,Fe)O_3} perovskite multi-cation solution can be defined as \ce{(Ca,Sr)(Co,Fe,Ni)(O)_3} or its oxygen non-stoichiometry can be defined as \ce{(Ca,Sr)(Co,Fe)(O,Va)_3}. %or among different sublattices 

We have identified two bottlenecks in CEF that limit its application in describing defective compounds. Firstly, there lacks a clear connection between defect formation energies and the CEF parameters of Gibbs energy. While previous studies \cite{HILLERT2001161, Rogal2014, Oates1995, Chen1998, LI20131213} have hinted at connections between defect formation energies and Gibbs energies of end members, a systematic relationship remains elusive. Consequently, most studies use a standard ``assessment" approach, fitting parameters to available experimental and computational data. This approach limits model development to specific material and thermodynamic ranges (temperatures, compositions) where data is accessible, hindering the creation of first-principles models. Even studies using first-principles DFT to compute defect formation energies typically use these values merely as starting points for fitting assessments to experimental data \cite{117, Peters2019, 119, 120, LI20131213}, rather than as direct inputs into Gibbs energy formulation. Secondly, the number of parameters in the CEF Gibbs energy formulation grows exponentially with added chemical or charge complexity (e.g., doped impurities, alloying components, or charged defects). CEF includes all potential end members as independent parameters, making it excessively intricate for multi-component and chemically complex compounds, as well as for describing charge carriers such as holes and free electrons. This computational complexity has constrained existing models of defective semiconductors and insulators to a limited set of binary examples where experimental data exists, so the model can ultimately be fit to experimental data (e.g., GaAs \cite{Chen1996}, CdTe \cite{Chen1998}, \ce{UO2} \cite{123}, PbSe \cite{Peters2019}, PbX (X=S,Te) \cite{119,120}, ZnS \cite{124}, ZnO \cite{LI20131213}).
%Because of this, CEF has only been applied to a few binary semiconductors with dilute point defects (e.g., GaAs\cite{Oates1995,Chen1996}, CdTe\cite{Chen1998}, \ce{UO2}\cite{123}, PbSe\cite{Peters2019}, PbX (where X = S, Te) \cite{119,120}, and ZnS\cite{124}). 

To address the existing limitations of CEF, we theoretically derive the Defect Energy Formalism (DEF) as a special case of CEF for CALPHAD thermodynamics of dilute defects. For formulating the Gibbs energy of a defective compound, DEF applies two main principles pertinent to dilute defects:

First is the linear mapping of defect formation energies along chemical composition, inspired by an earlier study by Anand \textit{et al} \cite{Anand2021Visualizing}, showing that the projection of a defective compound energy to its constituent species on the convex hull reflects the defect formation energy. Here, we evolve this concept to establish the physics-based relationship between DEF end members and defect formation energies. Additionally, we show that the Boltzmann statistics of dilute charge carriers naturally arise in the DEF formalism, similar to the CEF formalism. Figure \ref{fig:DEFconcept} illustrates the concept of defining DEF end-members from the projection of the defect formation energy on the convex hull to its constituent species. This concept underlies the principle of linear mapping, as implemented in DEF and detailed in this work.

Second is applying the superposition principle to the Gibbs energy of DEF end-members containing multiple defects, offering a computationally efficient framework by reducing the number of independent end-members in the sublattice model. In a system with $s$ sublattices, each hosting $N_s$ constituents, the number of CEF end-members equals $\prod_s N_s$, whereas for DEF it equals $\sum_s N_s$. In simpler terms, DEF simplifies the complexity of CEF parameters from a combinatorial factor of single-defect end-members to a summation. This reduction stems naturally from the superposition principles applicable to dilute defects, unlike CEF, which handles arbitrary constituent mixing on each sublattice. A DEF end-member with multiple defects describes a compound hosting non-interacting, isolated defects, unlike CEF, where an end-member with multiple defects corresponds to ``fictitious" end members with unrealistically high concentrations of interacting defects (see Figure \ref{fig:DEFconcept}).

Recently, we introduced DEF and demonstrated its application, confirming its feasibility \cite{Adekoya2024,ADEKOYA2024100109}. Here, we present a formal derivation of the DEF applicable to any type of point defects (vacancies, interstitials, and anti-sites), offering guidelines for constructing DEF models for any materials and combinations of dilute point defects. The organization of this paper is as follows: In section \ref{sec:DEFneutral}, we derive the DEF formulation for defective compounds with neutral defects, detailing the derivation of DEF end-member Gibbs energy in section \ref{sec:convexdefect} and the construction of DEF sublattice models in sections \ref{sec:2sub} and \ref{sec:3sub}. In section \ref{sec:DEFcharged}, we derive the DEF formulation for defective compounds with charged defects, followed by constructing a DEF sublattice model in section \ref{sec:2subcharge}. Section \ref{sec:generalrecipe} presents general guidelines for constructing DEF for any compounds and combinations of defects. 

\section{Defect Energy Formalism for Neutral Defects}\label{sec:DEFneutral}
The DEF construct is a modified version of the standard CEF, seen as a special case with specific constraints on the mixing behavior of its constituents. A phase in DEF is divided into one, two or more sublattices, labeled $s$, each containing $N_s$ constituents. For instance, (A,C)$_p$(B,D,E)$_q$ represents a typical two-sublattice model where A and C mix on the first sublattice and B, D, and E mix on the second, with $p$ and $q$ as stoichiometric coefficients for a formula unit containing $p+q$ atoms. In a typical CEF model, each sublattice can mix multiple primary constituents at arbitrary ranges. In DEF, however, defect concentrations remain in the dilute range, so defects are considered secondary constituents. Here, we focus on DEF models where each sublattice hosts one primary constituent (an atomic species) and multiple defects (e.g., vacancies, anti-sites, or interstitials) as secondary constituents, such as (A,Vac,B)$_p$(B,A)$_q$ for A-vacancy and B-antisite, and A-antisite. End-members represent phases with only one constituent per sublattice. Within the DEF construct, one end-member corresponds to defect-free ordered compounds with each sublattice hosting primary constituents (e.g., (A)$_p$(B)$_q$), while others are defective compounds with at least one defect constituent on a sublattice (e.g., (Vac)$_p$(B)$_q$). Despite appearing as nonphysical due to the notation showing high defect concentrations (e.g., (Vac)$_p$(B)$_q$), defective end-members in DEF actually correspond to physical compounds with dilute defect concentrations (see Figure \ref{fig:DEFconcept}). 
%The construction of DEF is distinct from a standard CEF (with defect constituents) in two ways: First, we show that DEF is constructed based on only a fraction of CEF end-members by reducing the number of end-members from $\prod_s N_s$, as in traditional CEF, to $\sum_s N_s$. %Second, we show that
%Although DEF can include multiple primary constituents with full mixing alongside secondary defects, this paper specifically examines constructions with one primary constituent per sublattice. Distinct from CEF, DEF results in all real end-members, with no fictitious ones. 

The constitution of a DEF phase is described by the site fractions of each constituent $J$ on each sublattice $s$, denoted by $y^s_J$. The summation of constituents' site fractions on each sublattice yields 1, or $\sum_J y^s_J =1$. The content of each component $I$ per formula unit is then related to the site fractions on individual sublattices according to the following equation (see equation 4 in Ref.~\cite{HILLERT2001161}): %(I = A,B) 
\begin{equation}\label{eq:XtoY}
    X_I = \frac{\sum_s {a^s y^s_{I}}}{\sum_s {a^s \left(1-y^s_{\text{vac}}\right)}}
\end{equation}
where $a^s$ is the stoichiometry coefficient of sublattice $s$. The composition $X\textsubscript{I}$ denotes the composition of a component per formula unit of atoms and not per site. Therefore, it directly relates to the composition space in a typical convex hull or phase diagram. We refer to the composition space as the $X$-space and the constitution of DEF site fractions as the $Y$-space.

The Gibbs energy per formula unit of the DEF phase, $G_{\text{uf}}$, is defined by the surface of reference energy $G_{\text{uf}}^{\text{s.r.}}$ along with the ideal mixing entropy, following the CEF (see equations 1 and 2 in Ref.\cite{HILLERT2001161}). Here, $G_{\text{uf}}^{\text{s.r.}}$ is a linear interpolation of end-member Gibbs energies, $^0G_{\text{end}}$, as follows:
\begin{align}\label{eq:GibbsDEF}
    \begin{split}
        & G_{\text{fu}} = G_{\text{fu}}^{\text{s.r.}} + kT \sum_s \sum_J  a^s y^s_J \ln(y^s_J) \\ 
        & G_{\text{fu}}^{\text{s.r.}} = \sum_{\text{end-members}} {{}^0G_{\text{end}} \prod_s y^s_J}
    \end{split}
\end{align}
where $k$ and $T$ denote the Boltzmann constant and temperature, respectively. In the $G_{\text{uf}}^{\text{s.r.}}$ formula, the summation runs over all end members, and the product runs over all sublattices in the DEF model. For each end-member, $J$ consists only of the constituents corresponding to that end-member. In DEF, we do not include the excess Gibbs energy term, commonly denoted as ${}^E G$ in a typical CEF model. This is because at dilute defect concentrations, the primary constituent and secondary defects mix like an ideal solution, where each added defect reduces the number of solvent atoms, making chemical activity proportional to chemical composition (i.e., Raoult's law for dilute solutions). Therefore, the only parameters of the Gibbs energy are the end-members, ${}^0G_{\text{end}}$. 

The grand canonical formation energy for a neutral defect is defined as 
\begin{align}\label{eq:defect_orig}
    \Delta H_{\text{d}} = E_{\text{def}} - E_{\text{pristine}} - \sum_i\Delta N_i \mu_i
\end{align}
where $E_{\text{def}}$ and $E_{\text{pristine}}$ denote the energies of the defective and pristine structures, respectively, $\Delta N_i$ is the number of atoms of species $i$ added to or removed from the defective structure (e.g., +1 for interstitials, -1 for vacancies) and $\mu_i$ is the chemical potential of the species $i$. The defect formation energy depends on the chemical equilibrium condition through the last term, which measures the chemical work associated with the exchange of an atomic species between the host compound and the chemical reservoir. The chemical potential of species $i$ is equal in both the host compound and the chemical reservoir, as dictated by the condition of chemical equilibrium. %For instance, if the AB compound has reached an equilibrium involving the exchange of A atoms to form A-vacancies with an A-rich phase, such as a pure A phase, the chemical potential is determined by the coexistence condition of A and AB. 

We relate the grand-canonical formation energy from equation \ref{eq:defect_orig} to the DEF end-member Gibbs energy, ${}^0G_{\text{end}}$, from equation \ref{eq:GibbsDEF}, inspired by Anand \textit{et al.}'s graphical representation of defect formation energy on the convex hull \cite{Anand2021Visualizing}.
They showed that projecting the defective compound formation energy relative to the (extended) convex hull onto its constituent species gives the defect formation energy of equation \ref{eq:defect_orig}. 
Their derivation applies to any chemical equilibrium condition, such as the coexistence of defective \AB with A-vacancy and pure A. In the CALPHAD framework, equilibrium is determined by common tangents on the convex hull. Therefore, the chemical potential of A-vacancy formation in a binary \AB compound (without other defects) is derived from the B-rich condition, indicated by the common tangent at the defective compound's composition.  In section \ref{sec:convexdefect}, we show that when the 
\textit{absolute defect energy} is projected onto the convex hull's relevant endpoint and adjusted to a unified reference energy, it becomes independent of the chemical potential. This energy, independent of chemical potential, directly relates to the DEF end-member Gibbs energy.
In sections \ref{sec:2sub} and \ref{sec:3sub}, we demonstrate the construction of DEF for various defective compounds with neutral point defects. 
\subsection{From defect formation energies on the convex hull to DEF end-member}\label{sec:convexdefect}
\begin{figure*}[tp]
    \centering
    \includegraphics[width=0.75\textwidth]{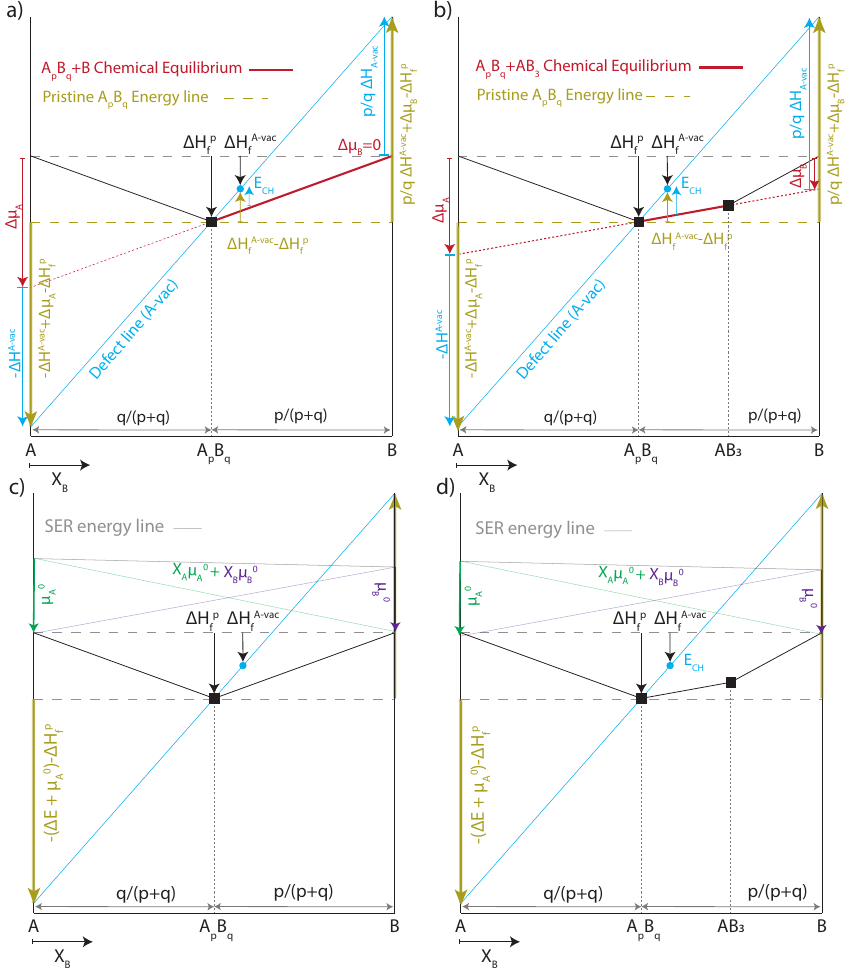}
    \caption{Formation energy per atom convex hull. a,b) Graphical representation of an A-vacancy formation energy and chemical potential change for model (a)A-\AB-B and (b) A-\AB-\ce{AB3}-B convex hulls. The equilibrium chemical condition is determined by the common tangent line at the defective compound composition, \AB+B coexistence line for (a) and \AB+\ce{AB3} coexistence line for (b). The formation energy of the defect-free \AB and defective \AB are shown by the black square and blue circle, respectively. $E_{\text{CH}}$ denotes the convex-hull distance, defined as the distance of the defective structure above the convex hull (see Ref.\cite{Anand2021Visualizing}). The common tangent lines determining the chemical potential condition are shown by bold red lines. Defect lines connecting the defective and pristine formation energies are shown by blue lines. The projection of $E_{\text{CH}}$ and common tangent line onto the A-end give the grand-canonical defect formation energy and chemical potential of A, respectively. The absolute defect energy, $\hd-\hp$, and its projections on the A-end and B-end are shown by yellow arrows. c,d) Graphical representation of adjusting the projected absolute defect energies to the common SER reference line for (c) A-\AB-B  and (d) A-\AB-\ce{AB3}-B convex hulls. The projected absolute energies and the adjustment energies are all independent of the convex hull shape and, therefore, the equilibrium chemical conditions determined by it.}
    \label{fig:comm_ref_line}
\end{figure*}

In this section, we convert the defect formation energy, projected onto the convex hull endpoint, to the DEF Gibbs energy of defective end members. Anand \textit{et al.} illustrated how to obtain the grand-canonical defect formation energy, $\Delta H_{\text{d}}$ from equation \ref{eq:defect_orig}, by projecting the defective structure's formation energy distance relative to the (extended) convex hull onto the relevant endpoint \cite{Anand2021Visualizing}(see Figure \ref{fig:comm_ref_line}(a) and (b)). Here, instead of a general chemical equilibrium condition, we impose the chemical equilibrium condition from the convex hull's common tangent so that no extended convex hull is considered. Also, we project the absolute defect energy instead of the convex hull distance and unify all energies, including formation energies of defective and pristine compounds and the absolute defect energy using a common reference energy. This unification is crucial for mapping defect formation energies onto DEF end members, showing that the DEF defective end-member Gibbs energy is directly related to a chemical-potential-independent absolute defect energy. 

Considering \AB with p + q atoms in its primitive cell, the formation energies of the pristine and defective compounds ($\hp$ and $\hd$) are defined with reference to their pure states. We reformulate the defect formation energy of equation \ref{eq:defect_orig} in terms of $\hp$ and $\hd$. 
If the total energies of the defective ($E_{\text{def}}$) and pristine ($E_{\text{pristine}}$) structures are calculated for a supercell with $l$ times more atoms than the primitive cell , the formation energies (per atom) of the defective and pristine structures are defined as 
\begin{align}\label{eq:formationenergies}
\begin{split}
    \Delta H^\text{d}_{\text{f}}  &= \frac{E_{\text{def}} - lp \mu^0_A -lq \mu^0_B - \sum \Delta N_i \mu^0_i}{l(p+q)+\sum \Delta N_i}  \\
    \Delta H^\text{p}_{\text{f}}  &= \frac{E_{\text{pristine}} - lp \mu^0_A -lq \mu^0_B }{l(p+q)}  \\
\end{split}
\end{align}
where $\mu^0_A$ and $\mu^0_B$ are the chemical potential (reference energy) of elements A and B in their standard state. $\mu^0_i$ is the reference chemical potential of species $i$ added or removed to form the defect, for example, A for an A interstitial in \AB. Rewriting the defect formation energy of equation \ref{eq:defect_orig} in terms of the formation energies of the defective and pristine structures yields the following equation:
\begin{align}\label{eq:general_defect}
\begin{split}
   \Delta H_{\text{d}}  & = \left(E_{\text{def}}- \sum_i\Delta N_i \mu_i\right) - E_{\text{pristine}} \\
   = &  \left( l(p+q) + \sum \Delta N_i \right) \left[\hd -\hp\right] \\ &+ \sum \Delta N_i\hp  - \sum \Delta N_i \Delta \mu_i
   \end{split}
\end{align}
Anand \textit{et al.} used a different formulation of equation \ref{eq:general_defect} (Eq. 10 in Ref. \cite{Anand2021Visualizing}) to show that the projection of the convex hull distance of a defective compound, $E_{\text{CH}}$ in Figure \ref{fig:comm_ref_line}, to the corresponding end on the convex hull ($i$ end) is equal to the defect formation energy $\Delta H_{\text{d}}$. This $\Delta H_{\text{d}}$ is chemical potential dependent. The equilibrium chemical potential is determined by the common tangent at the defective phase composition. As shown in Figure \ref{fig:comm_ref_line} (a) and (b) changes in the convex hull and the resulting common tangent affect both $\Delta H_{\text{d}}$ and $\Delta \mu_i$. However, DEF end-member Gibbs energies ${}{}^0G_{\text{end}}$ must be chemical-potential-independent and well-defined regardless of Gibbs energy changes in competing phases. Therefore, instead of projecting $E_{\text{CH}}$, which varies with the environment's chemical potential, we project the chemical-potential independent value $\hd-\hp$, we call the absolute defect energy (see Figure \ref{fig:comm_ref_line} (a,b)). Note that $\hd-\hp$ measures the absolute defect energy of the defective compound relative to the pristine compound, setting the defect-free \AB{} as the reference state, unlike $E_{\text{CH}}$, which measures the distance between the defective compound and the chemical potential energy line (convex hull common tangent). By rearranging equation \ref{eq:general_defect}, we project $\hd-\hp$ onto its corresponding endpoint as follows
\begin{align}\label{eq:general_defect_ref_pristine}
\begin{split}
   &\Delta H_{\text{d}} + \sum \Delta N_i \Delta \mu_i - \sum \Delta N_i\hp \\
   &=  \overbrace{\left( l(p+q) + \sum \Delta N_i \right)}^{f_i^d} \left[\hd -\hp\right] 
   \end{split}
\end{align}
As shown in Ref.\cite{Anand2021Visualizing}, $f_i^d$ is the projection factor that projects $E_{\text{CH}}$, and similarly $\hd-\hp$, to the $i$-end on the convex hull. Figure \ref{fig:comm_ref_line} illustrates this projection for a defective \AB with an A-vacancy. According to equation \ref{eq:general_defect_ref_pristine}, the projection on the A-end is $-(\Delta H_{\text{A-vac}} - \Delta \mu_A +\hp)$ with $\Delta N_A=-1$. The negative sign in front of the parenthesis arises from the A-vacancy projection factor, $f_i^{\text{\tiny A-vac}}$, which inverts the absolute defect energy direction when projected onto the opposite side relative to the \AB composition (see Figure \ref{fig:comm_ref_line}). As we elaborate later in section \ref{sec:2sub}, the A-vacancy defective compound corresponds to Vac:B DEF end-member and must be projected to the B-end. Accordingly, the B-end projection is $\frac{p}{q}(\Delta H_{\text{A-vac}}-\Delta \mu_A+\hp) = \frac{p}{q}(\Delta H_{\text{A-vac}})+\Delta \mu_B-\hp$ (see Figure \ref{fig:comm_ref_line}(a,b)). Note that $\Delta \mu_A$ and $\Delta \mu_B$ are related according to $\frac{p}{p+q} \Delta \mu_A +\frac{q}{p+q} \Delta \mu_B = \hp$. 
%The DEF end-member Gibbs energy associated with A-vacancy is Vac:B, which is the projection on the B-end. In CALPHAD, 
Although $\Delta H_{\text{d}}$ and $\sum \Delta N_i \Delta \mu_i$ (left hand side of equation \ref{eq:general_defect_ref_pristine}) vary with chemical potential, their sum does not. Replacing $\Delta H_{\text{d}}$ from equation \ref{eq:defect_orig} into equation \ref{eq:general_defect_ref_pristine} results in $\Delta H_{\text{d}} + \sum \Delta N_i \Delta \mu_i = E_{\text{def}} - E_{\text{pristine}} -\sum \Delta N_i \mu_i^0 = \Delta E_{\text{def}} -\sum \Delta N_i \mu_i^0$, a value that is independent of the chemical potential of the coexistence condition. $\Delta E_{\text{def}}$ is the difference between the total energy of the defective and pristine structures, commonly obtained through DFT total energy calculations. $\Delta E_{\text{def}}$ is chemical-potential-independent and needs to be calculated only once for a defective structure, regardless of its equilibrium conditions or coexistence with competing phases. 
\begin{figure*}[tph!]
    \centering
    \includegraphics[width=0.75\textwidth]{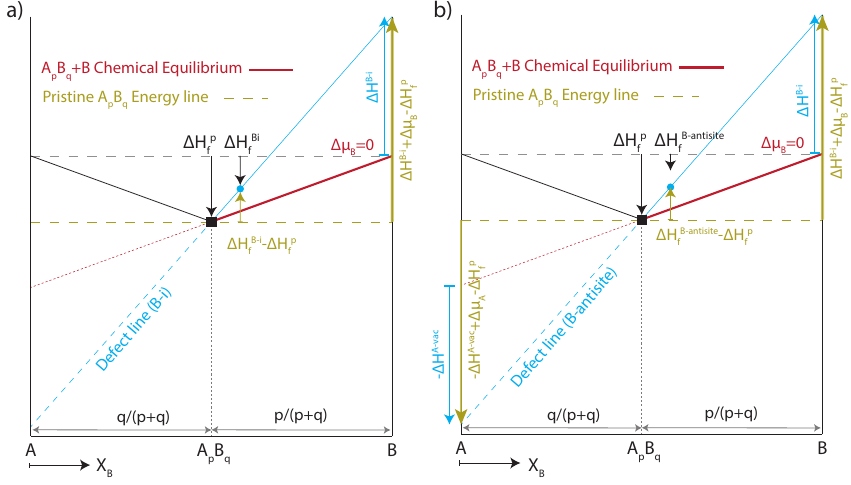}
    \caption{Formation energy per atom convex hull. Graphical representation of the absolute defect energy for a) B-interstitial and b) B-antisite in an A-\AB-B convex hull.}
    \label{fig:antisite}
\end{figure*}

Aside from projecting to the relevant end-point, we need to unify the reference energy between $\hp$, $\hd$, and the projection of equation \ref{eq:general_defect_ref_pristine}. The reference state for $\hp$ and $\hd$ is the stable element reference (SER), as shown in Figure \ref{fig:comm_ref_line}(c,d)) and equation \ref{eq:formationenergies}. Accordingly, the projection of $\hd-\hp$ (equation \ref{eq:general_defect_ref_pristine}), referenced to the pristine compound, must be adjusted to the SER. %the defective DEF end-members' Gibbs energies, which were projected with respect to the pristine phase, must be adjusted to the SER. 
Additionally, we must convert the projection on the convex hull from per atom values to per formula unit values to describe the DEF end-member Gibbs energy (see equation \ref{eq:GibbsDEF}). These adjustments will convert the B-end projection of \AB with an A-vacancy to the DEF Vac:B end-member, \edit{denoted by ${}^0G_{\text{Vac:B}} $, according to} %the projected formation energy difference on the B-end of the convex hull to the DEF Vac:B end-member, as shown below. We provide more details on the connection between convex hull B-end and DEF Vac:B end-member in the following sections.  
\begin{align}\label{eq:projectiontoDEF}
\begin{split}
   {}^0G_{\text{Vac:B}} &=q[\overbrace{\frac{p}{q}(\underbrace{\Delta H_{\text{A-vac}}-\Delta \mu_A}_{\Delta E_{\text{A-vac}}+\mu_A^0} + \hp)}^{\text{\scriptsize{B-end projection of A-vac}}} + \overbrace{\hp + \mu_B^0}^{\text{\scriptsize{Adjustment to SER}}}] \\ 
   &= p\Delta E_{\text{A-vac}} + \overbrace{p\mu_A^0+q\mu_B^0+(p+q)\hp}^{{}{}^0G_{A:B}}
   \end{split}
\end{align}

The conversion approach between the absolute defect energy and DEF end-member Gibbs energy, detailed in this section, applies to any type of point defect. For example, a B-interstitial defect in \AB (i.e., $\Delta N_B = +1$) \edit{can be represented by} a three-sublattice model \ce{(A)_p(B)_q(Vac,B)_m}, \edit{where $p$ ,$q$ and $m$ are the stoichiometric coefficients of the sublattices}. The DEF end-member \edit{${}^0G_{\text{A:B:B}}$} corresponds to the B-end projection of the absolute defect energy $\hd-\hp$, which according to equation \ref{eq:general_defect_ref_pristine} is $\Delta H_{\text{\ce{B^i}}} + \Delta \mu_B -\hp = \Delta E_{\text{\ce{B^i}}} -\mu_B^0 -\hp$ (see Figure \ref{fig:antisite}(a)). Adjusting to SER and per formula unit results in $m(\Delta E_{\text{\ce{B^i}}} -\mu_B^0-\hp + \hp + \mu_B^0) = m\Delta E_{\text{\ce{B^i}}}$.
\begin{figure}[bp]
    \centering
    \includegraphics[width=0.4\textwidth]{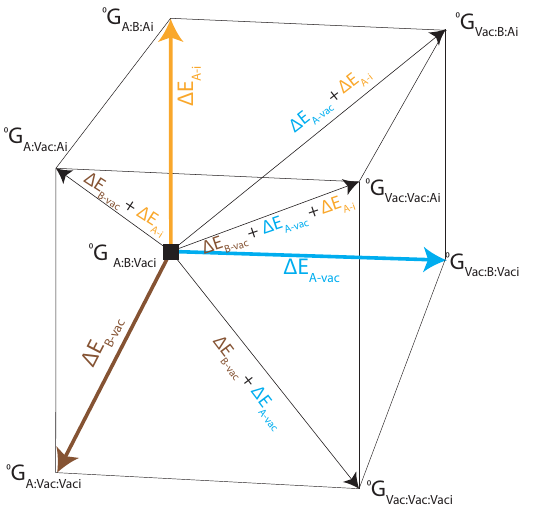}
    \caption{\textbf{Superposition principle underlying DEF}. Superposition of the projected absolute defect energies to obtain DEF end-member Gibbs energies, illustrated for (A,Vac)(B,Vac)(Vac\textsuperscript{i},A\textsuperscript{i}). As detailed in section \ref{sec:3sub}, the axes of the constitutional cube are formed by the A-vacancy, B-vacancy, and A-interstitial defect lines, shown by the blue, brown, and orange lines, respectively. The Gibbs energy of the DEF end-members on the constitutional cube is obtained from the superposition of the projected absolute defect energies, illustrated by defect basis vectors, onto the pristine end-member Gibbs energy, illustrated by the black square point.}
    \label{fig:superposition}
\end{figure}

We construct all defective DEF end-members with respect to the defect-free compound \AB, ensuring that the mapping between the convex hull composition and DEF constitution space is physical in the dilute defect range. As detailed in sections \ref{sec:2sub} and \ref{sec:3sub}, the pristine compound forms the origin of the DEF constitution space. In other words, the projected absolute defect energies measure the energy of defective end-members with respect to the pristine compound. Accordingly, the Gibbs energy of the pristine end-member must be added to all defective end-members. Therefore, for B-interstitial defects in \AB, ${}^0G_{\text{A:B:B\textsuperscript{i}}} = {}^0G_{\text{A:B:Vac}} +m\Delta E_{\text{\ce{B^i}}}$. The case of vacancies is slightly different. For example, for a defective \AB with A-vacancy defects, the defective end-member can be described as ${}^0G_{\text{Vac:B}} = {}^0G_{\text{A:B}} - {}^0G_{\text{A:Vac}} = {}^0G_{\text{A:B}} -{}^0G_{\text{A:B}} +{}^0G_{\text{Vac:B}}$. Note that the second equation gives the projection on the B-end; thus, the first two terms cancel out. The first equation gives the flipped projection on A-end for A-vacancy detailed in Ref.\cite{Anand2021Visualizing} (see Figure \ref{fig:comm_ref_line}). 

%antisite
For a B-antisite defect, the defective end-member energy can be described as
\begin{align}\label{eq:antisite}
\begin{split}
        {}^0G_{\text{B:B}} &= {}^0G_{\text{A:B}} -  {}^0G_{\text{A:Vac}} + {}^0G_{\text{B:Vac}} \\
        &= \overbrace{{}^0G_{\text{A:B}}}^{\text{\scriptsize{DEF origin}}} +\overbrace{(-{}^0G_{\text{A:B}} + {}^0G_{\text{Vac:B}})}^{\text{\scriptsize{B-end projection of A-vac}}} \\
        &+ \overbrace{{}^0G_{\text{B:Vac}}}^{\text{\scriptsize{B-end projection of B-interstitial}}} \\
       &= {}^0G_{\text{A:B}} + p\Delta E_{\text{\ce{Vac_A}}} + p\Delta E_{\text{\ce{B^i}}}
\end{split}
\end{align}
Note that ${}^0G_{\text{Vac:B}}$ indicates the end-member containing A-vacancy, while ${}^0G_{\text{B:Vac}}$ shows the end-member containing B-interstitial. Therefore, a B-antisite defect can be considered a combination of these two individual defects (see Figure \ref{fig:antisite}(b)).

%multiple defects 
For DEF end-members containing multiple defects, the superposition principle can be directly applied to the projected absolute defect energies. For example, for the Vac:Vac
end-member, the absolute defect energy projections on the B-end and A-end are simply added so that:
\begin{align}\label{eq:vacvac}
\begin{split}
        {}{}^0G_{\text{Vac:Vac}} &= \overbrace{{}{}^0G_{\text{A:B}}}^{\text{\scriptsize{DEF origin}}} + \overbrace{(-{}^0G_{\text{A:B}} + {}{}^0G_{\text{Vac:B}})}^{\text{\scriptsize{B-end projection of A-vac}}} \\ &+ \overbrace{(-{}{}^0G_{\text{A:B}} + {}{}^0G_{\text{A:Vac}})}^{\text{\scriptsize{A-end projection of B-vac}}} \\
        &= {}{}^0G_{\text{A:B}} + p\Delta E_{\text{A-vac}} + q\Delta E_{\text{B-vac}}
        \end{split}
\end{align}
Similarly, the Vac:Vac:A\textsuperscript{i} in the three-sublattice model can be described as (see Figure \ref{fig:superposition}) 
\begin{align}\label{eq:vacvacint}
\begin{split}
         &{}^0G_{\text{Vac:Vac:A\textsuperscript{i}}} = \overbrace{{}{}^0G_{\text{A:B}}}^{\text{\scriptsize{DEF origin}}} + \overbrace{(-{}^0G_{\text{A:B}} + {}^0G_{\text{Vac:B}})}^{\text{\scriptsize{B-end projection of A-vac}}}  \\ &+ \overbrace{(-{}^0G_{\text{A:B}} + {}{}^0G_{\text{A:Vac}})}^{\text{\scriptsize{A-end projection of B-vac}}} + \overbrace{{}{}^0G_{\text{Vac:Vac:A\textsuperscript{i}}}}^{\text{\scriptsize{A-end projection of B-interstitial}}} \\
         & = {}{}^0G_{\text{A:B}} + p\Delta E_{\text{A-vac}} + q\Delta E_{\text{B-vac}} + m\Delta E_{\text{\ce{B^i}}}
\end{split}
\end{align}

The general derivation of the DEF end-member Gibbs energy aligns with our earlier derivation of DEF in Ref.\cite{Adekoya2024} as shown in Appendix \ref{sec:equivalence} for an example vacancy defect. 

\begin{figure}[tbph]
    \centering
    \includegraphics[width=0.5\textwidth]{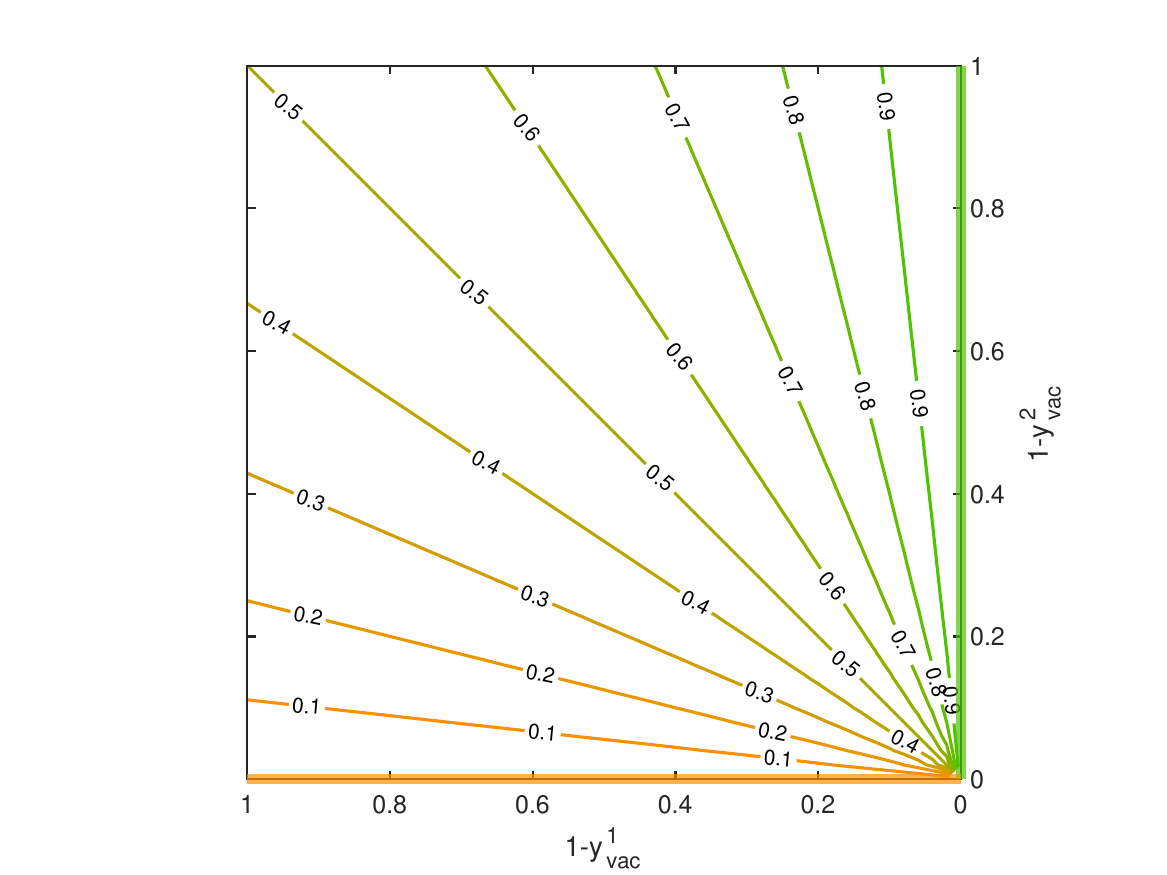}
    \caption{Mapping of the constitutional square for the (A,Vac)(B,Vac) DEF model, formed by ($1-\yt$)-($1-\yu$) axes,  into the mole fraction composition $X_B$. $X_B=1$ and $X_B=0$ lines are shown by thick green and orange lines, respectively.}
    \label{fig:map}
\end{figure}
\subsection{AB compound with dilute A-vacancy and B-vacancy}\label{sec:2sub}
\begin{figure*}[tpbh]
    \centering
    \includegraphics[width=\textwidth]{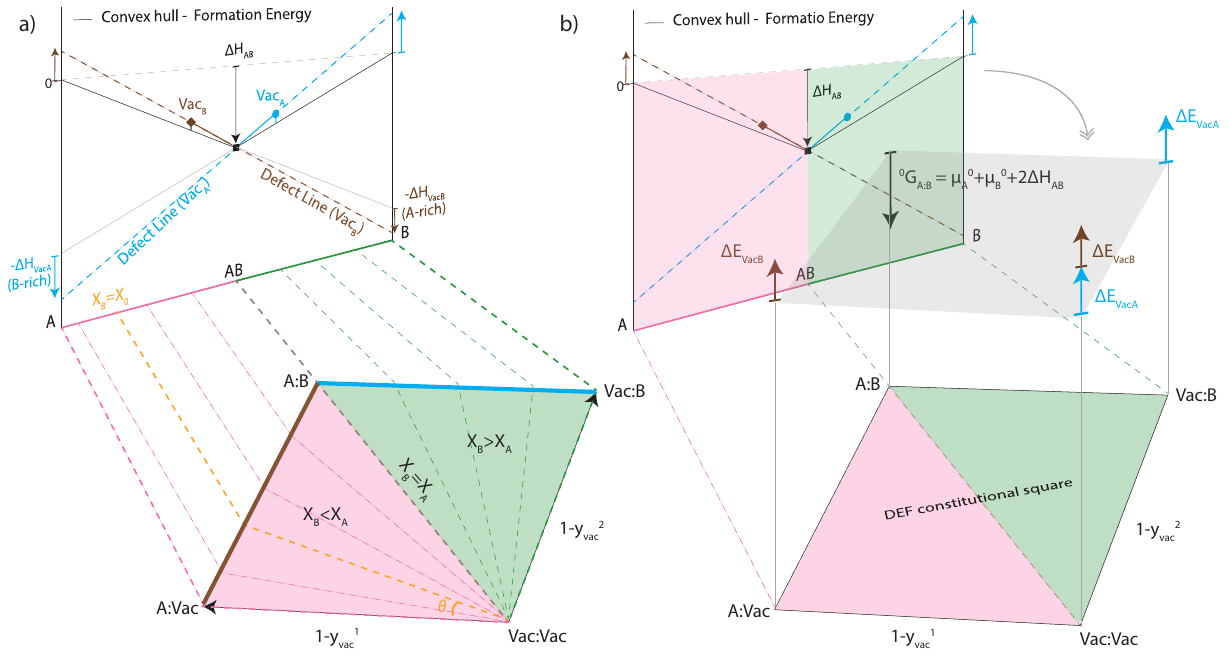}
    \caption{Mapping the composition and formation energy from a model convex hull (A-AB-B) to the constitutional square of a two-sublattice DEF model for A-vacancy and B-vacancy, (A, Vac)(B, Vac).Formation energies of defect-free AB ($\Delta H_{AB}$), defective AB with A-vacancy, and defective AB with B-vacancy are represented by the black square, blue circle, and brown diamond, respectively. The defect lines for A-vacancy and B-vacancy are indicated by the blue and brown lines, respectively. (a) Graphic representation of the mapping between $X_B$ on the convex hull and the constitutional square of the two-sublattice model. Physically relevant $X_B$ values for A-vacancy lie on the AB-B side (B-rich), colored green, and for B-vacancy lie on the A-AB side (A-rich), colored pink. The corresponding areas on the constitutional square for A-vacancy and B-vacancy are colored green and pink, respectively. (b) Graphic representation of the mapping between absolute defect energies, $\Delta E_{\text{VacA}}$ and $\Delta E_{\text{VacB}}$, and DEF end-members' Gibbs energies. These defect energies measure the Gibbs energy of defective end-members relative to the pristine end-member A:B, as detailed in Section \ref{sec:convexdefect}.} %result from unifying the reference energy lines on the convex hull, shown as the common energy reference line (see details in Section \ref{sec:convexdefect}). %The convex hull distance $E_{\text{CH}}$ for A-vacancy and B-vacancy are shown by solid vertical blue and brown lines. The chemical potentials of A and B for formation of A-vacancy and B-vacancy are determined from the intercepts of the common tangents on the A and B end axes, respectively. A-vacancy and B-vacancy formation energies, $\Delta H_{\text{VacA}}$ and $\Delta H_{\text{VacB}}$, are illustrated as the intercept between the defect line and the common tangent line on the A and B end points, respectively. 
    \label{fig:main-2sub}
\end{figure*}

In this section, we first derive the mapping between the composition space in a standard convex hull, $X$, and the constitutional space defined by the site fractions of constituents in the DEF sublattice model, $Y$, for the AB binary compound with dilute neutral vacancies, described by a two-sublattice model as (A,Vac)(B,Vac). The constitutional square consists of two axes, $\yt$ and $\yu$, denoting the site fractions of vacancies in the first and second sublattices, respectively. 
Within each sublattice, the total site fractions of distinct constituents equal one. Hence the site fractions of A and B, $\yA$ and $\yB$, follow $1-\yt = \yA$ and $1-\yu = \yB$. Therefore, the constitutional space of the two-sublattice model has two degrees of freedom. On the other hand, the composition of mole fractions of components in the $X$ space for a binary system has one degree of freedom. The $X$-composition of each component A or B per formula unit (or per mole of formula unit) can be related to the DEF site fractions using the following equations (see equation \ref{eq:XtoY}), 
\begin{align}\label{eq:map}
\begin{split}
X_A & = \frac{\yA}{\yA+\yB} = \frac{1-\yt}{(1-\yt)+(1-\yu)} \\
X_B &= \frac{\yB}{\yA+\yB}  =  \frac{1-\yu}{(1-\yt)+(1-\yu)} \\
\end{split}
\end{align}
which provides the functional mapping between $X$ and $Y$. As shown in Figure \ref{fig:main-2sub}(a), the mapping expands the 1-dimensional $X$-space, associated with two defects lines, A-vacancy, and B-vacancy, into the 2-dimensional constitutional square $Y$. In other words, the non-parallel A-vacancy and B-vacancy lines form the orthogonal basis vectors for the Y-space. %For sites fractions of $\yt=1,\yu=1$, the mapping of equation \ref{eq:map} suggests that there exists no physical correspondence in the $X$-space, which results for $X_A$ and $X_B$ to be ill-defined (more details below). 

As shown in Figure \ref{fig:map}, each line on the ($1-\yt$)-($1-\yu$) constitutional square maps into one point on $X_B$. The line at $\yt=1$ corresponds to $X_B=1$ and the line at $\yu=1$ corresponds to $X_B=0$. The diagonal line at $\yt=\yu$ corresponds to $X_B=\frac{1}{2}$. On these three lines exist 4 special points, constituting the end-members at $\yt=1,\yu=0$, corresponding to the Vac:B end-member, $\yt=0,\yu=1$, corresponding to the A:Vac end-member, $\yt=0,\yu=0$, corresponding to the A:B end-member, and $\yt=1,\yu=1$, corresponding to the Vac:Vac end-member (see Figure \ref{fig:main-2sub} and \ref{fig:map}). Note that at $\yt=1,\yu=1$, $X_B$ is ill-defined and this point sits at the intersection of all constant $X_B$ contour lines (see Figure \ref{fig:map}). %Therefore, one can conclude that the Vac:Vac end-member can be associated to any arbitrary $X_B$ value. %However, as detailed below, we identify two points on $X_B$ that have a physical correspondence to the Vac:Vac end-member.

%mapping X-Y
As illustrated in Figure \ref{fig:main-2sub}(a), the AB-B segment of $X_B$, which includes the A-vacancy defect line extending from pristine AB to B (AB$\rightarrow$B) corresponds to the top-right triangular region within the $Y$-square. This region includes any combination of $\yt$ and $\yu$ as long as $\yt \geq \yu$. However, only the top edge of the $Y$-square (or line $\yu=0$) corresponds to the A-vacancy defect line on the convex hull as it physically connects to an A-vacancy, forming \ce{A_{1-$\delta$}B} in dilute ranges, without B-vacancy defects. Therefore, we map the A-vacancy line on the AB$\rightarrow$B segment of $X_B$ to the $\yu=0$ line in the constitutional square, as shown by the blue line in Figure \ref{fig:main-2sub}(a). Similarly, the A-AB segment of $X_B$, which includes the B-vacancy defect line extending from pristine AB to A, corresponds to the bottom-left triangular region within the $Y$-square. The only line on this triangle with a one-on-one mapping to B-vacancy without an A-vacancy is $\yt=0$, forming the second basis vector of the $Y$ constitutional square (\edit{brown} line in Figure \ref{fig:main-2sub}(a)). The vertices of the constitutional square, $Y$, constitute the DEF end-members, for which there is a one-on-one correspondence to $X$ according to equation \ref{eq:map}; A:B defined at $\yt=\edit{0},\yu=0$ (the origin of the $Y$-space) maps to $X_B=\frac{1}{2}$ or AB compound composition, Vac:B defined at $\yt=1,\yu=0$ maps to $X_B=1$ or B-end of the convex hull, and A:Vac defined at $\yt=0,\yu=1$ maps to $X_B=0$ or A-end of the convex hull. $X_B$ for Vac:Vac ($\yt=1,\yu=1$) is ill-defined. However, the Vac:Vac point in $Y$ sits at the intersections of all $X$-constant contours (or level lines) (see Figure \ref{fig:map}) and, therefore, can be arbitrarily defined as the summation of any $X$ point. We define Vac:Vac as the sum of A:B, A:Vac and Vac:B which aligns with the superposition principle in the context of dilute defects (see equation \ref{eq:vacvac}).

As the mapping between $X_B$ and DEF end-members is established, we can relate DEF end-member Gibbs energies (per formula unit for AB) to defect absolute energies according to the procedure detailed in Section \ref{sec:convexdefect}, and shown in Figure \ref{fig:main-2sub}(b) %the DEF end-member Gibbs energies (per formula unit for AB) are related to the chemical-potential-independent defect formation energies (per atom), as projected on the A-end and B-end of the convex hull and unified to a common reference state, as the following: 

{\allowdisplaybreaks
\begin{align}\label{eq:DEFmapping2}
%\begin{split}
&{}^0G_{\text{A:B}}  = \mu_A^0 + \mu_B^0+ 2\Delta H_{\text{f}}^{\text{AB}} \nonumber\\
    & {}^0G_{\text{A:Vac}}  = {}^0G_{\text{A:B}} + \Delta E_{\text{B-vacancy}} \nonumber \\ %+ \mu_A^0 + \mu_B^0+ 2\Delta H_{\text{f}}^{\text{AB}} 
& {}^0G_{\text{Vac:B}}  = {}^0G_{\text{A:B}} + \Delta E_{\text{A-vacancy}} \nonumber\\ %+ \mu_A^0 + \mu_B^0+ 2 \Delta H_{\text{f}}^{\text{AB}}\\
& {}^0G_{\text{Vac:Vac}}  = {}^0G_{\text{A:B}} + \Delta E_{\text{B-vacancy}} + \Delta E_{\text{A-vacancy}}  %+ 2\mu_A^0 + 2\mu_B^0+ 4\Delta H_{\text{f}}^{\text{AB}}\\
%\end{split}
\end{align}
}

The DEF end-members Gibbs energies for a general binary compound \AB is related to the absolute defect energies by 
\begin{align}\label{eq:DEFmapping2_pq}
\begin{split}
&{}^0G_{\text{A:B}}  = p\mu_A^0 + q\mu_B^0+ (p+q)\Delta H_{\text{f}}^{\text{AB}} \\
    & {}^0G_{\text{A:Vac}}  =  {}^0G_{\text{A:B}}  + q\Delta E_{\text{B-vacancy}} \\
    & {}^0G_{\text{Vac:B}}  = {}^0G_{\text{A:B}} + p \Delta E_{\text{A-vacancy}}\\
    & {}^0G_{\text{Vac:Vac}}  = {}^0G_{\text{A:B}}  +  p\Delta E_{\text{A-vacancy}} +q \Delta E_{\text{B-vacancy}} \\
    \end{split}
\end{align}

The above equations follow the superposition principle underlying the thermodynamic conditions for individual isolated defects at dilute ranges, aligned with the DEF construct for dilute defects. For example, the Gibbs energy of Vac:Vac end-member (an end-member containing multiple defects) is the sum of the formation energies of both A-vacancy and B-vacancy (see equation \ref{eq:vacvac}).  This represents a defective compound containing both A-vacancy and B-vacancy defects at dilute concentrations. 
The DEF end-member Gibbs energies from equation \ref{eq:DEFmapping2_pq} can be substituted into equation \ref{eq:GibbsDEF} to obtain the Gibbs energy per formula unit of the defective compound. 

\begin{figure*}[tp]
    \centering
    \includegraphics[width=\textwidth]{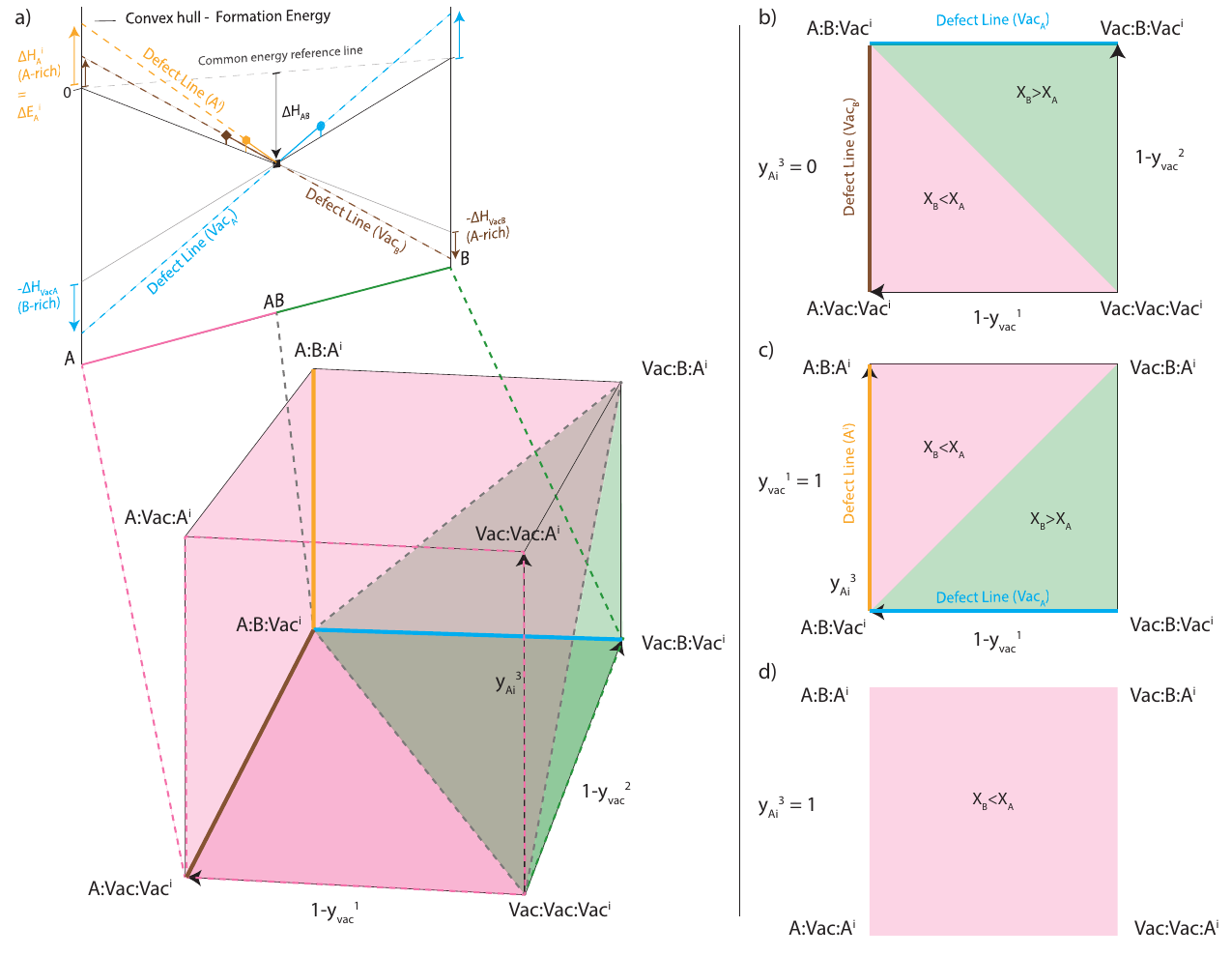}
    \caption{Mapping the composition and formation energy from a model convex hull (A-AB-B) to the constitutional cube of a three-sublattice DEF model for A-vacancy, B-vacancy, and A-interstitial. Formation energies of defect-free AB ($\Delta H_{AB}$) , defective AB with A-vacancy, defective AB with B-vacancy, and defective AB with A-interstitial are shown by the black square, blue circle, brown diamond, and orange hexagon, respectively. The defect lines for A-vacancy, B-vacancy, and A-interstitial are shown by the blue, brown, and orange lines, respectively. (a) Graphic representation of the mapping between $X_B$ on the convex hull and the constitutional cube of the three-sublattice model (or $Y$-space cube). The $X_B$ composition on the AB-B and A-AB sides maps into the green and pink regions of the $Y$-space cube, respectively. Lines on the constitutional cube that are physically relevant to the individual defect lines on the convex hull are shown by the respective colors of the defect lines, i.e., blue, brown, and orange for A-vacancy, B-vacancy, and A-interstitial. Projections of the constitutional cube into planes defined by (b) $\yi=0$, (c) $\yt=1$ and (d) $\yi=1$.} %$E_{\text{eCH}}$ for A-vacancy, B-vacancy, and A-interstitial are shown by solid vertical blue, brown, and orange lines. The chemical potentials of A and B for the formation of A-vacancy and A-interstitial and B-vacancy are determined from the intercepts of the common tangents on the A and B end axes, respectively. Defect formation energies, $\Delta H_{\text{def}}$, are illustrated as the intercept between the defect line and the common tangent line on the A end for A-vacancy and A-interstitial and on the B end for B-vacancy. Chemical-potential-independent defect energies, $\Delta E_{\text{def}}$, are also shown for each defect. 
    \label{fig:main-3sub}
\end{figure*}
\subsection{AB compound with dilute A-vacancy, B-vacancy, and A-interstitial}\label{sec:3sub}
Here, we describe the DEF for the binary AB compound with dilute ranges of neutral vacancies and a self-interstitial defect, described by a three-sublattice model as (A,Vac)(B,Vac)(Vac\textsuperscript{i},A\textsuperscript{i}). The first and second sublattices contain the A and B substitutional sites, hosting A, B, or vacancy defects, while the third sublattice contains interstitial sites, primarily occupied by vacancies and host secondary A-interstitial defects. The constitutional space (or $Y$-space) with 3 sublattices forms a cube with eight end-members (or eight vertices). As shown in Figure \ref{fig:main-3sub}(a), the constitutional cube is formed by three axes (or three degrees of freedom), $\yt$, $\yu$, and $\yi$, denoting the site fractions of A- and B- vacancies and A-interstitial in the first, second, and third sublattices, respectively. Each of the non-parallel defect lines (A-vacancy, B-vacancy, or A-interstitial), all intersecting at the AB point on the convex hull, form one axis of the constitutional cube in the $Y$-space (see Figure \ref{fig:main-3sub}(a)). The point of intersection, or the A:B:Vac\textsuperscript{i} serves as the origin for the $Y$-constitutional cube. %, and, as detailed below, its corresponding Gibbs energy is $2\Delta H_{\text{AB}}$.  

\begin{figure}[tbph]
    \centering
    \includegraphics[width=0.45\textwidth]{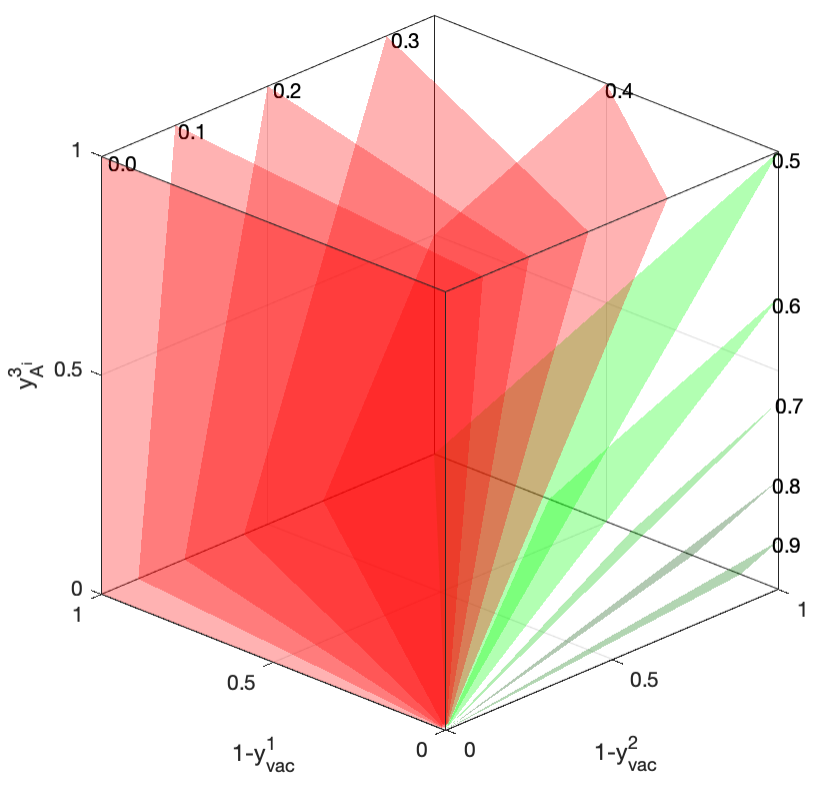}
    \caption{Mapping of the constitutional cube for the (A,Vac)(B,Vac)(Vac\textsuperscript{i},A\textsuperscript{i}) DEF model, formed by ($1-\yt$)-($1-\yu$)-$\yi$, into the mole fraction composition $X_B$. The contour level planes in the constitutional cube are labeled with corresponding $X_B$ values. Level planes for $X_B < 0.5$ and $X_B \geq 0.5$ are colored red and green, respectively.}
    \label{fig:map3}
\end{figure}

The $X$-composition of each component A or B per formula unit is related to the DEF site fractions using the following equations (see equation \ref{eq:XtoY}). 
\begin{align}\label{eq:map3}
\begin{split}
    X_A & = \frac{\yA + \yi}{\yA+\yB+\yi} = \frac{(1-\yt) + \yi}{(1-\yt)+(1-\yu)+\yi} \\
    X_B &= \frac{\yB}{\yA+\yB+\yi}  =  \frac{1-\yu}{(1-\yt)+(1-\yu)+\yi} \\
\end{split}
\end{align}
which provide the mapping between the $X$- and $Y$- spaces. The mapping expands the 1-dimensional $X$-space associated with three defects lines, A-vacancy, B-vacancy, and A-interstitial, into a 3-dimensional constitutional cube, as shown in Figure \ref{fig:main-3sub}(a). Figure \ref{fig:map3} illustrates the $X_B$-constant contours (or level planes) on the $Y$-space (or constitutional cube of the DEF model).  
As shown in Figure \ref{fig:map3}, each plane on the ($1-\yt$)-($1-\yu$)-$\yi$ constitutional cube maps into one point on $X_B$. The face plane at $\yu =1 $ corresponds to $X_B=0$, and the body diagonal plane, formed by $\yt=\yu$ and $1-\yu=\yi$ lines, corresponds to $X_B=\frac{1}{2}$. The plane corresponding to $X_B=1$ collapses into the line at $\yt=1$ and $\yi=0$. Note that the plane at $\yi =0$ replicates the composition square for the (A,Vac)(B,Vac) model detailed in section \ref{sec:2sub}. 

To map the defect formation energy from the convex hull to the corresponding end-member Gibbs energy, ${}^0G_{\text{end}}$ , on the constitutional cube, we first identify the physically relevant lines on the $Y$-cube to the $X_B$ values for each defect line. As shown in Figure \ref{fig:main-3sub}(a), both the B-vacancy and A-interstitial defect lines lie on the A-AB side of $X_B$ on the convex hull and, as expected by the $X$-$Y$ mapping of equation \ref{eq:map3}, their physically relevant $Y$-lines span over the $X_B < X_A$ region on the $Y$-cube (colored red in Figure \ref{fig:main-3sub}(a)). On the other hand, the A-vacancy defect line lies on the AB-B side of $X_B$, and thus its physically relevant $Y$-line maps into the $X_B > X_A$ region on the $Y$-cube (colored green in Figure \ref{fig:main-3sub}(a)). As shown in Figure\ref{fig:main-3sub}(a), the lines at $\yu=0$ and $\yi=0$, $\yt=0$ and $\yi=0$, and $\yu=0$ and $\yt=0$, corresponds to the physically relevant $Y$-value for A-vacancy, B-vacancy, and A-interstitial defect lines, respectively. As shown in Figure \ref{fig:main-3sub}(a), these are the basis vectors (axes) on the $Y$-constitutional cube, and each axis represents a single defect type at different concentrations in the dilute ranges of isolated defects. As shown in Figure \ref{fig:main-3sub}(a), the origin for the $Y$-constitutional cube, or the A:B:Vac\textsuperscript{i} end-member, corresponds to the defect-free AB compound with the formation energy of $\Delta H_{\text f}^{\text{AB}}$. Other end-members are obtained by extending the origin along the three axes of $Y$-cube, corresponding to A:B:A\textsuperscript{i} for A-interstitial axis, Vac:B:Vac\textsuperscript{i} for A-vacancy axis, and A:Vac:Vac\textsuperscript{i} for B-vacancy axis. %As shown in Figure \ref{fig:superposition}, we apply the superposition principle of end-member Gibbs energies to identify ${}^0G_{\text{end}}$ for these three end members as ${}^0G_{\text{A:B}} + \Delta E_{\text{A-interstitial}}$, ${}^0G_{\text{A:B}} + \Delta E_{\text{A-vacancy}}$, and ${}^0G_{\text{A:B}} + \Delta E_{\text{B-vacancy}}$, respectively. Note that $\Delta E_{\text{d}}$ denotes the chemical-potential-independent defect energies as defined in equation \ref{eq:formationchargedunitfied} (see details in Section \ref{sec:convexdefect}). The superposition principle of defect formation energies can also be applied to identify ${}^0G_{\text{end}}$ for other end-members as the following: 
The DEF end-member Gibbs energies for the AB binary compound with A-vacancy, B-vacancy, and A-interstitial defects are 
\begin{align}\label{eq:DEFmapping3}
%\begin{split}
&{}^0G_{\text{A:B:Vac\textsuperscript{i}}}  = \mu_A^0+\mu_B^0+2 \Delta H_{\text{f}}^{\text{AB}} \nonumber\\
    & {}^0G_{\text{A:Vac:Vac\textsuperscript{i}}}  = {}^0G_{\text{A:B:Vac\textsuperscript{i}}}+ \Delta E_{\text{B-vacancy}} \nonumber\\
    & {}^0G_{\text{Vac:B:Vac\textsuperscript{i}}}  ={}^0G_{\text{A:B:Vac\textsuperscript{i}}}+ \Delta E_{\text{A-vacancy}} \nonumber \\
    & {}^0G_{\text{Vac:Vac:Vac\textsuperscript{i}}}  = {}{}^0G_{\text{A:B:Vac\textsuperscript{i}}}+ \Delta E_{\text{B-vacancy}} + \Delta E_{\text{A-vacancy}} \nonumber \\
&{}^0G_{\text{A:B:A\textsuperscript{i}}}  = {}{}^0G_{\text{A:B:Vac\textsuperscript{i}}}+ \Delta E_{A^i}\nonumber \\
    & {}^0G_{\text{A:Vac:A\textsuperscript{i}}}  = {}^0G_{\text{A:B:Vac\textsuperscript{i}}} + \Delta E_{\text{B-vacancy}} + \Delta E_{A^i}\nonumber \\
    & {}^0G_{\text{Vac:B:A\textsuperscript{i}}}  ={}^0G_{\text{A:B:Vac\textsuperscript{i}}} + \Delta E_{\text{A-vacancy}}+ \Delta E_{A^i}\nonumber \\
    & {}^0G_{\text{Vac:Vac:A\textsuperscript{i}}}  = {}^0G_{\text{A:B:Vac\textsuperscript{i}}}  + \Delta E_{\text{B-vacancy}}\nonumber \\ &\hspace{2cm} + \Delta E_{\text{A-vacancy}} + \Delta E_{A^i} 
%\end{split}
\end{align}

For a general binary compound \AB with a DEF sublattie model of (A,Vac)$_p$(B,Vac)$_q$(Vac\textsuperscript{i},A\textsuperscript{i})$_m$, the end-member Gibbs energies are
{\allowdisplaybreaks
\begin{align}\label{eq:DEFmapping3_pq}
%\begin{split}
&{}^0G_{\text{A:B:Vac\textsuperscript{i}}}  = p\mu_A^0+q\mu_B^0+(p+q) \Delta H_{\text{f}}^{\text{AB}} \nonumber \\
    & {}^0G_{\text{A:Vac:Vac\textsuperscript{i}}}  = {}^0G_{\text{A:B:Vac\textsuperscript{i}}} + q\Delta E_{\text{B-vacancy}} \nonumber \\
    & {}^0G_{\text{Vac:B:Vac\textsuperscript{i}}}  ={}^0G_{\text{A:B:Vac\textsuperscript{i}}}+ p\Delta E_{\text{A-vacancy}} \nonumber \\
    & {}^0G_{\text{Vac:Vac:Vac\textsuperscript{i}}}  = {}^0G_{\text{A:B:Vac\textsuperscript{i}}} + q\Delta E_{\text{B-vacancy}} + p\Delta E_{\text{A-vacancy}} \nonumber \\
&{}^0G_{\text{A:B:A\textsuperscript{i}}}  = {}^0G_{\text{A:B:Vac\textsuperscript{i}}} + m\Delta E_{A^i} \nonumber \\
    & {}^0G_{\text{A:Vac:A\textsuperscript{i}}}  = {}^0G_{\text{A:B:Vac\textsuperscript{i}}} + q\Delta E_{\text{B-vacancy}} + m\Delta E_{A^i} \nonumber \\
    & {}^0G_{\text{Vac:B:A\textsuperscript{i}}}  ={}^0G_{\text{A:B:Vac\textsuperscript{i}}} + p\Delta E_{\text{A-vacancy}} + m\Delta E_{A^i}  \nonumber \\
    & {}^0G_{\text{Vac:Vac:A\textsuperscript{i}}}  = {}^0G_{\text{A:B:Vac\textsuperscript{i}}} + q\Delta E_{\text{B-vacancy}}  \nonumber  \\ &\hspace{2cm}  + p\Delta E_{\text{A-vacancy}} + m\Delta E_{A^i}
%\end{split}
\end{align}
}
A similar DEF model for the B-interstitial defect, (A,Vac)(B,Vac)(Vac\textsuperscript{i},B\textsuperscript{i}) is illustrated in Appendix Figure \ref{fig:Bi}. The general recipe for constructing a typical DEF model for neutral defects is provided in section \ref{sec:generalrecipe}. 
%end members Gibbs energy 
\section{Defect Energy Formalism for Charged Defects}\label{sec:DEFcharged}
For a compound with charged defects, the concentration of defects is controlled by the Fermi level through the charge neutrality condition
\begin{equation}\label{eq:chargenutrality}
    0 = n - p - \sum_{d,q} q_d c_d 
    %N^d_\text{site}\exp{-{\Delta H_{\text{def}}}/{kT}}    
\end{equation}
where $n$, $p$ and $c_d$ are the concentration of free electrons, holes, and defect $d$ with a net charge of $q_d$ (e.g., +2 for a doubly ionized donor with 2 electrons removed or -2 for an acceptor with 2 electrons added), respectively. $k$ and $T$ are the Boltzmann constant and temperature. $n$ and $p$ are a function of the Fermi level, while $c_d$ is a function of both the Fermi level and chemical potential. The concentration of electrons and holes for non-degenerate semiconductors, \edit{where the distance between the Fermi level and the top of the valence band/bottom of the conduction band is much larger than $kT$ }, follows the Boltzmann distribution \cite{Sze2006} %(i.e., $-\frac{E_c-E_f}{kT} < 0 $)
\begin{align}\label{eq:carrierconc}
\begin{split}
   n & = N_c \exp{\left(-\frac{E_c-E_f}{kT}\right)} \\
   %\int_{E_g}^{\infty} g(E) \frac{1}{1+\exp{\frac{(E-E_f)}{kT}}} dE \\
   p &= N_v \exp{\left(-\frac{E_f-E_v}{kT}\right)} 
   %\int_{-\infty}^{0} g(E) \frac{1}{1+\exp{\frac{(E_f-E)}{kT}}} dE \\
\end{split}
\end{align}
where $N_v$ ($N_c$) is the effective density of states in the valence (conduction) band, and $E_f$, $E_c$, and $E_v$ denote the Fermi level, the conduction band minimum, and the valance band maximum energies, respectively. The effective density of states (per volume) in the valence and conduction band are given by $N_v = 2\left(\frac{2\pi m^*_h kT}{h^2}\right)^{3/2}$ and $N_c = 2\left(\frac{2\pi m^*_e kT}{h^2}\right)^{3/2}$, where $h$ is the Plank's constant, $m^*_h$ and $m^*_e$ are the effective masses of holes and electrons at the valence band and conduction band edges, respectively. Multiplying $N_v$ and $N_c$ in the volume per formula unit of the compound, $V_0$, counts the effective densities per formula unit (e.g., see Ref.~\cite{Chen1998,Peters2019}), which we consider in this work. %$M_c$ is the number of equivalent minima in the conduction band. --> I dont think I need Mc for non-degenerate  
%denotes the electronic density of states and $E_g$ is the band gap of the defect-free compound. Energy values, $E$, are relative to the valence band maximum. 
%[\textcolor{red}{add the volume term from Chen 1998}] 
Note that according to equation \ref{eq:carrierconc}, the product of $n$ and $p$ is constant and independent of $E_f$, given by $np = N_c N_v \exp{\left(-\frac{E_g}{kT}\right)}$, where $E_g = E_c-E_v$ is the band gap of the defect-free compound. % (a.k.a. the mass-action law)

The concentration of defect $d$ in the dilute range, $c_d$, is given by the Arrhenius relation \cite{Freysoldt2014}
\begin{equation}
    c_d^q  = c_0 \exp{\left(-\frac{\Delta H_{\text{d}}^q}{kT}\right)}
\end{equation}
where $c_0$ denotes the concentration of possible defect sites in the host compound and $\Delta H_{\text{d}}^{\text{q}}$ denotes the grand-canonical formation energy for the defect $d$ with a net charge of $q$, which in addition to chemical potential depends on $E_f$ according to the following equation.
\begin{equation}\label{eq:formationcharged}
    \Delta H_d^q  =  E_{\text{def}}^q - E_{\text{pristine}} - \sum\Delta N_i \mu_i +qE_f 
\end{equation}
Here, $E_{\text{def}}^q$ and $E_{\text{pristine}}$ denote the energies of the defective and pristine structures, respectively, where the defective structure has a net charge of $q$. $\Delta N_i$ is the number of atoms of species $i$ added to or removed from the defective structure (e.g., +1 for interstitials, -1 for vacancies) and $\mu_i$ is the chemical potential of the species $i$. %Note that $E_{\text{pristine}}$ and $\mu_i$ are independent of $E_f$ while $E_{\text{def}}$ associates with the defected structure with a net charge of $q_d$ (see Figure \ref{fig:charge-Ef}). 

\begin{figure*}[tp]
    \centering
    \includegraphics[width=0.9\textwidth]{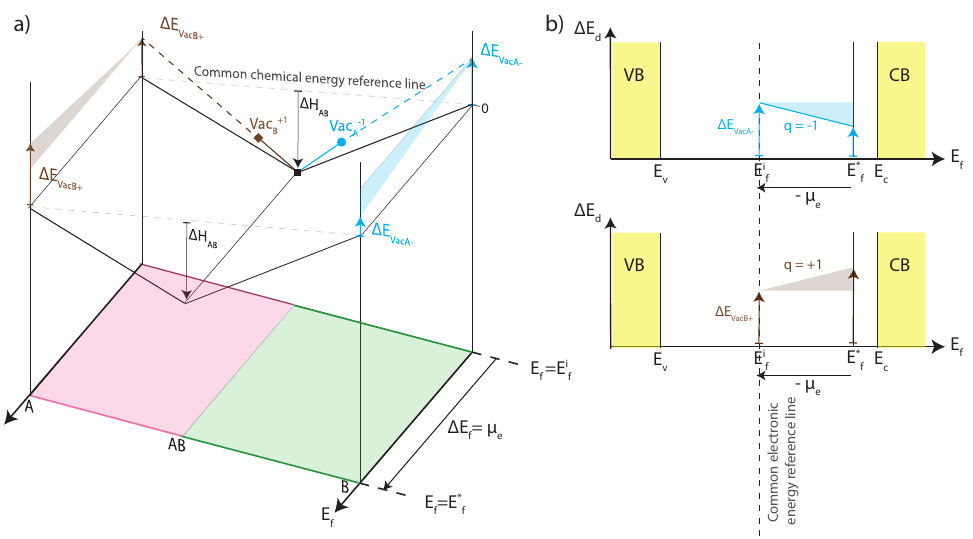}
    \caption{Formation energy convex hull in the composition-Fermi level space ($X-E_f$). a) Graphical illustration of a typical A-AB-B convex hull along the Fermi level and chemical composition. The A-AB-A convex hull, including the A+AB and AB+B common tangent lines, is independent of the Fermi level because it consists of charge neutral compounds, resulting in a flat extrusion of the A-AB-B hull along $E_f$. In contrast, the energy of a defected compound with a net charge $q$ depends on $E_f$ and varies with a slope of $q$ along $E_f$, as shown by shaded triangles. The absolute defect energy of defected AB ($\Delta E_{\text{d}}$) are shown for A-vacancy with $q=-1$ and B-vacancy with $q=+1$. b) Graphical representation of unifying the electronic energy reference for charge carriers and ionic defects. The convex hull at the equilibrium Fermi level ($E_f^*$) is shifted to the intrinsic Fermi level ($E_f^i$). The chemical-potential-independent absolute defect energy is shifted by $-q\mu_e$ (see equation \ref{eq:formationchargedunitfied}).}
    \label{fig:charge-Ef}
\end{figure*}
The equilibrium Fermi level, $E^*_f$, is uniquely determined by solving the charge neutrality condition of equation \ref{eq:chargenutrality} for $E_f$. As shown in Figure \ref{fig:charge-Ef}(a), the equilibrium Fermi level can be associated to a plane on the formation energy convex hull in the composition-Fermi level space ($X-E_f$). The formation energy of defect $d$ varies with $E_f$ with a slope of $q$ according to equation \ref{eq:formationcharged}. Therefore, the charge state of defect $d$ is implicitly determined by $E_f^*$, where different charge states of a given defect $d$ enter the charge neutrality condition of equation \ref{eq:chargenutrality}. The equilibrium Fermi level $E^*_f$ is a measure of the chemical potential of electrons at equilibrium, $\mu_e$ (i.e., $\mu_e=E^*_f$) \cite{Seebauer2009}, defined as the Gibbs energy required to add or remove an electron from the compound. 

%%%%%%
% standard electron chemical potential
%%%%%%
We define the standard chemical potential for electrons and holes, $\mu_e^0$ and $\mu_h^0$, respectively, for the Fermi level of the intrinsic case or a defect-free compound. Therefore, $\mu_e^0$ and $\mu_h^0$ represent the Gibbs energy change for creation of an electron and hole through the electron-hole pair reaction ($\text{e}\textsuperscript{-} + \text{h}\textsuperscript{+} \longleftrightarrow 0$) in the absence of any ionic defect. The electron-hole pair reaction implies that $\mu_e^0+\mu_h^0=0$ at equilibrium. The intrinsic Fermi level is simply determined through $n = p$ (see equation \ref{eq:chargenutrality}), resulting in $E_f^i=\frac{E_v+E_c}{2} - \frac{kT}{2} \ln(\frac{N_c}{N_v})$. We consider the intrinsic Fermi level as the reference electronic energy in the DEF framework because it corresponds to a defect-free compound with a net charge of zero. This choice for the reference electronic state is consistent with the choice for the reference chemical state corresponding to the defect-free compound (see section \ref{sec:convexdefect}). Setting $E_f^i=0$ results in $E_c = \frac{E_g}{2} + \frac{kT}{2} \ln\left(\frac{N_c}{N_v}\right)$ and $E_v = -\frac{E_g}{2} + \frac{kT}{2} \ln\left(\frac{N_c}{N_v}\right)$. %Replacing the resulting $E_c$ and $E_v$ in equation \ref{eq:carrierconc} gives the standard chemical potential for electrons and holes as $\mu_e^0 = $
The standard chemical potential of an electron is the Gibbs energy change due to adding an electron to the conduction band minimum, with the energy term equal to $\frac{E_g}{2} + \frac{kT}{2} \ln\left(\frac{N_c}{N_v}\right)$ and the entropy contribution term equal to $-kT\ln N_c$, corresponding to the selection of an electron site among $N_c$ available sites. Therefore,  $\mu_e^0= \frac{E_g}{2} + \frac{kT}{2} \ln\left(\frac{N_c}{N_v}\right) - kT \ln N_c$. Similarly, the standard chemical potential for a hole is equal to the energy for removing an electron from the valence band with an entropy term corresponding to selection of a candidate hole site among $N_v$ available sites, resulting in $\mu_h^0 = \frac{E_g}{2} - \frac{kT}{2} \ln\left(\frac{N_c}{N_v}\right) - kT\ln N_v$. 

For the general case of a defected compound (non-intrinsic, non-degenerate), the shift in the equilibrium Fermi level $E_f^*$ with respect to the intrinsic Fermi level $E_f^i$, $\Delta E_f$, measures the chemical potential of electrons ($\Delta E_f = \mu_e$) according to the following equation (rearrangement of equation \ref{eq:carrierconc})
{\allowdisplaybreaks
\begin{align}\label{eq:refrencemue}
  %\begin{split}
    & n  = N_c \exp{\left(-\frac{E_c-(E_f^i+\Delta E_f)}{kT}\right)} \nonumber \\ & = N_c \exp{\left(-\frac{(E_c-E_f^i)}{kT}\right)\exp\left(-\frac{\mu_e}{kT}\right)}  \nonumber \\
\rightarrow &  \mu_e  = \left[\frac{E_g}{2} + \frac{kT}{2} \ln\left(\frac{N_c}{N_v}\right) - kT\ln(Nc)\right] +kT \ln n \nonumber \\
\rightarrow &  \mu_e  = \mu_e^0 +kT \ln n
   %p &= N_v \exp{\left(-\frac{E_f-E_v}{kT}\right)} 
%\end{split}  
\end{align}
}
%Note that the concentration of free electrons in the general case is not the same as the concentration of holes. 
Similarly, the chemical potential of holes is given by ($\Delta E_f = -\mu_h$) 
\begin{align}\label{eq:refrencemue2}
  %\begin{split}
    & p  = N_v \exp{\left(-\frac{(E_f^i+\Delta E_f)-E_v}{kT}\right)} \nonumber \\ &= N_v \exp{\left(-\frac{(E_f^i-E_v)}{kT}\right)\exp\left(\frac{\mu_h}{kT}\right)} \nonumber \\
\rightarrow & \mu_h  = \left[\frac{E_g}{2} - \frac{kT}{2} \ln\left(\frac{N_c}{N_v}\right) - kT\ln(Nv)\right] +kT \ln p \nonumber \\
\rightarrow &  \mu_h  = \mu_h^0 +kT \ln p
   %p &= N_v \exp{\left(-\frac{E_f-E_v}{kT}\right)} 
%\end{split}  
\end{align}
Note that according to the above equation, $\mu_e + \mu_h =0$ is always satisfied, considering that $np = N_c N_v \exp{\left(-\frac{E_g}{kT}\right)}$. 

The DEF construction for defected phases with charged defects includes an auxiliary sublattice to host electronic constituents, including free electrons and holes. Including the auxiliary sublattice besides the regular atomic (or ionic) sublattices is essential for determining the concentration of charge carriers, as is usually desired in the thermodynamic description of semiconductors. Existing studies in the literature have used the CEF both with or without including the charge carrier sublattice. The charge carrier sublattice represents the electron reservoir and thus facilitates the exchange of electrons to or from the regular atomic sublattices in a grand-canonical description. The number of available sites (per formula unit of the compound) on the free electron or hole sublattice are $N_c$ and $N_v$, respectively. Therefore, one may consider a DEF model with separate sublattices for free electrons and holes such as (A,Vac\textsuperscript{-1})(B,A\textsuperscript{+1})(Vac,e\textsuperscript{-})$_{N_c}$(Vac,h\textsuperscript{+})$_{N_v}$. However, this representation makes the modeling complex as $N_c$ and $N_v$ can vary with compositions or by adding new chemical components to the system. Therefore, as recommended by other CEF studies \cite{Chen1996,Chen1998,HILLERT2001161}, we set the number of sites in the free electron and hole sublattices equal. Chen and Hillert \cite{Chen1996} compensate for the incorrect number of sites in the charge carrier sublattices by subtracting the terms $RT\ln N_c$ and $RT\ln N_v$ from the Gibbs formation energy of electrons and holes, respectively (see section 4 in Ref. \cite{Chen1996} for details). As shown in equation \ref{eq:refrencemue}, these correction terms have emerged in the derivation of $\mu_e^0$ and $\mu_h^0$, and are inherently embedded in the DEF formalism as detailed below. Therefore, the DEF model can be simplified to host free electrons and holes on a common sublattice as (A,Vac\textsuperscript{-1})(B,A\textsuperscript{+1})(Vac,e\textsuperscript{-},h\textsuperscript{+}). % to host the free electrons and holes on a common sublattice %(as detailed in the following section)

The Gibbs energy per formula unit of the compound with charged defects is defined similar to a compound with neutral defects consisting of the surface of reference energy, $G^{\text{s.r.}}$, and the ideal mixing term for each sublattice. The additional Gibbs energy terms due to the charge carrier sublattice can be formulated as the following (after section 4 in Ref. \cite{Chen1996})
\begin{align}
%\begin{split}
\label{eq:GibbsDEFcharge}
     G_{\text{fu}} =& G_{\text{fu}}^{\text{at}} + G_{\text{fu}}^{\text{ax}} \nonumber\\
      = & G_{\text{fu}}^{\text{at}} + \overbrace{ \sum\sum \yii\yjj}^{=1} [({}^0G_{\text{i:j:e\textsuperscript{-}}}- {}^0G_{\text{i:j:vac}}) \ye \nonumber \\
       & \hspace{1.5cm}+ ({}^0G_{\text{i:j:h\textsuperscript{+}}}- {}^0G_{\text{i:j:vac}}) \yh  ] \nonumber\\
      & + kT  \left( \ye \ln(\ye) + \yh \ln(\yh) + \yvac \ln(\yvac) \right) \nonumber\\
      =& G_{\text{fu}}^{\text{at}} + \ye \underbrace{\left[({}^0G_{\text{i:j:e\textsuperscript{-}}}- {}^0G_{\text{i:j:vac}}) + kT \ln (\ye) \right]}_{\mu_e}  \nonumber \\  & +  \yh \underbrace{\left[ ({}{}^0G_{\text{i:j:h\textsuperscript{+}}}- {}^0G_{\text{i:j:vac}}) 
      + kT \ln (\yh)\right]}_{\mu_h} \nonumber\\ & + kT  \yvac \ln(\yvac)
%\end{split}
\end{align}

where $G_{\text{fu}}^{\text{at}}$ is the Gibbs energy for a DEF model that only consists of the atomic sublattices (see equation \ref{eq:GibbsDEF}) and $G_{\text{fu}}^{\text{ax}}$ is the additional Gibbs energy due to the charge carrier (or auxiliary) sublattice. $i$ and $j$ run over end-members that are formed by the atomic sublattices only and thus $\sum\sum \yii\yjj = 1 $. Additionally, in deriving equation \ref{eq:GibbsDEFcharge}, terms such as ${}^0{G}_{\text{i:j:vac}} \times\yii\yjj \yvac$ are rearranged as ${}^0G_{\text{i:j:vac}} \times \left[\yii\yjj (1-\ye-\yh)\right] = {}^0G_{\text{i:j}}\times \yii\yjj (1) - {{}^0G}_{\text{i:j:vac}} \times\yii\yjj (\ye+\yh)$, where the first term is incorporated into $G_{\text{fu}}^{\text{at}}$ and the second term is embedded into $G_{\text{fu}}^{\text{ax}}$. $G_{\text{fu}}^{\text{at}}$ is the Gibbs energy for a DEF without considering the auxiliary charge carrier sublattice, and is given by $G_{\text{fu}}^{\text{at}}= \sum\sum  {}^0G_{\text{i:j}} \yii\yjj + kT \sum_s \sum_k y_k \ln(y_k)$. $i$ and $j$ run over end-members that are formed by the atomic sublattices only, $s$ only runs over atomic sublattices, and $k$ runs over constituents of sublattice $s$. In the DEF formalism, Gibbs energy of defective end-members describing ionic defects with a non-zero net charge are parameterized as the sum of neutral defect formation energy and the ionization energy (as detailed below).

Within $G_{\text{fu}}^{\text{ax}}$ formulation, the chemical potential of free electrons ($\mu_e$) and holes ($\mu_h$) naturally arise, directly linking the DEF end-members with charge carriers to the chemical potentials of electrons and holes. For example, the term $({}^0G_{\text{i:j:h\textsuperscript{+}}}- {}^0G_{\text{i:j:vac}}) + kT \ln (\yh)$ in the DEF Gibbs energy corresponds to $\mu_h^0 +kT \ln p$ in equation \ref{eq:refrencemue}. In deriving this equality, we use the following equation 
\begin{align}
    \begin{split}
    & \frac{p}{N_v} = \left(\frac{p}{N_v} N_v\right) \frac{1}{N_v} = \left(\frac{\yh}{\underbrace{\yh+\ye+\yvac}_{1}} \right) \frac{1}{N_v} \\
    & kT \ln(\frac{p}{N_v}) = kT \ln \yh - kT\ln N_v \rightarrow kT\ln p = kT \ln \yh
\end{split}
\end{align}
The direct connection between the chemical potential of electrons and holes and the DEF Gibbs energy formulation implies that the end-member Gibbs energy value, $(^0G_{\text{i:j:h\textsuperscript{+}}}- ^0G_{\text{i:j:vac}})$, associated with formation of a hole, is $\mu_h^0$. The mixing entropy term associated with the vacant sites on the auxiliary sublattice (last term in equation \ref{eq:GibbsDEFcharge}) approaches zero for low concentration of charge carriers, when $\yvac \rightarrow 1$ ($\lim_{X \to 0} (1-X)\ln(1-X) = -X$), and thus can be neglected.

Considering separate sublattices for free electrons and holes results in the same DEF Gibbs energy formulation as using a common sublattice for both. For separate sublattices, additional terms of the form $(^0G_{\text{i:j:e\textsuperscript{-}:h\textsuperscript{+}}} - ^0G_{\text{i:j:vac:vac}})$ appear in the DEF Gibbs energy of equation \ref{eq:GibbsDEFcharge}. This term represents the Gibbs energy for an electron-hole pair formation ($\text{e}\textsuperscript{-} + \text{h}\textsuperscript{+} \longleftrightarrow 0$), which is equal to zero at thermal equilibrium, making the Gibbs energy formulation the same as a single sublattice hosting electrons and holes. Additionally, terms of the form $\yii \yjj \yvac \yaddvac{}^0G_{\text{i:j:vac:vac}} $ appear that can be rearranged as $ \yii \yjj (1-\ye) (1-y^\text{4}_{\text{h\textsubscript{+}}}){}^0G_{\text{i:j:vac:vac}} = \yii \yjj {}^0G_{\text{i:j}}   - \yii \yjj \ye {}^0G_{\text{i:j:vac:vac}} -\yii \yjj y^\text{4}_{\text{h\textsubscript{+}}} {}^0G_{\text{i:j:vac:vac}} + \yii \yjj (\ye y^\text{3}_{\text{h\textsubscript{+}}}) {}^0G_{\text{i:j:vac:vac}} $, to give the same formulation as for the common auxiliary sublattice in equation \ref{eq:GibbsDEFcharge}. Note that the last term can be neglected because both $\ye$ and $\yh$ are small in the dilute range.  

%[ionization of atomic defects - what are the Gibbs energy terms]
Within the DEF Gibbs energy formulation of equation \ref{eq:GibbsDEFcharge}, energy contributions associated with ionized atomic defects must be relative to the reference electronic energy state, same as charge carriers, which we defined as the intrinsic Fermi level, $E_f^i$. Similar to unifying the chemical energy reference line for neutral defects (see section \ref{sec:convexdefect}), the formation energy of equation \ref{eq:formationcharged} must be shifted by $-q(E_f^*-E_f^i)$ to unify the electronic energy reference line (see Figure \ref{fig:charge-Ef}(b)). Accordingly, the absolute defect energy of a defect $d$ with a net charge of $q$ with a unified chemical and electronic reference state is defined as
\begin{align}
\begin{split}
    &\underbrace{\Delta H_d^q + \sum N_i \mu_i}_{\Delta E_d^q} - q\underbrace{(E_f^*-E_f^i)}_{\mu_e} =  E_{\text{def}}^q - E_{\text{pristine}} +q(E_f^i) \\
    & = \overbrace{\underbrace{E_{\text{def}}^q - E_{\text{def}}^{\emptyset}}_{\text{ionization energy}} + \underbrace{E_{\text{def}}^{\emptyset} - E_{\text{pristine}}}_{\text{neutral defect energy}}}^{\Delta E_d^q} \\ & \hspace{3cm}+q(E_v + \frac{E_g}{2} - \frac{kT}{2} \ln(\frac{N_c}{N_v}))
\end{split}\label{eq:formationchargedunitfied}
\end{align}
where $E_{\text{def}}^{\emptyset}$ denotes the energy of the charge-neutral defective compound. Typical DFT total energy calculations directly compute the sum of neutral defect and ionization energies. The intrinsic Fermi level is formulated in terms of $E_v$ and $E_g$, typically obtained from DFT calculations. The defect energy on the right hand side of equation \ref{eq:formationchargedunitfied} is independent of the equilibrium Fermi level and chemical potential. 

The chemical-potential and Fermi-level-independent defect energy of equation \ref{eq:formationchargedunitfied} is directly connected to the DEF end-member Gibbs energy. For example, consider the (A,Vac\textsuperscript{-1})(B,A\textsuperscript{+1})(Vac,e\textsuperscript{-},h\textsuperscript{+}) model, with a cationic A-vacancy on the first atomic sublattice and an anionic A-antisite on the second. The end-member A:A\textsuperscript{+1}:Vac associates with a defective compound with dilute anionic A-antisite defects and its Gibbs energy equals ${}^0G_{A:A\textsuperscript{+1}:Vac} = {}^0G_{A:B:Vac} + (E_{\text{A-antisite}}^+ - E_{\text{pristine}}) +(E_v + \frac{E_g}{2} - \frac{kT}{2} \ln(\frac{N_c}{N_v}))$. The superposition principle is applied to derive the Gibbs energy of end-members that include both charged ionic defects in the atomic sublattices and charge carriers in the auxiliary sublattice. For example, ${}^0G_{A:A\textsuperscript{+1}:e\textsuperscript{-}} = {}^0G_{A:B:Vac} + (E_{\text{A-antisite}}^+ - E_{\text{pristine}})+(E_v + \frac{E_g}{2} - \frac{kT}{2} \ln(\frac{N_c}{N_v})) + \mu_e^0 = {}^0G_{A:B:Vac} + \Delta E_{\text{A-antisite}}^+ + (E_v + E_g)-kT\ln N_c$. 
%%%%%%%%%%%%%%%%%%%%%%%%%%%%%%%%%%%%%%%%%%%%%%%%%
\subsection{AB compound with dilute charged A-vacancy and B-vacancy}\label{sec:2subcharge}
\begin{figure*}[t]
    \centering
    \includegraphics[width=0.8\textwidth]{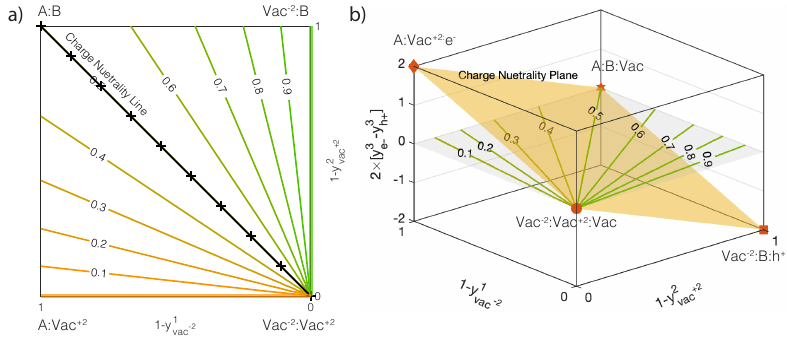}
    \caption{The DEF constitutional space with and without the auxiliary charge carrier sublattice. (a) The constitutional square for the (A,Vac\textsuperscript{-2})(B,Vac\textsuperscript{+2}) DEF model, mapped to the binary system composition, $X_B$. The $X_B=1$ and $X_B=0$ lines are shown in thick green and orange, respectively. Charge neutrality is only satisfied along the hatched black line. b)  The constitutional cube for the (A,Vac\textsuperscript{-2})(B,Vac\textsuperscript{+2})(Vac,e\textsuperscript{-},h\textsuperscript{+}) DEF model, with basis axes ($1-\yt$), ($1-\yu$), ($\ye-\yh$). Charge neutrality is satisfied across the orange plane. Contour lines for constant $X_B$ values are projected on the $(\ye$-$\yh) = 0$ plane. The top ($(\ye$-$\yh) = 1$) and bottom ($(\ye$-$\yh) = -1$) planes contain end-members with free electrons and holes, respectively. The four corners of the charge neutrality plane, marked in red, are physically connected to the absolute defect energy projections on the convex hull at equilibrium Fermi energy.} 
    \label{fig:chargeneutralityplane}
\end{figure*}

In the following section, we elaborate on mapping the composition and defect formation energies onto the constitutional space and Gibbs energy of the DEF model for a compound with charged defects. To better understand the importance of the auxiliary charge carrier sublattice in the DEF model, we first consider a DEF model without the auxiliary sublattice for the AB binary compound with dilute charged vacancies, described by two ionic (or atomic) sublattices (A,Vac\textsuperscript{-2})(B,Vac\textsuperscript{+2}). The first and second sublattices contain A and B substitutional sites, respectively, hosting A, B, or charged vacancy defects. As shown in Figure \ref{fig:chargeneutralityplane}(a), the constitutional space (or $Y$-space) forms a square with four end-members, described by two axes, $\yt$ and $\yu$, representing the site fractions of A and B vacancies in the first and second sublattices, respectively. Similar to neutral defects, each of the non-parallel defect lines (A-vacancy and B-vacancy) intersecting at the AB point on the convex hull forms one axis of the $Y$ constitutional square. Unlike neutral defects, the charge neutrality condition imposes an additional constraint on the mapping between the composition space ($X$) and the constitutional space of DEF ($Y$). 
%The point of intersection, or the A:B serves as the origin for the Y-composition square, and as detailed below its corresponding Gibbs energy is $\Delta H_{\text{AB}}$.  

The $X$-composition  of A or B relates to the DEF site fractions as per equation \ref{eq:map} similar to neutral defects. However, enforcing charge neutrality constrains the $Y$-space square into the neutrality line, where $\yt=\yu$, as shown in Figure \ref{fig:chargeneutralityplane}(a). %and according to the following equation.
%\begin{equation}\label{eq:map2chargenutrality}
%    %0 = n - p + \sum_d q_d c_{0,d}\exp{-\Delta H_{\text{def}}}
%    \yt = \yu
%\end{equation}
Note that A:B and Vac\textsuperscript{-2}:Vac\textsuperscript{+2} end-members are charge-neutral, while Vac\textsuperscript{-2}:B and A:Vac\textsuperscript{+2} have a net charge. The neutrality line restricts the mapping between $X$ and $Y$ to a single point, $X_B=\frac{1}{2}$, extending along the two charge-neutral end members. No other point on the constitutional square, including Vac\textsuperscript{-2}:B and A:Vac\textsuperscript{+2}, corresponds to a physically relevant $X_B$ value for A-vacancy and B-vacancy defect lines. As suggested by Rogal \textit{et al} \cite{Rogal2014}, one can use the (A,Vac\textsuperscript{-2})(B,Vac\textsuperscript{+2}) model in a standard CEF framework to describe the Gibbs energy of defected compounds (see section 3.2 in Ref. \cite{Rogal2014}). However, there is no direct connection between the formation energies of charged defects projected on the convex hull ends and the end-members of the (A,Vac\textsuperscript{-2})(B,Vac\textsuperscript{+2}) model, as we aim to develop in DEF. The relevant convex hull for defect formation energy projections is at the equilibrium Fermi level, $E_f^*$, where charge carriers exist at concentrations different from an intrinsic compound ($\Delta \mu_e \neq 0$, see Figure \ref{fig:charge-Ef}(a)). Rogal \textit{et al.} \cite{Rogal2014} propose determining the Gibbs energy of charged end-members by considering their charge-neutral combinations (Kröger-Vink approach). However, this requires data for neutral combinations, which is computationally demanding and becomes intractable with several competing combinations. %To avoid this, a grand-canonical approach is needed \cite{Rogal2014}.
%, where the charge state of a defect is determined through exchange of electrons with a reservoir controlled by the Fermi level.
%
%Additionally, a CEF model that is constrained to charge neutrality lines (or planes or hyper-planes for higher number of sublattices or defects) cannot be utilized to calculate the concentration of charge defects. To clarify this limitation, let us look at another example. Consider  

\begin{figure*}[tbph]
    \centering
    \includegraphics[width=0.75\textwidth]{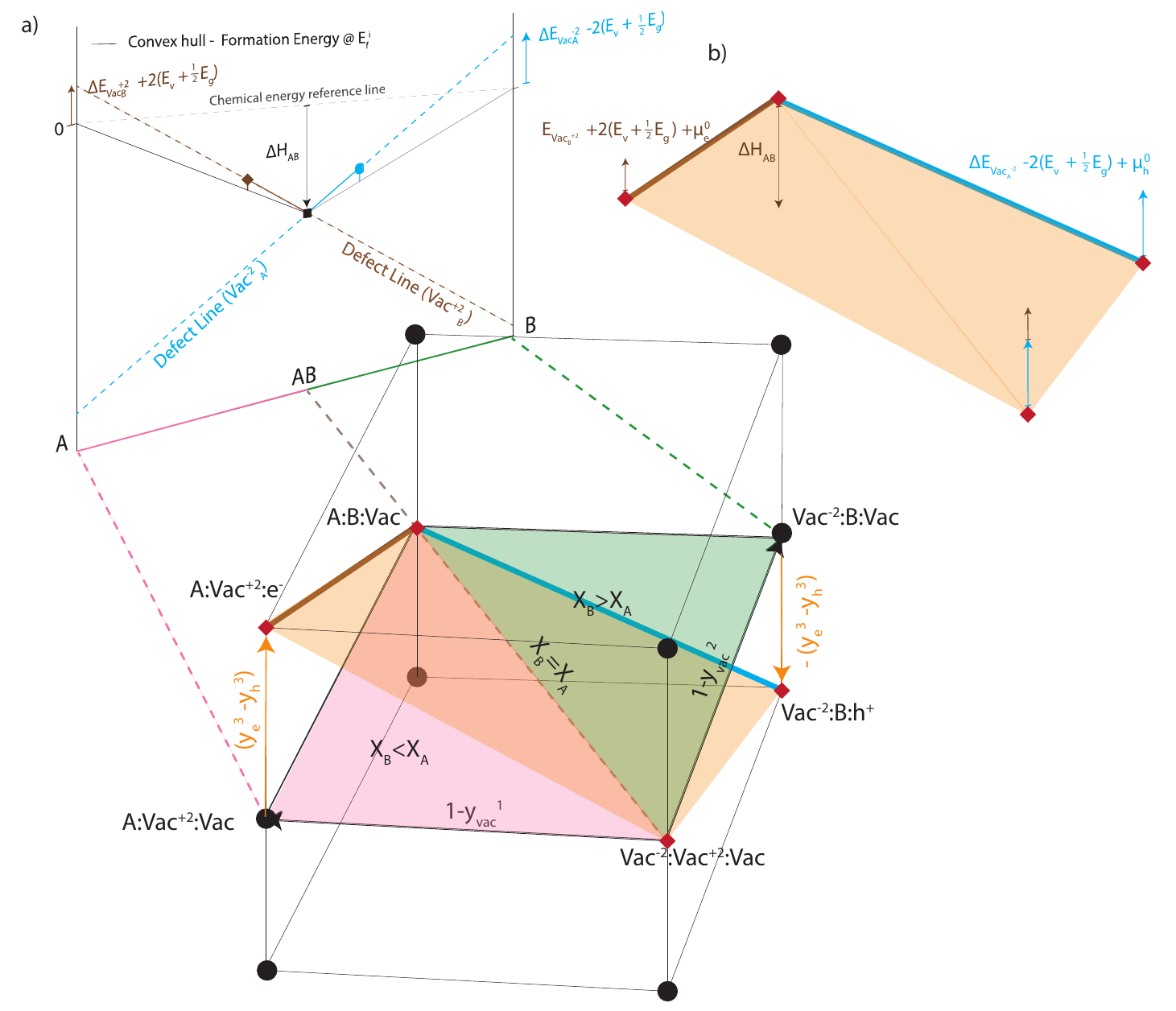}
    \caption{Mapping the composition and formation energy from a model convex hull (A-AB-B) to the constitutional space of a three-sublattice DEF model (A,Vac\textsuperscript{-2})(B,Vac\textsuperscript{+2})(Vac,e\textsuperscript{-},h\textsuperscript{+}). Formation energies of the defect-free AB, defective AB with A-vacancy, and defective AB B-vacancy are represented by the black square, blue circle, and brown diamond, respectively. The defect lines for A-vacancy and B-vacancy are shown by the blue and brown lines, respectively. The defective compound contains charged defects, determined by the equilibrium Fermi level. (a) Graphic representation of the mapping between $X_B$ composition on the convex hull and the constitutional space ($Y$-space). The $X_B$ on the AB-B and A-AB sides maps into the green and pink regions of the $Y$-space, respectively. The charge neutrality plane is illustrated in orange. Lines in the $Y$-space that are physically relevant to the defect lines on the convex hull are shown in blue and brown. DEF end-members are depicted by black circles on the constitutional cube. Red marks indicate end-members directly connected to the formation energies on the convex hull. (b) Graphic representation of mapping the chemical potential and Fermi-level-independent defect energies onto DEF end-members' Gibbs energies at the corners of the charge neutrality plane.} %Chemical-potential and Fermi-level-independent defect energies ($\Delta E_d$) are illustrated as the intercepts between the defect line and the common tangent line.
    \label{fig:main2subcharge}
\end{figure*}

Instead of the Kröger-Vink approach, we define an auxiliary sublattice to host charge carriers, including free electrons and holes. This method, used in the literature \cite{Chen1996,Chen1998,Peters2019,119,120,123,124}, represents the electron reservoir in the grand canonical description. In section \ref{sec:DEFcharged}, we establish a direct connection between the DEF end-members, the absolute defect energy of isolated charged defects, and the chemical potentials of charge carriers. The DEF model with the auxiliary sublattice is described as (A,Vac\textsuperscript{-2})(B,Vac\textsuperscript{+2})(Vac,e\textsuperscript{-},h\textsuperscript{+}), where the third sublattice can host vacancies, free electrons, or holes. The auxiliary sublattice adds a new degree of freedom, $\ye-\yh$, representing the excess electrons as the difference between the site fractions of electrons and holes. Note that the concentrations of electrons and holes are interdependent through $np = N_c N_v \exp{\left(-\frac{E_g}{kT}\right)}$ (see equation \ref{eq:carrierconc}). The resulting constitutional space is a cube, as shown in Figure \ref{fig:chargeneutralityplane}(b). The charge neutrality condition constrains the $Y$-space cube to the neutrality plane, given by $\ye-\yh+2\times\yt-2\times\yu=0$ (see equation \ref{eq:chargenutrality}). The projection of the charge neutrality plane onto the $(\ye-\yh)=0$ plane includes any combination of  $\yt$ and $\yu$. Thus, adding the charge carrier (auxiliary) sublattice allows exploring any $\yt$ and $\yu$ combination, subject to the constraint that their net charge is balanced by free electrons and holes, as defined by the charge neutrality plane. This enables direct mapping of the complete range of $X_B$ values on the convex hull to the DEF $Y$-space, as shown in Figure \ref{fig:main2subcharge}. 
%We refer to the charge carrier sublattice as an auxiliary sublattice because ...[it site fraction of electrons and holes does not directly enter the chemical compositions - instead it indirectly determines the charge state of defects - this can influence the concentration of defects - their site fraction at equilibrium ]. This auxiliary axis only facilitates the connection between physical charged defect formation energies and the DEF end-members.   %Note that the site fraction of holes do not introduce a new degree of freedom as the concentration of holes and electrons are related through the following relationship %Considering the electronic entities e\textsuperscript{-} and h\textsuperscript{+} as electronic defects, the constitutional space for DEF.  %of the three sublattice model, including the charge carrier sublattice,

%[one paragraph here - end members ye=1 ln(n)=0 -> me = me0 - we have 12 end-members although we know some are redundant] -3 planes plane =1,0=-1
As shown in Figure \ref{fig:main2subcharge}(a), The constitutional cube has 12 end-members at the corners of the three planes defined by $(\ye$-$\yh) = {-1,0,1}$. The top plane corresponds to $\ye=1$ and $\yh=0$, and the bottom plane to $\ye=0$ and $\yh=1$, each allowing any combination of $\yt$ and $\yu$. The auxiliary sublattice on the top plane is occupied by free electrons ($\ye=1$), and on the bottom plane by holes ($\yh=1$). Free electron occupation implies $\ye=1$ or $n=1$, resulting in $kT \ln n =0$ or $\mu_e = \mu_e^0$ (see equation \ref{eq:refrencemue2}), making the Gibbs energy of top plane end-members include $\mu_e^0$ for free electron formation. The same applies to the bottom plane for holes. The middle plane has end-members with no charge carriers. Four of the 12 end-members correspond to physically relevant dilute defects: A:B:Vac and Vac\textsuperscript{-2}:Vac\textsuperscript{+2}:Vac on the middle plane, A:Vac\textsuperscript{+2}:e\textsuperscript{-} on the top plane, and Vac\textsuperscript{-2}:B:h\textsuperscript{+} on the bottom plane. For example, the A:Vac\textsuperscript{+2}:e\textsuperscript{-} end-member describes the creation of a positively charged ionic defect by exchanging free electrons with the reservoir in the grand-canonical description. As shown in Figure \ref{fig:main2subcharge}(b), the chemical-potential Fermi-level-independent defect energies of equation \ref{eq:formationchargedunitfied} are directly mapped to these four end-members. The Gibbs energy of other end-members is obtained by decomposing the Gibbs energy of multi-defected, physically connected end-members using the superposition principle. For example, the multi-defected A:Vac\textsuperscript{+2}:e\textsuperscript{-} end-member can be decomposed into two independent defective end-members: A:Vac\textsuperscript{+2}:Vac, which includes the charged ionic defect, and A:B:e\textsuperscript{-}, which includes the free electron.

The end-member Gibbs energy for a general DEF model (A,Vac\textsuperscript{-q})$_{\text{$\alpha$}}$(B,Vac\textsuperscript{+q})$_{\text{$\beta$}}$(Vac,e\textsuperscript{-})(Vac,h\textsuperscript{+}), assuming $m^*_e \approx m^*_h \rightarrow N_c \approx N_v$, is formulated based on the absolute defect energies of the defective and pristine compounds (per atom) as follows
\begin{widetext}
\begin{align}
\begin{split}\label{eq:DEFmapping2charge}
&{}^0G_{\text{A:B:Vac}}  = \alpha\mu_A^0 + \beta\mu_B^0+ (\alpha+\beta)\Delta H_{\text{f}}^{\text{AB}}  \\
&{}^0G_{\text{A:B:e\textsuperscript{-}}}  = {}^0G_{\text{A:B:Vac}} + \left(\frac{E_g}{2}-kT\ln N_c\right)\\
&{}^0G_{\text{A:B:h\textsuperscript{+}}}  = {}^0G_{\text{A:B:Vac}}+ \left(\frac{E_g}{2}-kT\ln N_v\right)  \\
&{}^0G_{\text{Vac\textsuperscript{-q}:B:Vac}}  = {}^0G_{\text{A:B:Vac}} + \alpha\left( \Delta E_{Vac_A^{-q}} -q(E_v + \frac{E_g}{2})\right) \\
&{}^0G_{\text{Vac\textsuperscript{-q}:B:e\textsuperscript{-}}}  = {}^0G_{\text{A:B:Vac}} + \alpha\left( \Delta E_{Vac_A^{-q}} -q(E_v + \frac{E_g}{2})\right) + \left(\frac{E_g}{2}-kT\ln N_c\right) \\
&{}^0G_{\text{Vac\textsuperscript{-q}:B:h\textsuperscript{+}}} = {}^0G_{\text{A:B:Vac}} + \alpha\left( \Delta E_{Vac_A^{-q}} -q(E_v + \frac{E_g}{2})\right) + \left(\frac{E_g}{2}-kT\ln N_v\right) \\
&{}^0G_{\text{A:Vac\textsuperscript{+q}:Vac}} = {}^0G_{\text{A:B:Vac}} + \beta\left(\Delta E_{Vac_B^{+q}} +q(E_v + \frac{E_g}{2})\right) \\
&{}^0G_{\text{A:Vac\textsuperscript{+q}:e\textsuperscript{-}}} = {}^0G_{\text{A:B:Vac}} + 
 \beta\left( \Delta E_{Vac_B^{+q}}+q(E_v + \frac{E_g}{2})\right) + \left(\frac{E_g}{2}-kT\ln N_c\right)\\
&{}^0G_{\text{A:Vac\textsuperscript{+q}:h\textsuperscript{+}}} = {}^0G_{\text{A:B:Vac}} + 
 \beta\left( \Delta E_{Vac_B^{+q}}+q(E_v + \frac{E_g}{2})\right) + \left(\frac{E_g}{2}-kT\ln N_v\right) \\
&{}^0G_{\text{Vac\textsuperscript{-q}:Vac\textsuperscript{+q}:Vac}} = {}^0G_{\text{A:B:Vac}} + \alpha\left(\Delta E_{Vac_A^{-q}} -q(E_v+\frac{E_g}{2})\right) + \beta\left(\Delta E_{Vac_B^{+q}} +q(E_v+\frac{E_g}{2})\right)  \\
&{}^0G_{\text{Vac\textsuperscript{-q}:Vac\textsuperscript{+q}:e\textsuperscript{-}}} = {}^0G_{\text{A:B:Vac}} + \alpha\left(\Delta E_{Vac_A^{-q}} -q(E_v+\frac{E_g}{2})\right) + \beta\left(\Delta E_{Vac_B^{+q}} +q(E_v+\frac{E_g}{2})\right) + \left(\frac{E_g}{2}-kT\ln N_c\right)\\
&{}^0G_{\text{Vac\textsuperscript{-q}:Vac\textsuperscript{+q}:h\textsuperscript{+}}} = {}^0G_{\text{A:B:Vac}} + \alpha\left(\Delta E_{Vac_A^{-q}} -q(E_v+\frac{E_g}{2})\right) + \beta\left(\Delta E_{Vac_B^{+q}} +q(E_v+\frac{E_g}{2})\right) + \left(\frac{E_g}{2}-kT\ln N_v\right) \\
\end{split}
\end{align}
\end{widetext}
The DEF end-member Gibbs energies from equation \ref{eq:DEFmapping2charge} can be plugged into equation \ref{eq:GibbsDEFcharge} to obtain the Gibbs energy per formula unit of the defective compound. %The general recipe for constructing a typical DEF model for defective compounds with charged defects is provided in section \ref{sec:generalrecipe}.
%[talk about the physical meaning of the marked four end-members on the charge neutrality plane - they connect to the convex hull at Ef*.]
%
%[instead of superposition we apply the decomposition of combined defect into isolated defects.]  
%where $\mu_e^0 \approx E_g - kT \ln(N_c)$ and $\mu_h^0 \approx E_g - kT \ln(N_v)$ (see equation \ref{eq:refrencemue2}). The formation energy of charged ionic defects is obtained from equation \ref{eq:formationchargedunitfied}. 
\section{The General Recipe For Constructing Defect Energy Formalism}\label{sec:generalrecipe}
The general procedure for constructing a typical DEF model for neutral defects is as follows:
\begin{itemize}
\setlength\itemsep{0em}
    \item \textit{Defect lines and defect formation energy projections on the convex hull}: Plot the desired defective compounds on the formation energy convex hull. Each defective compound corresponds to a point on the convex hull diagram. The line connecting this point to the defect-free compound point is the defect line (see Ref. \cite{Anand2021Visualizing}). Draw the $N_d$ non-parallel defect lines for $N_d$ different defects. 
    \item \textit{DEF constitutional $Y$-space}: Construct the constitutional space for the DEF model using the $N_d$ defect lines as the orthogonal basis. The origin represents the defect-free compound, and each basis vector corresponds to a defect line. Each axis indicates the site fraction of the corresponding defect on its host sublattice. The end-members are points with site fraction values of 0 or 1 along each axis. Note that the $N_d$ degrees of freedom, associated with fractions of defects (or secondary constituents), can be hosted on $N_d$ sublattices or fewer. For example, two defects (e.g., A-vacancy and B-antisite) can be defined on the first sublattice.
    \item \textit{DEF end-members' Gibbs energy}: The Gibbs energy of the origin point in the $Y$-space equals the formation energy of the defect-free compound, which also serves as the reference state for chemical energy. The Gibbs energy of other end-members, formed by translating the origin using defect line vectors, is the sum of the defect-free compound Gibbs energy and the absolute defect energy of the corresponding defects. Note that the number of independent DEF parameters equals $N_d$. 
\end{itemize}

The general procedure for constructing a typical DEF model for charged defects is as follows:
\begin{itemize}
\setlength\itemsep{0em}
    \item \textit{Identify the intrinsic Fermi level}: The intrinsic Fermi level $E_f^i$ is typically identified from DFT calculations of the valence band maximum and band gap energies using the formula $E_f^i = E_v + E_g/2 - kT/2 \ln(\frac{N_c}{N_v})$. 
    \item \textit{Ionic defect lines projections on the convex hull}: The ionic defect lines with charge $q$ are constructed on the convex hull associated with $E_f^i$. The chemical-potential-independent defect energies projected on the convex hull are adjusted by adding $qE_f^i$ to yield the chemical-potential-Fermi-level-independent defect energies (see equations \ref{eq:general_defect_ref_pristine} and \ref{eq:formationchargedunitfied}). 
    \item \textit{DEF construction of atomic and auxiliary sublattices}: The DEF model is decomposed into atomic and charge carrier sublattices. The origin of the constitutional space represents the defect-free compound, with each basis vector corresponding to a defect line. An additional basis axis describes the net concentration of free electrons and holes on the auxiliary sublattice. This model consists of $N_d + 1$ degrees of freedom: $N_d$ for the number of ionic defects and one for the charge carriers. 
    \item \textit{DEF end-members Gibbs energy}: The Gibbs energy of end-members formed by the atomic sublattices is connected to their corresponding ionic defect energies on the convex hull using the same procedure as for neutral defects. The defect-free compound (i.e., the origin) serves as the reference state for chemical and electronic energies. Note that defect formation energies include the additional electronic work term, $q_dE_f^i$. The Gibbs energy of end-members on the auxiliary sublattice is connected to the intrinsic chemical potential of electrons and holes in a non-degenerate, defect-free compound. The reference chemical potentials of electrons and holes, $\mu_e^0$ and $\mu_h^0$, are added to the Gibbs energy of end-members containing free electrons and holes. 
\end{itemize}

\edit{\section{Discussion}}\label{sec:discuss}
\edit{The primary scope of this work is to present a formal theoretical derivation of the Defect Energy Formalism (DEF), enabling the integration of DFT-calculated defect energies into the CALPHAD framework for thermodynamic modeling. DEF defines explicit connections between absolute defect energies—independent of chemical potential or Fermi level—and the Gibbs energy parameters of defective compounds. This enables a first-principles model for dilute point defects, eliminating the need for fitting experimental or simulation data. DEF also simplifies the inherent complexity of CEF by superimposing defect energies, making it suitable for modeling multi-component and chemically complex systems.}

\edit{The methodology outlined in this paper is developed specifically for dilute point defects, where defect-defect interactions are negligible. For more complicated systems with defect complexes or higher defect concentrations, however, modifications to the formalism would be required to account for additional physical effects, such as the defect binding energies. These adjustments are beyond the scope of this work but will be explored in future research as we aim to expand the DEF framework to more complicated defect scenarios.}

\edit{It is also crucial to note that the accuracy of DEF is inherently tied to the precision of defect energies derived from DFT calculations. Supercell size effects can introduce significant errors due to spurious interactions between defects and their periodic images, including quantum-mechanical wave-function overlap and electrostatic interactions in the case of charged defects, which do not always scale predictably. Without precise DFT results, the reliability of DEF outputs, particularly phase diagram predictions, can be compromised. As emphasized by Freysoldt et al. \cite{PhysRevB.105.014103}, appropriate correction schemes are necessary to address these supercell effects. The choice of a correction approach should be tailored to the material system, defect type, and dimensionality. By ensuring the accuracy of DFT-calculated defect energies through proper treatment of supercell size effects, DEF can be reliably applied to generate phase diagrams that align with experimental data.}

\edit{In the current DEF formulation, configurational entropy effects are naturally incorporated, particularly through the second term in equation \ref{eq:GibbsDEF}, which accounts for the ideal entropy of mixing based on the assumption of random defect distribution within each sublattice. However, defect formation entropies, which arise from vibrational or electronic changes induced by defects, are not yet fully considered. These effects can be integrated into the framework by adding appropriate $TS$ terms to capture the additional entropy contributions. Future work will focus on developing a more rigorous formalism to incorporate these entropy effects, enhancing the accuracy of the temperature dependence in DEF. While configurational entropy remains the dominant factor, accounting for these other contributions will enable a more comprehensive treatment of temperature effects. }

\section*{Acknowledgements}
We extend our gratitude to Prof. G. Jeffrey Snyder for conceptualizing the idea of Defect Energy Formalism and providing critical insights pivotal to the successful derivation of this work. S. K. acknowledges support from the National Science Foundation (NSF) under Award Number DMR-1954621. A.H.A acknowledges support from the U.S. Department of Commerce, National Institute of Standards and Technology as part of the Center for Hierarchical Materials Design (CHiMaD) under Award No. 70NANB19H005. 
\section*{Competing Interests} 
All authors declare no competing financial or non-financial interests. 

\onecolumngrid
%\bibliography{ref_DEF_paper.bib}

\begin{thebibliography}{75}%
\makeatletter
\providecommand \@ifxundefined [1]{%
 \@ifx{#1\undefined}
}%
\providecommand \@ifnum [1]{%
 \ifnum #1\expandafter \@firstoftwo
 \else \expandafter \@secondoftwo
 \fi
}%
\providecommand \@ifx [1]{%
 \ifx #1\expandafter \@firstoftwo
 \else \expandafter \@secondoftwo
 \fi
}%
\providecommand \natexlab [1]{#1}%
\providecommand \enquote  [1]{``#1''}%
\providecommand \bibnamefont  [1]{#1}%
\providecommand \bibfnamefont [1]{#1}%
\providecommand \citenamefont [1]{#1}%
\providecommand \href@noop [0]{\@secondoftwo}%
\providecommand \href [0]{\begingroup \@sanitize@url \@href}%
\providecommand \@href[1]{\@@startlink{#1}\@@href}%
\providecommand \@@href[1]{\endgroup#1\@@endlink}%
\providecommand \@sanitize@url [0]{\catcode `\\12\catcode `\$12\catcode
  `\&12\catcode `\#12\catcode `\^12\catcode `\_12\catcode `\%12\relax}%
\providecommand \@@startlink[1]{}%
\providecommand \@@endlink[0]{}%
\providecommand \url  [0]{\begingroup\@sanitize@url \@url }%
\providecommand \@url [1]{\endgroup\@href {#1}{\urlprefix }}%
\providecommand \urlprefix  [0]{URL }%
\providecommand \Eprint [0]{\href }%
\providecommand \doibase [0]{http://dx.doi.org/}%
\providecommand \selectlanguage [0]{\@gobble}%
\providecommand \bibinfo  [0]{\@secondoftwo}%
\providecommand \bibfield  [0]{\@secondoftwo}%
\providecommand \translation [1]{[#1]}%
\providecommand \BibitemOpen [0]{}%
\providecommand \bibitemStop [0]{}%
\providecommand \bibitemNoStop [0]{.\EOS\space}%
\providecommand \EOS [0]{\spacefactor3000\relax}%
\providecommand \BibitemShut  [1]{\csname bibitem#1\endcsname}%
\let\auto@bib@innerbib\@empty
%</preamble>
\bibitem [{\citenamefont {Pei}\ \emph {et~al.}(2011)\citenamefont {Pei},
  \citenamefont {Shi}, \citenamefont {LaLonde}, \citenamefont {Wang},
  \citenamefont {Chen},\ and\ \citenamefont {Snyder}}]{1}%
  \BibitemOpen
  \bibfield  {author} {\bibinfo {author} {\bibfnamefont {Y.}~\bibnamefont
  {Pei}}, \bibinfo {author} {\bibfnamefont {X.}~\bibnamefont {Shi}}, \bibinfo
  {author} {\bibfnamefont {A.}~\bibnamefont {LaLonde}}, \bibinfo {author}
  {\bibfnamefont {H.}~\bibnamefont {Wang}}, \bibinfo {author} {\bibfnamefont
  {L.}~\bibnamefont {Chen}}, \ and\ \bibinfo {author} {\bibfnamefont {G.~J.}\
  \bibnamefont {Snyder}},\ }\href {\doibase 10.1038/nature09996} {\bibfield
  {journal} {\bibinfo  {journal} {Nature}\ }\textbf {\bibinfo {volume} {473}},\
  \bibinfo {pages} {66} (\bibinfo {year} {2011})}\BibitemShut {NoStop}%
\bibitem [{\citenamefont {Jood}\ \emph {et~al.}(2020)\citenamefont {Jood},
  \citenamefont {Male}, \citenamefont {Anand}, \citenamefont {Matsushita},
  \citenamefont {Takagiwa}, \citenamefont {Kanatzidis}, \citenamefont
  {Snyder},\ and\ \citenamefont {Ohta}}]{2}%
  \BibitemOpen
  \bibfield  {author} {\bibinfo {author} {\bibfnamefont {P.}~\bibnamefont
  {Jood}}, \bibinfo {author} {\bibfnamefont {J.~P.}\ \bibnamefont {Male}},
  \bibinfo {author} {\bibfnamefont {S.}~\bibnamefont {Anand}}, \bibinfo
  {author} {\bibfnamefont {Y.}~\bibnamefont {Matsushita}}, \bibinfo {author}
  {\bibfnamefont {Y.}~\bibnamefont {Takagiwa}}, \bibinfo {author}
  {\bibfnamefont {M.~G.}\ \bibnamefont {Kanatzidis}}, \bibinfo {author}
  {\bibfnamefont {G.~J.}\ \bibnamefont {Snyder}}, \ and\ \bibinfo {author}
  {\bibfnamefont {M.}~\bibnamefont {Ohta}},\ }\href {\doibase
  10.1021/jacs.0c07067} {\bibfield  {journal} {\bibinfo  {journal} {Journal of
  the American Chemical Society}\ }\textbf {\bibinfo {volume} {142}},\ \bibinfo
  {pages} {15464} (\bibinfo {year} {2020})},\ \bibinfo {note} {pMID:
  32786772},\ \Eprint
  {http://arxiv.org/abs/https://doi.org/10.1021/jacs.0c07067}
  {https://doi.org/10.1021/jacs.0c07067} \BibitemShut {NoStop}%
\bibitem [{\citenamefont {Seebauer}\ and\ \citenamefont {Kratzer}(2006)}]{3}%
  \BibitemOpen
  \bibfield  {author} {\bibinfo {author} {\bibfnamefont {E.~G.}\ \bibnamefont
  {Seebauer}}\ and\ \bibinfo {author} {\bibfnamefont {M.~C.}\ \bibnamefont
  {Kratzer}},\ }\href {\doibase https://doi.org/10.1016/j.mser.2006.01.002}
  {\bibfield  {journal} {\bibinfo  {journal} {Materials Science and
  Engineering: R: Reports}\ }\textbf {\bibinfo {volume} {55}},\ \bibinfo
  {pages} {57} (\bibinfo {year} {2006})}\BibitemShut {NoStop}%
\bibitem [{\citenamefont {Zheng}\ \emph {et~al.}(2021)\citenamefont {Zheng},
  \citenamefont {Slade}, \citenamefont {Hu}, \citenamefont {Tan}, \citenamefont
  {Luo}, \citenamefont {Luo}, \citenamefont {Xu}, \citenamefont {Yan},\ and\
  \citenamefont {Kanatzidis}}]{4}%
  \BibitemOpen
  \bibfield  {author} {\bibinfo {author} {\bibfnamefont {Y.}~\bibnamefont
  {Zheng}}, \bibinfo {author} {\bibfnamefont {T.~J.}\ \bibnamefont {Slade}},
  \bibinfo {author} {\bibfnamefont {L.}~\bibnamefont {Hu}}, \bibinfo {author}
  {\bibfnamefont {X.~Y.}\ \bibnamefont {Tan}}, \bibinfo {author} {\bibfnamefont
  {Y.}~\bibnamefont {Luo}}, \bibinfo {author} {\bibfnamefont {Z.-Z.}\
  \bibnamefont {Luo}}, \bibinfo {author} {\bibfnamefont {J.}~\bibnamefont
  {Xu}}, \bibinfo {author} {\bibfnamefont {Q.}~\bibnamefont {Yan}}, \ and\
  \bibinfo {author} {\bibfnamefont {M.~G.}\ \bibnamefont {Kanatzidis}},\ }\href
  {\doibase 10.1039/D1CS00347J} {\bibfield  {journal} {\bibinfo  {journal}
  {Chem. Soc. Rev.}\ }\textbf {\bibinfo {volume} {50}},\ \bibinfo {pages}
  {9022} (\bibinfo {year} {2021})}\BibitemShut {NoStop}%
\bibitem [{\citenamefont {Slade}\ \emph {et~al.}(2021)\citenamefont {Slade},
  \citenamefont {Anand}, \citenamefont {Wood}, \citenamefont {Male},
  \citenamefont {Imasato}, \citenamefont {Cheikh}, \citenamefont {{Al Malki}},
  \citenamefont {Agne}, \citenamefont {Griffith}, \citenamefont {Bux},
  \citenamefont {Wolverton}, \citenamefont {Kanatzidis},\ and\ \citenamefont
  {Snyder}}]{5}%
  \BibitemOpen
  \bibfield  {author} {\bibinfo {author} {\bibfnamefont {T.~J.}\ \bibnamefont
  {Slade}}, \bibinfo {author} {\bibfnamefont {S.}~\bibnamefont {Anand}},
  \bibinfo {author} {\bibfnamefont {M.}~\bibnamefont {Wood}}, \bibinfo {author}
  {\bibfnamefont {J.~P.}\ \bibnamefont {Male}}, \bibinfo {author}
  {\bibfnamefont {K.}~\bibnamefont {Imasato}}, \bibinfo {author} {\bibfnamefont
  {D.}~\bibnamefont {Cheikh}}, \bibinfo {author} {\bibfnamefont {M.~M.}\
  \bibnamefont {{Al Malki}}}, \bibinfo {author} {\bibfnamefont {M.~T.}\
  \bibnamefont {Agne}}, \bibinfo {author} {\bibfnamefont {K.~J.}\ \bibnamefont
  {Griffith}}, \bibinfo {author} {\bibfnamefont {S.~K.}\ \bibnamefont {Bux}},
  \bibinfo {author} {\bibfnamefont {C.}~\bibnamefont {Wolverton}}, \bibinfo
  {author} {\bibfnamefont {M.~G.}\ \bibnamefont {Kanatzidis}}, \ and\ \bibinfo
  {author} {\bibfnamefont {G.~J.}\ \bibnamefont {Snyder}},\ }\href {\doibase
  https://doi.org/10.1016/j.joule.2021.03.009} {\bibfield  {journal} {\bibinfo
  {journal} {Joule}\ }\textbf {\bibinfo {volume} {5}},\ \bibinfo {pages} {1168}
  (\bibinfo {year} {2021})}\BibitemShut {NoStop}%
\bibitem [{\citenamefont {Jiang}\ \emph {et~al.}(2014)\citenamefont {Jiang},
  \citenamefont {He}, \citenamefont {Zhu}, \citenamefont {Fu}, \citenamefont
  {Liu}, \citenamefont {Hu},\ and\ \citenamefont {Zhao}}]{6}%
  \BibitemOpen
  \bibfield  {author} {\bibinfo {author} {\bibfnamefont {G.}~\bibnamefont
  {Jiang}}, \bibinfo {author} {\bibfnamefont {J.}~\bibnamefont {He}}, \bibinfo
  {author} {\bibfnamefont {T.}~\bibnamefont {Zhu}}, \bibinfo {author}
  {\bibfnamefont {C.}~\bibnamefont {Fu}}, \bibinfo {author} {\bibfnamefont
  {X.}~\bibnamefont {Liu}}, \bibinfo {author} {\bibfnamefont {L.}~\bibnamefont
  {Hu}}, \ and\ \bibinfo {author} {\bibfnamefont {X.}~\bibnamefont {Zhao}},\
  }\href {\doibase https://doi.org/10.1002/adfm.201400123} {\bibfield
  {journal} {\bibinfo  {journal} {Advanced Functional Materials}\ }\textbf
  {\bibinfo {volume} {24}},\ \bibinfo {pages} {3776} (\bibinfo {year}
  {2014})},\ \Eprint
  {http://arxiv.org/abs/https://onlinelibrary.wiley.com/doi/pdf/10.1002/adfm.201400123}
  {https://onlinelibrary.wiley.com/doi/pdf/10.1002/adfm.201400123} \BibitemShut
  {NoStop}%
\bibitem [{\citenamefont {Ganose}\ \emph {et~al.}(2022)\citenamefont {Ganose},
  \citenamefont {Scanlon}, \citenamefont {Walsh},\ and\ \citenamefont
  {Hoye}}]{7}%
  \BibitemOpen
  \bibfield  {author} {\bibinfo {author} {\bibfnamefont {A.~M.}\ \bibnamefont
  {Ganose}}, \bibinfo {author} {\bibfnamefont {D.~O.}\ \bibnamefont {Scanlon}},
  \bibinfo {author} {\bibfnamefont {A.}~\bibnamefont {Walsh}}, \ and\ \bibinfo
  {author} {\bibfnamefont {R.~L.~Z.}\ \bibnamefont {Hoye}},\ }\href {\doibase
  10.1038/s41467-022-32131-4} {\bibfield  {journal} {\bibinfo  {journal}
  {Nature Communications}\ }\textbf {\bibinfo {volume} {13}},\ \bibinfo {pages}
  {4715} (\bibinfo {year} {2022})}\BibitemShut {NoStop}%
\bibitem [{\citenamefont {Huang}\ \emph {et~al.}(2018)\citenamefont {Huang},
  \citenamefont {Yin},\ and\ \citenamefont {He}}]{8}%
  \BibitemOpen
  \bibfield  {author} {\bibinfo {author} {\bibfnamefont {Y.}~\bibnamefont
  {Huang}}, \bibinfo {author} {\bibfnamefont {W.-J.}\ \bibnamefont {Yin}}, \
  and\ \bibinfo {author} {\bibfnamefont {Y.}~\bibnamefont {He}},\ }\href
  {\doibase 10.1021/acs.jpcc.7b10045} {\bibfield  {journal} {\bibinfo
  {journal} {The Journal of Physical Chemistry C}\ }\textbf {\bibinfo {volume}
  {122}},\ \bibinfo {pages} {1345} (\bibinfo {year} {2018})},\ \Eprint
  {http://arxiv.org/abs/https://doi.org/10.1021/acs.jpcc.7b10045}
  {https://doi.org/10.1021/acs.jpcc.7b10045} \BibitemShut {NoStop}%
\bibitem [{\citenamefont {Park}\ \emph {et~al.}(2018)\citenamefont {Park},
  \citenamefont {Kim}, \citenamefont {Xie},\ and\ \citenamefont {Walsh}}]{9}%
  \BibitemOpen
  \bibfield  {author} {\bibinfo {author} {\bibfnamefont {J.~S.}\ \bibnamefont
  {Park}}, \bibinfo {author} {\bibfnamefont {S.}~\bibnamefont {Kim}}, \bibinfo
  {author} {\bibfnamefont {Z.}~\bibnamefont {Xie}}, \ and\ \bibinfo {author}
  {\bibfnamefont {A.}~\bibnamefont {Walsh}},\ }\href {\doibase
  10.1038/s41578-018-0026-7} {\bibfield  {journal} {\bibinfo  {journal} {Nature
  Reviews Materials}\ }\textbf {\bibinfo {volume} {3}},\ \bibinfo {pages} {194}
  (\bibinfo {year} {2018})}\BibitemShut {NoStop}%
\bibitem [{\citenamefont {Vidal}\ \emph {et~al.}(2012)\citenamefont {Vidal},
  \citenamefont {Lany}, \citenamefont {d'Avezac}, \citenamefont {Zunger},
  \citenamefont {Zakutayev}, \citenamefont {Francis},\ and\ \citenamefont
  {Tate}}]{10}%
  \BibitemOpen
  \bibfield  {author} {\bibinfo {author} {\bibfnamefont {J.}~\bibnamefont
  {Vidal}}, \bibinfo {author} {\bibfnamefont {S.}~\bibnamefont {Lany}},
  \bibinfo {author} {\bibfnamefont {M.}~\bibnamefont {d'Avezac}}, \bibinfo
  {author} {\bibfnamefont {A.}~\bibnamefont {Zunger}}, \bibinfo {author}
  {\bibfnamefont {A.}~\bibnamefont {Zakutayev}}, \bibinfo {author}
  {\bibfnamefont {J.}~\bibnamefont {Francis}}, \ and\ \bibinfo {author}
  {\bibfnamefont {J.}~\bibnamefont {Tate}},\ }\href {\doibase
  10.1063/1.3675880} {\bibfield  {journal} {\bibinfo  {journal} {Applied
  Physics Letters}\ }\textbf {\bibinfo {volume} {100}},\ \bibinfo {pages}
  {032104} (\bibinfo {year} {2012})},\ \Eprint
  {http://arxiv.org/abs/https://doi.org/10.1063/1.3675880}
  {https://doi.org/10.1063/1.3675880} \BibitemShut {NoStop}%
\bibitem [{\citenamefont {Geisz}\ and\ \citenamefont {Friedman}(2002)}]{11}%
  \BibitemOpen
  \bibfield  {author} {\bibinfo {author} {\bibfnamefont {J.~F.}\ \bibnamefont
  {Geisz}}\ and\ \bibinfo {author} {\bibfnamefont {D.~J.}\ \bibnamefont
  {Friedman}},\ }\href {\doibase 10.1088/0268-1242/17/8/305} {\bibfield
  {journal} {\bibinfo  {journal} {Semiconductor Science and Technology}\
  }\textbf {\bibinfo {volume} {17}},\ \bibinfo {pages} {769} (\bibinfo {year}
  {2002})}\BibitemShut {NoStop}%
\bibitem [{\citenamefont {Tongay}\ \emph {et~al.}(2013)\citenamefont {Tongay},
  \citenamefont {Suh}, \citenamefont {Ataca}, \citenamefont {Fan},
  \citenamefont {Luce}, \citenamefont {Kang}, \citenamefont {Liu},
  \citenamefont {Ko}, \citenamefont {Raghunathanan}, \citenamefont {Zhou},
  \citenamefont {Ogletree}, \citenamefont {Li}, \citenamefont {Grossman},\ and\
  \citenamefont {Wu}}]{12}%
  \BibitemOpen
  \bibfield  {author} {\bibinfo {author} {\bibfnamefont {S.}~\bibnamefont
  {Tongay}}, \bibinfo {author} {\bibfnamefont {J.}~\bibnamefont {Suh}},
  \bibinfo {author} {\bibfnamefont {C.}~\bibnamefont {Ataca}}, \bibinfo
  {author} {\bibfnamefont {W.}~\bibnamefont {Fan}}, \bibinfo {author}
  {\bibfnamefont {A.}~\bibnamefont {Luce}}, \bibinfo {author} {\bibfnamefont
  {J.~S.}\ \bibnamefont {Kang}}, \bibinfo {author} {\bibfnamefont
  {J.}~\bibnamefont {Liu}}, \bibinfo {author} {\bibfnamefont {C.}~\bibnamefont
  {Ko}}, \bibinfo {author} {\bibfnamefont {R.}~\bibnamefont {Raghunathanan}},
  \bibinfo {author} {\bibfnamefont {J.}~\bibnamefont {Zhou}}, \bibinfo {author}
  {\bibfnamefont {F.}~\bibnamefont {Ogletree}}, \bibinfo {author}
  {\bibfnamefont {J.}~\bibnamefont {Li}}, \bibinfo {author} {\bibfnamefont
  {J.~C.}\ \bibnamefont {Grossman}}, \ and\ \bibinfo {author} {\bibfnamefont
  {J.}~\bibnamefont {Wu}},\ }\href {\doibase 10.1038/srep02657} {\bibfield
  {journal} {\bibinfo  {journal} {Scientific Reports}\ }\textbf {\bibinfo
  {volume} {3}},\ \bibinfo {pages} {2657} (\bibinfo {year} {2013})}\BibitemShut
  {NoStop}%
\bibitem [{\citenamefont {Mora-Ser{\'o}}(2020)}]{13}%
  \BibitemOpen
  \bibfield  {author} {\bibinfo {author} {\bibfnamefont {I.}~\bibnamefont
  {Mora-Ser{\'o}}},\ }\href {\doibase 10.1038/s41560-020-0621-x} {\bibfield
  {journal} {\bibinfo  {journal} {Nature Energy}\ }\textbf {\bibinfo {volume}
  {5}},\ \bibinfo {pages} {363} (\bibinfo {year} {2020})}\BibitemShut {NoStop}%
\bibitem [{\citenamefont {Park}\ and\ \citenamefont {Walsh}(2019)}]{14}%
  \BibitemOpen
  \bibfield  {author} {\bibinfo {author} {\bibfnamefont {J.-S.}\ \bibnamefont
  {Park}}\ and\ \bibinfo {author} {\bibfnamefont {A.}~\bibnamefont {Walsh}},\
  }\href {\doibase 10.1038/s41560-019-0329-y} {\bibfield  {journal} {\bibinfo
  {journal} {Nature Energy}\ }\textbf {\bibinfo {volume} {4}},\ \bibinfo
  {pages} {95} (\bibinfo {year} {2019})}\BibitemShut {NoStop}%
\bibitem [{\citenamefont {Walsh}\ and\ \citenamefont {Zunger}(2017)}]{15}%
  \BibitemOpen
  \bibfield  {author} {\bibinfo {author} {\bibfnamefont {A.}~\bibnamefont
  {Walsh}}\ and\ \bibinfo {author} {\bibfnamefont {A.}~\bibnamefont {Zunger}},\
  }\href {\doibase 10.1038/nmat4973} {\bibfield  {journal} {\bibinfo  {journal}
  {Nature Materials}\ }\textbf {\bibinfo {volume} {16}},\ \bibinfo {pages}
  {964} (\bibinfo {year} {2017})}\BibitemShut {NoStop}%
\bibitem [{\citenamefont {G.~Pacchioni}(85 0)}]{16}%
  \BibitemOpen
  \bibfield  {author} {\bibinfo {author} {\bibfnamefont {D.~L.~G.}\
  \bibnamefont {G.~Pacchioni}, \bibfnamefont {L.~Skuja}},\ }\href@noop {}
  {\emph {\bibinfo {title} {Defects in \ce{SiO2} and Related Dielectrics:
  Science and Technology}}}\ (\bibinfo  {publisher} {Springer Dordrecht},\
  \bibinfo {year} {{2012, ISBN 978-0-7923-6685-0}})\BibitemShut {NoStop}%
\bibitem [{\citenamefont {Raghavachari}\ \emph {et~al.}(2002)\citenamefont
  {Raghavachari}, \citenamefont {Ricci},\ and\ \citenamefont {Pacchioni}}]{17}%
  \BibitemOpen
  \bibfield  {author} {\bibinfo {author} {\bibfnamefont {K.}~\bibnamefont
  {Raghavachari}}, \bibinfo {author} {\bibfnamefont {D.}~\bibnamefont {Ricci}},
  \ and\ \bibinfo {author} {\bibfnamefont {G.}~\bibnamefont {Pacchioni}},\
  }\href {\doibase 10.1063/1.1423664} {\bibfield  {journal} {\bibinfo
  {journal} {The Journal of Chemical Physics}\ }\textbf {\bibinfo {volume}
  {116}},\ \bibinfo {pages} {825} (\bibinfo {year} {2002})},\ \Eprint
  {http://arxiv.org/abs/https://doi.org/10.1063/1.1423664}
  {https://doi.org/10.1063/1.1423664} \BibitemShut {NoStop}%
\bibitem [{\citenamefont {Mills}(1980)}]{18}%
  \BibitemOpen
  \bibfield  {author} {\bibinfo {author} {\bibfnamefont {D.~L.}\ \bibnamefont
  {Mills}},\ }\href {\doibase 10.1063/1.327548} {\bibfield  {journal} {\bibinfo
   {journal} {Journal of Applied Physics}\ }\textbf {\bibinfo {volume} {51}},\
  \bibinfo {pages} {5864} (\bibinfo {year} {1980})},\ \Eprint
  {http://arxiv.org/abs/https://doi.org/10.1063/1.327548}
  {https://doi.org/10.1063/1.327548} \BibitemShut {NoStop}%
\bibitem [{\citenamefont {Ma}\ and\ \citenamefont {Rohlfing}(2008)}]{19}%
  \BibitemOpen
  \bibfield  {author} {\bibinfo {author} {\bibfnamefont {Y.}~\bibnamefont
  {Ma}}\ and\ \bibinfo {author} {\bibfnamefont {M.}~\bibnamefont {Rohlfing}},\
  }\href {\doibase 10.1103/PhysRevB.77.115118} {\bibfield  {journal} {\bibinfo
  {journal} {Phys. Rev. B}\ }\textbf {\bibinfo {volume} {77}},\ \bibinfo
  {pages} {115118} (\bibinfo {year} {2008})}\BibitemShut {NoStop}%
\bibitem [{\citenamefont {Girard}\ \emph {et~al.}(2019)\citenamefont {Girard},
  \citenamefont {Alessi}, \citenamefont {Richard}, \citenamefont
  {Martin-Samos}, \citenamefont {{De Michele}}, \citenamefont {Giacomazzi},
  \citenamefont {Agnello}, \citenamefont {Francesca}, \citenamefont {Morana},
  \citenamefont {Winkler}, \citenamefont {Reghioua}, \citenamefont {Paillet},
  \citenamefont {Cannas}, \citenamefont {Robin}, \citenamefont {Boukenter},\
  and\ \citenamefont {Ouerdane}}]{20}%
  \BibitemOpen
  \bibfield  {author} {\bibinfo {author} {\bibfnamefont {S.}~\bibnamefont
  {Girard}}, \bibinfo {author} {\bibfnamefont {A.}~\bibnamefont {Alessi}},
  \bibinfo {author} {\bibfnamefont {N.}~\bibnamefont {Richard}}, \bibinfo
  {author} {\bibfnamefont {L.}~\bibnamefont {Martin-Samos}}, \bibinfo {author}
  {\bibfnamefont {V.}~\bibnamefont {{De Michele}}}, \bibinfo {author}
  {\bibfnamefont {L.}~\bibnamefont {Giacomazzi}}, \bibinfo {author}
  {\bibfnamefont {S.}~\bibnamefont {Agnello}}, \bibinfo {author} {\bibfnamefont
  {D.~D.}\ \bibnamefont {Francesca}}, \bibinfo {author} {\bibfnamefont
  {A.}~\bibnamefont {Morana}}, \bibinfo {author} {\bibfnamefont
  {B.}~\bibnamefont {Winkler}}, \bibinfo {author} {\bibfnamefont
  {I.}~\bibnamefont {Reghioua}}, \bibinfo {author} {\bibfnamefont
  {P.}~\bibnamefont {Paillet}}, \bibinfo {author} {\bibfnamefont
  {M.}~\bibnamefont {Cannas}}, \bibinfo {author} {\bibfnamefont
  {T.}~\bibnamefont {Robin}}, \bibinfo {author} {\bibfnamefont
  {A.}~\bibnamefont {Boukenter}}, \ and\ \bibinfo {author} {\bibfnamefont
  {Y.}~\bibnamefont {Ouerdane}},\ }\href {\doibase
  https://doi.org/10.1016/j.revip.2019.100032} {\bibfield  {journal} {\bibinfo
  {journal} {Reviews in Physics}\ }\textbf {\bibinfo {volume} {4}},\ \bibinfo
  {pages} {100032} (\bibinfo {year} {2019})}\BibitemShut {NoStop}%
\bibitem [{\citenamefont {Bourrellier}\ \emph {et~al.}(2016)\citenamefont
  {Bourrellier}, \citenamefont {Meuret}, \citenamefont {Tararan}, \citenamefont
  {Stéphan}, \citenamefont {Kociak}, \citenamefont {Tizei},\ and\
  \citenamefont {Zobelli}}]{21}%
  \BibitemOpen
  \bibfield  {author} {\bibinfo {author} {\bibfnamefont {R.}~\bibnamefont
  {Bourrellier}}, \bibinfo {author} {\bibfnamefont {S.}~\bibnamefont {Meuret}},
  \bibinfo {author} {\bibfnamefont {A.}~\bibnamefont {Tararan}}, \bibinfo
  {author} {\bibfnamefont {O.}~\bibnamefont {Stéphan}}, \bibinfo {author}
  {\bibfnamefont {M.}~\bibnamefont {Kociak}}, \bibinfo {author} {\bibfnamefont
  {L.~H.~G.}\ \bibnamefont {Tizei}}, \ and\ \bibinfo {author} {\bibfnamefont
  {A.}~\bibnamefont {Zobelli}},\ }\href {\doibase 10.1021/acs.nanolett.6b01368}
  {\bibfield  {journal} {\bibinfo  {journal} {Nano Letters}\ }\textbf {\bibinfo
  {volume} {16}},\ \bibinfo {pages} {4317} (\bibinfo {year} {2016})},\ \bibinfo
  {note} {pMID: 27299915},\ \Eprint
  {http://arxiv.org/abs/https://doi.org/10.1021/acs.nanolett.6b01368}
  {https://doi.org/10.1021/acs.nanolett.6b01368} \BibitemShut {NoStop}%
\bibitem [{\citenamefont {Tuller}\ and\ \citenamefont {Bishop}(2011)}]{22}%
  \BibitemOpen
  \bibfield  {author} {\bibinfo {author} {\bibfnamefont {H.~L.}\ \bibnamefont
  {Tuller}}\ and\ \bibinfo {author} {\bibfnamefont {S.~R.}\ \bibnamefont
  {Bishop}},\ }\href {\doibase 10.1146/annurev-matsci-062910-100442} {\bibfield
   {journal} {\bibinfo  {journal} {Annual Review of Materials Research}\
  }\textbf {\bibinfo {volume} {41}},\ \bibinfo {pages} {369} (\bibinfo {year}
  {2011})},\ \Eprint
  {http://arxiv.org/abs/https://doi.org/10.1146/annurev-matsci-062910-100442}
  {https://doi.org/10.1146/annurev-matsci-062910-100442} \BibitemShut {NoStop}%
\bibitem [{\citenamefont {Brecher}\ \emph {et~al.}(1990)\citenamefont
  {Brecher}, \citenamefont {Wei},\ and\ \citenamefont {Rhodes}}]{23}%
  \BibitemOpen
  \bibfield  {author} {\bibinfo {author} {\bibfnamefont {C.}~\bibnamefont
  {Brecher}}, \bibinfo {author} {\bibfnamefont {G.~C.}\ \bibnamefont {Wei}}, \
  and\ \bibinfo {author} {\bibfnamefont {W.~H.}\ \bibnamefont {Rhodes}},\
  }\href {\doibase https://doi.org/10.1111/j.1151-2916.1990.tb09784.x}
  {\bibfield  {journal} {\bibinfo  {journal} {Journal of the American Ceramic
  Society}\ }\textbf {\bibinfo {volume} {73}},\ \bibinfo {pages} {1473}
  (\bibinfo {year} {1990})},\ \Eprint
  {http://arxiv.org/abs/https://ceramics.onlinelibrary.wiley.com/doi/pdf/10.1111/j.1151-2916.1990.tb09784.x}
  {https://ceramics.onlinelibrary.wiley.com/doi/pdf/10.1111/j.1151-2916.1990.tb09784.x}
  \BibitemShut {NoStop}%
\bibitem [{\citenamefont {Tuller}(2003)}]{24}%
  \BibitemOpen
  \bibfield  {author} {\bibinfo {author} {\bibfnamefont {H.~L.}\ \bibnamefont
  {Tuller}},\ }\href {\doibase https://doi.org/10.1016/S0013-4686(03)00352-9}
  {\bibfield  {journal} {\bibinfo  {journal} {Electrochimica Acta}\ }\textbf
  {\bibinfo {volume} {48}},\ \bibinfo {pages} {2879} (\bibinfo {year}
  {2003})},\ \bibinfo {note} {electrochemistry in Molecular and Microscopic
  Dimensions}\BibitemShut {NoStop}%
\bibitem [{\citenamefont {Aidhy}\ \emph {et~al.}(2015)\citenamefont {Aidhy},
  \citenamefont {Sachan}, \citenamefont {Zarkadoula}, \citenamefont
  {Pakarinen}, \citenamefont {Chisholm}, \citenamefont {Zhang},\ and\
  \citenamefont {Weber}}]{25}%
  \BibitemOpen
  \bibfield  {author} {\bibinfo {author} {\bibfnamefont {D.~S.}\ \bibnamefont
  {Aidhy}}, \bibinfo {author} {\bibfnamefont {R.}~\bibnamefont {Sachan}},
  \bibinfo {author} {\bibfnamefont {E.}~\bibnamefont {Zarkadoula}}, \bibinfo
  {author} {\bibfnamefont {O.}~\bibnamefont {Pakarinen}}, \bibinfo {author}
  {\bibfnamefont {M.~F.}\ \bibnamefont {Chisholm}}, \bibinfo {author}
  {\bibfnamefont {Y.}~\bibnamefont {Zhang}}, \ and\ \bibinfo {author}
  {\bibfnamefont {W.~J.}\ \bibnamefont {Weber}},\ }\href {\doibase
  10.1038/srep16297} {\bibfield  {journal} {\bibinfo  {journal} {Scientific
  Reports}\ }\textbf {\bibinfo {volume} {5}},\ \bibinfo {pages} {16297}
  (\bibinfo {year} {2015})}\BibitemShut {NoStop}%
\bibitem [{\citenamefont {Zhang}\ \emph {et~al.}(2020)\citenamefont {Zhang},
  \citenamefont {Tao}, \citenamefont {Xie}, \citenamefont {Wang}, \citenamefont
  {Zou}, \citenamefont {Chen}, \citenamefont {Wang}, \citenamefont {Jia},\ and\
  \citenamefont {Wang}}]{26}%
  \BibitemOpen
  \bibfield  {author} {\bibinfo {author} {\bibfnamefont {Y.}~\bibnamefont
  {Zhang}}, \bibinfo {author} {\bibfnamefont {L.}~\bibnamefont {Tao}}, \bibinfo
  {author} {\bibfnamefont {C.}~\bibnamefont {Xie}}, \bibinfo {author}
  {\bibfnamefont {D.}~\bibnamefont {Wang}}, \bibinfo {author} {\bibfnamefont
  {Y.}~\bibnamefont {Zou}}, \bibinfo {author} {\bibfnamefont {R.}~\bibnamefont
  {Chen}}, \bibinfo {author} {\bibfnamefont {Y.}~\bibnamefont {Wang}}, \bibinfo
  {author} {\bibfnamefont {C.}~\bibnamefont {Jia}}, \ and\ \bibinfo {author}
  {\bibfnamefont {S.}~\bibnamefont {Wang}},\ }\href {\doibase
  https://doi.org/10.1002/adma.201905923} {\bibfield  {journal} {\bibinfo
  {journal} {Advanced Materials}\ }\textbf {\bibinfo {volume} {32}},\ \bibinfo
  {pages} {1905923} (\bibinfo {year} {2020})},\ \Eprint
  {http://arxiv.org/abs/https://onlinelibrary.wiley.com/doi/pdf/10.1002/adma.201905923}
  {https://onlinelibrary.wiley.com/doi/pdf/10.1002/adma.201905923} \BibitemShut
  {NoStop}%
\bibitem [{\citenamefont {Li}\ \emph {et~al.}(2019)\citenamefont {Li},
  \citenamefont {Hussain}, \citenamefont {Cui}, \citenamefont {Li},\ and\
  \citenamefont {Yang}}]{27}%
  \BibitemOpen
  \bibfield  {author} {\bibinfo {author} {\bibfnamefont {P.}~\bibnamefont
  {Li}}, \bibinfo {author} {\bibfnamefont {F.}~\bibnamefont {Hussain}},
  \bibinfo {author} {\bibfnamefont {P.}~\bibnamefont {Cui}}, \bibinfo {author}
  {\bibfnamefont {Z.}~\bibnamefont {Li}}, \ and\ \bibinfo {author}
  {\bibfnamefont {J.}~\bibnamefont {Yang}},\ }\href {\doibase
  10.1103/PhysRevMaterials.3.115402} {\bibfield  {journal} {\bibinfo  {journal}
  {Phys. Rev. Mater.}\ }\textbf {\bibinfo {volume} {3}},\ \bibinfo {pages}
  {115402} (\bibinfo {year} {2019})}\BibitemShut {NoStop}%
\bibitem [{\citenamefont {Defferriere}\ \emph {et~al.}(2021)\citenamefont
  {Defferriere}, \citenamefont {Kalaev}, \citenamefont {Rupp},\ and\
  \citenamefont {Tuller}}]{28}%
  \BibitemOpen
  \bibfield  {author} {\bibinfo {author} {\bibfnamefont {T.}~\bibnamefont
  {Defferriere}}, \bibinfo {author} {\bibfnamefont {D.}~\bibnamefont {Kalaev}},
  \bibinfo {author} {\bibfnamefont {J.~L.~M.}\ \bibnamefont {Rupp}}, \ and\
  \bibinfo {author} {\bibfnamefont {H.~L.}\ \bibnamefont {Tuller}},\ }\href
  {\doibase https://doi.org/10.1002/adfm.202005640} {\bibfield  {journal}
  {\bibinfo  {journal} {Advanced Functional Materials}\ }\textbf {\bibinfo
  {volume} {31}},\ \bibinfo {pages} {2005640} (\bibinfo {year} {2021})},\
  \Eprint
  {http://arxiv.org/abs/https://onlinelibrary.wiley.com/doi/pdf/10.1002/adfm.202005640}
  {https://onlinelibrary.wiley.com/doi/pdf/10.1002/adfm.202005640} \BibitemShut
  {NoStop}%
\bibitem [{\citenamefont {Li}\ \emph {et~al.}(2020)\citenamefont {Li},
  \citenamefont {Shu}, \citenamefont {Hu}, \citenamefont {Ran}, \citenamefont
  {Li}, \citenamefont {Zheng},\ and\ \citenamefont {Long}}]{29}%
  \BibitemOpen
  \bibfield  {author} {\bibinfo {author} {\bibfnamefont {J.}~\bibnamefont
  {Li}}, \bibinfo {author} {\bibfnamefont {C.}~\bibnamefont {Shu}}, \bibinfo
  {author} {\bibfnamefont {A.}~\bibnamefont {Hu}}, \bibinfo {author}
  {\bibfnamefont {Z.}~\bibnamefont {Ran}}, \bibinfo {author} {\bibfnamefont
  {M.}~\bibnamefont {Li}}, \bibinfo {author} {\bibfnamefont {R.}~\bibnamefont
  {Zheng}}, \ and\ \bibinfo {author} {\bibfnamefont {J.}~\bibnamefont {Long}},\
  }\href {\doibase https://doi.org/10.1016/j.cej.2019.122678} {\bibfield
  {journal} {\bibinfo  {journal} {Chemical Engineering Journal}\ }\textbf
  {\bibinfo {volume} {381}},\ \bibinfo {pages} {122678} (\bibinfo {year}
  {2020})}\BibitemShut {NoStop}%
\bibitem [{\citenamefont {Marrocchelli}\ \emph {et~al.}(2012)\citenamefont
  {Marrocchelli}, \citenamefont {Bishop}, \citenamefont {Tuller},\ and\
  \citenamefont {Yildiz}}]{30}%
  \BibitemOpen
  \bibfield  {author} {\bibinfo {author} {\bibfnamefont {D.}~\bibnamefont
  {Marrocchelli}}, \bibinfo {author} {\bibfnamefont {S.~R.}\ \bibnamefont
  {Bishop}}, \bibinfo {author} {\bibfnamefont {H.~L.}\ \bibnamefont {Tuller}},
  \ and\ \bibinfo {author} {\bibfnamefont {B.}~\bibnamefont {Yildiz}},\ }\href
  {\doibase https://doi.org/10.1002/adfm.201102648} {\bibfield  {journal}
  {\bibinfo  {journal} {Advanced Functional Materials}\ }\textbf {\bibinfo
  {volume} {22}},\ \bibinfo {pages} {1958} (\bibinfo {year} {2012})},\ \Eprint
  {http://arxiv.org/abs/https://onlinelibrary.wiley.com/doi/pdf/10.1002/adfm.201102648}
  {https://onlinelibrary.wiley.com/doi/pdf/10.1002/adfm.201102648} \BibitemShut
  {NoStop}%
\bibitem [{\citenamefont {Noureldine}\ \emph {et~al.}(2015)\citenamefont
  {Noureldine}, \citenamefont {Lardhi}, \citenamefont {Ziani}, \citenamefont
  {Harb}, \citenamefont {Cavallo},\ and\ \citenamefont {Takanabe}}]{31}%
  \BibitemOpen
  \bibfield  {author} {\bibinfo {author} {\bibfnamefont {D.}~\bibnamefont
  {Noureldine}}, \bibinfo {author} {\bibfnamefont {S.}~\bibnamefont {Lardhi}},
  \bibinfo {author} {\bibfnamefont {A.}~\bibnamefont {Ziani}}, \bibinfo
  {author} {\bibfnamefont {M.}~\bibnamefont {Harb}}, \bibinfo {author}
  {\bibfnamefont {L.}~\bibnamefont {Cavallo}}, \ and\ \bibinfo {author}
  {\bibfnamefont {K.}~\bibnamefont {Takanabe}},\ }\href {\doibase
  10.1039/C5TC03134F} {\bibfield  {journal} {\bibinfo  {journal} {J. Mater.
  Chem. C}\ }\textbf {\bibinfo {volume} {3}},\ \bibinfo {pages} {12032}
  (\bibinfo {year} {2015})}\BibitemShut {NoStop}%
\bibitem [{\citenamefont {Boehm}\ \emph {et~al.}(2005)\citenamefont {Boehm},
  \citenamefont {Bassat}, \citenamefont {Dordor}, \citenamefont {Mauvy},
  \citenamefont {Grenier},\ and\ \citenamefont {Stevens}}]{32}%
  \BibitemOpen
  \bibfield  {author} {\bibinfo {author} {\bibfnamefont {E.}~\bibnamefont
  {Boehm}}, \bibinfo {author} {\bibfnamefont {J.-M.}\ \bibnamefont {Bassat}},
  \bibinfo {author} {\bibfnamefont {P.}~\bibnamefont {Dordor}}, \bibinfo
  {author} {\bibfnamefont {F.}~\bibnamefont {Mauvy}}, \bibinfo {author}
  {\bibfnamefont {J.-C.}\ \bibnamefont {Grenier}}, \ and\ \bibinfo {author}
  {\bibfnamefont {P.}~\bibnamefont {Stevens}},\ }\href {\doibase
  https://doi.org/10.1016/j.ssi.2005.06.033} {\bibfield  {journal} {\bibinfo
  {journal} {Solid State Ionics}\ }\textbf {\bibinfo {volume} {176}},\ \bibinfo
  {pages} {2717} (\bibinfo {year} {2005})}\BibitemShut {NoStop}%
\bibitem [{\citenamefont {Rogal}\ \emph {et~al.}(2014)\citenamefont {Rogal},
  \citenamefont {Divinski}, \citenamefont {Finnis}, \citenamefont {Glensk},
  \citenamefont {Neugebauer}, \citenamefont {Perepezko}, \citenamefont
  {Schuwalow}, \citenamefont {Sluiter},\ and\ \citenamefont
  {Sundman}}]{Rogal2014}%
  \BibitemOpen
  \bibfield  {author} {\bibinfo {author} {\bibfnamefont {J.}~\bibnamefont
  {Rogal}}, \bibinfo {author} {\bibfnamefont {S.}~\bibnamefont {Divinski}},
  \bibinfo {author} {\bibfnamefont {M.}~\bibnamefont {Finnis}}, \bibinfo
  {author} {\bibfnamefont {A.}~\bibnamefont {Glensk}}, \bibinfo {author}
  {\bibfnamefont {J.}~\bibnamefont {Neugebauer}}, \bibinfo {author}
  {\bibfnamefont {J.}~\bibnamefont {Perepezko}}, \bibinfo {author}
  {\bibfnamefont {S.}~\bibnamefont {Schuwalow}}, \bibinfo {author}
  {\bibfnamefont {M.}~\bibnamefont {Sluiter}}, \ and\ \bibinfo {author}
  {\bibfnamefont {B.}~\bibnamefont {Sundman}},\ }\href {\doibase
  10.1002/pssb.201350155} {\bibfield  {journal} {\bibinfo  {journal} {Physica
  Status Solidi. B: Basic Research}\ }\textbf {\bibinfo {volume} {251}},\
  \bibinfo {pages} {97} (\bibinfo {year} {2014})},\ \bibinfo {note} {harvest
  NEO Article first published online: 27 12 2013}\BibitemShut {NoStop}%
\bibitem [{\citenamefont {Anand}\ \emph {et~al.}(2022)\citenamefont {Anand},
  \citenamefont {Toriyama}, \citenamefont {Wolverton}, \citenamefont {Haile},\
  and\ \citenamefont {Snyder}}]{57}%
  \BibitemOpen
  \bibfield  {author} {\bibinfo {author} {\bibfnamefont {S.}~\bibnamefont
  {Anand}}, \bibinfo {author} {\bibfnamefont {M.~Y.}\ \bibnamefont {Toriyama}},
  \bibinfo {author} {\bibfnamefont {C.}~\bibnamefont {Wolverton}}, \bibinfo
  {author} {\bibfnamefont {S.~M.}\ \bibnamefont {Haile}}, \ and\ \bibinfo
  {author} {\bibfnamefont {G.~J.}\ \bibnamefont {Snyder}},\ }\href {\doibase
  10.1021/accountsmr.2c00044} {\bibfield  {journal} {\bibinfo  {journal}
  {Accounts of Materials Research}\ }\textbf {\bibinfo {volume} {3}},\ \bibinfo
  {pages} {685} (\bibinfo {year} {2022})},\ \Eprint
  {http://arxiv.org/abs/https://doi.org/10.1021/accountsmr.2c00044}
  {https://doi.org/10.1021/accountsmr.2c00044} \BibitemShut {NoStop}%
\bibitem [{\citenamefont {Toriyama}\ \emph {et~al.}(2022)\citenamefont
  {Toriyama}, \citenamefont {Brod},\ and\ \citenamefont {Snyder}}]{58}%
  \BibitemOpen
  \bibfield  {author} {\bibinfo {author} {\bibfnamefont {M.~Y.}\ \bibnamefont
  {Toriyama}}, \bibinfo {author} {\bibfnamefont {M.~K.}\ \bibnamefont {Brod}},
  \ and\ \bibinfo {author} {\bibfnamefont {G.~J.}\ \bibnamefont {Snyder}},\
  }\href {\doibase https://doi.org/10.1002/cnma.202200222} {\bibfield
  {journal} {\bibinfo  {journal} {ChemNanoMat}\ }\textbf {\bibinfo {volume}
  {8}},\ \bibinfo {pages} {e202200222} (\bibinfo {year} {2022})},\ \Eprint
  {http://arxiv.org/abs/https://onlinelibrary.wiley.com/doi/pdf/10.1002/cnma.202200222}
  {https://onlinelibrary.wiley.com/doi/pdf/10.1002/cnma.202200222} \BibitemShut
  {NoStop}%
\bibitem [{\citenamefont {Freysoldt}\ \emph {et~al.}(2014)\citenamefont
  {Freysoldt}, \citenamefont {Grabowski}, \citenamefont {Hickel}, \citenamefont
  {Neugebauer}, \citenamefont {Kresse}, \citenamefont {Janotti},\ and\
  \citenamefont {Van~de Walle}}]{Freysoldt2014}%
  \BibitemOpen
  \bibfield  {author} {\bibinfo {author} {\bibfnamefont {C.}~\bibnamefont
  {Freysoldt}}, \bibinfo {author} {\bibfnamefont {B.}~\bibnamefont
  {Grabowski}}, \bibinfo {author} {\bibfnamefont {T.}~\bibnamefont {Hickel}},
  \bibinfo {author} {\bibfnamefont {J.}~\bibnamefont {Neugebauer}}, \bibinfo
  {author} {\bibfnamefont {G.}~\bibnamefont {Kresse}}, \bibinfo {author}
  {\bibfnamefont {A.}~\bibnamefont {Janotti}}, \ and\ \bibinfo {author}
  {\bibfnamefont {C.~G.}\ \bibnamefont {Van~de Walle}},\ }\href {\doibase
  10.1103/RevModPhys.86.253} {\bibfield  {journal} {\bibinfo  {journal} {Rev.
  Mod. Phys.}\ }\textbf {\bibinfo {volume} {86}},\ \bibinfo {pages} {253}
  (\bibinfo {year} {2014})}\BibitemShut {NoStop}%
\bibitem [{\citenamefont {Hine}\ \emph {et~al.}(2009)\citenamefont {Hine},
  \citenamefont {Frensch}, \citenamefont {Foulkes},\ and\ \citenamefont
  {Finnis}}]{52}%
  \BibitemOpen
  \bibfield  {author} {\bibinfo {author} {\bibfnamefont {N.~D.~M.}\
  \bibnamefont {Hine}}, \bibinfo {author} {\bibfnamefont {K.}~\bibnamefont
  {Frensch}}, \bibinfo {author} {\bibfnamefont {W.~M.~C.}\ \bibnamefont
  {Foulkes}}, \ and\ \bibinfo {author} {\bibfnamefont {M.~W.}\ \bibnamefont
  {Finnis}},\ }\href {\doibase 10.1103/PhysRevB.79.024112} {\bibfield
  {journal} {\bibinfo  {journal} {Phys. Rev. B}\ }\textbf {\bibinfo {volume}
  {79}},\ \bibinfo {pages} {024112} (\bibinfo {year} {2009})}\BibitemShut
  {NoStop}%
\bibitem [{\citenamefont {Ramprasad}\ \emph {et~al.}(2012)\citenamefont
  {Ramprasad}, \citenamefont {Zhu}, \citenamefont {Rinke},\ and\ \citenamefont
  {Scheffler}}]{53}%
  \BibitemOpen
  \bibfield  {author} {\bibinfo {author} {\bibfnamefont {R.}~\bibnamefont
  {Ramprasad}}, \bibinfo {author} {\bibfnamefont {H.}~\bibnamefont {Zhu}},
  \bibinfo {author} {\bibfnamefont {P.}~\bibnamefont {Rinke}}, \ and\ \bibinfo
  {author} {\bibfnamefont {M.}~\bibnamefont {Scheffler}},\ }\href {\doibase
  10.1103/PhysRevLett.108.066404} {\bibfield  {journal} {\bibinfo  {journal}
  {Phys. Rev. Lett.}\ }\textbf {\bibinfo {volume} {108}},\ \bibinfo {pages}
  {066404} (\bibinfo {year} {2012})}\BibitemShut {NoStop}%
\bibitem [{\citenamefont {Lany}\ and\ \citenamefont {Zunger}(2008)}]{54}%
  \BibitemOpen
  \bibfield  {author} {\bibinfo {author} {\bibfnamefont {S.}~\bibnamefont
  {Lany}}\ and\ \bibinfo {author} {\bibfnamefont {A.}~\bibnamefont {Zunger}},\
  }\href {\doibase 10.1103/PhysRevB.78.235104} {\bibfield  {journal} {\bibinfo
  {journal} {Phys. Rev. B}\ }\textbf {\bibinfo {volume} {78}},\ \bibinfo
  {pages} {235104} (\bibinfo {year} {2008})}\BibitemShut {NoStop}%
\bibitem [{\citenamefont {Lany}\ and\ \citenamefont {Zunger}(2010)}]{55}%
  \BibitemOpen
  \bibfield  {author} {\bibinfo {author} {\bibfnamefont {S.}~\bibnamefont
  {Lany}}\ and\ \bibinfo {author} {\bibfnamefont {A.}~\bibnamefont {Zunger}},\
  }\href {\doibase 10.1103/PhysRevB.81.113201} {\bibfield  {journal} {\bibinfo
  {journal} {Phys. Rev. B}\ }\textbf {\bibinfo {volume} {81}},\ \bibinfo
  {pages} {113201} (\bibinfo {year} {2010})}\BibitemShut {NoStop}%
\bibitem [{\citenamefont {Lany}\ and\ \citenamefont {Zunger}(2009)}]{56}%
  \BibitemOpen
  \bibfield  {author} {\bibinfo {author} {\bibfnamefont {S.}~\bibnamefont
  {Lany}}\ and\ \bibinfo {author} {\bibfnamefont {A.}~\bibnamefont {Zunger}},\
  }\href {\doibase 10.1088/0965-0393/17/8/084002} {\bibfield  {journal}
  {\bibinfo  {journal} {Modelling and Simulation in Materials Science and
  Engineering}\ }\textbf {\bibinfo {volume} {17}},\ \bibinfo {pages} {084002}
  (\bibinfo {year} {2009})}\BibitemShut {NoStop}%
\bibitem [{\citenamefont {Kattner}(2016)}]{101}%
  \BibitemOpen
  \bibfield  {author} {\bibinfo {author} {\bibfnamefont {U.~R.}\ \bibnamefont
  {Kattner}},\ }\href@noop {} {\enquote {\bibinfo {title} {{THE CALPHAD METHOD
  AND ITS ROLE IN MATERIAL AND PROCESS DEVELOPMENT}},}\ }\bibinfo
  {howpublished} {https://tecnologiammm.com.br/doi/10.4322/2176-1523.1059}
  (\bibinfo {year} {2016})\BibitemShut {NoStop}%
\bibitem [{\citenamefont {Kattner}(1997)}]{102}%
  \BibitemOpen
  \bibfield  {author} {\bibinfo {author} {\bibfnamefont {U.~R.}\ \bibnamefont
  {Kattner}},\ }\href {\doibase 10.1007/s11837-997-0024-5} {\bibfield
  {journal} {\bibinfo  {journal} {JOM}\ }\textbf {\bibinfo {volume} {49}},\
  \bibinfo {pages} {14} (\bibinfo {year} {1997})}\BibitemShut {NoStop}%
\bibitem [{\citenamefont {Hickel}\ \emph {et~al.}(2014)\citenamefont {Hickel},
  \citenamefont {Kattner},\ and\ \citenamefont {Fries}}]{103}%
  \BibitemOpen
  \bibfield  {author} {\bibinfo {author} {\bibfnamefont {T.}~\bibnamefont
  {Hickel}}, \bibinfo {author} {\bibfnamefont {U.~R.}\ \bibnamefont {Kattner}},
  \ and\ \bibinfo {author} {\bibfnamefont {S.~G.}\ \bibnamefont {Fries}},\
  }\href {\doibase https://doi.org/10.1002/pssb.201470107} {\bibfield
  {journal} {\bibinfo  {journal} {physica status solidi (b)}\ }\textbf
  {\bibinfo {volume} {251}},\ \bibinfo {pages} {9} (\bibinfo {year} {2014})},\
  \Eprint
  {http://arxiv.org/abs/https://onlinelibrary.wiley.com/doi/pdf/10.1002/pssb.201470107}
  {https://onlinelibrary.wiley.com/doi/pdf/10.1002/pssb.201470107} \BibitemShut
  {NoStop}%
\bibitem [{\citenamefont {Campbell}\ \emph {et~al.}(2014)\citenamefont
  {Campbell}, \citenamefont {Kattner},\ and\ \citenamefont {Liu}}]{104}%
  \BibitemOpen
  \bibfield  {author} {\bibinfo {author} {\bibfnamefont {C.~E.}\ \bibnamefont
  {Campbell}}, \bibinfo {author} {\bibfnamefont {U.~R.}\ \bibnamefont
  {Kattner}}, \ and\ \bibinfo {author} {\bibfnamefont {Z.-K.}\ \bibnamefont
  {Liu}},\ }\href {\doibase 10.1186/2193-9772-3-12} {\bibfield  {journal}
  {\bibinfo  {journal} {Integrating Materials and Manufacturing Innovation}\
  }\textbf {\bibinfo {volume} {3}},\ \bibinfo {pages} {158} (\bibinfo {year}
  {2014})}\BibitemShut {NoStop}%
\bibitem [{\citenamefont {Liu}\ and\ \citenamefont {Wang}(2016)}]{105}%
  \BibitemOpen
  \bibfield  {author} {\bibinfo {author} {\bibfnamefont {Z.-K.}\ \bibnamefont
  {Liu}}\ and\ \bibinfo {author} {\bibfnamefont {Y.}~\bibnamefont {Wang}},\
  }\href {\doibase 10.1017/CBO9781139018265} {\emph {\bibinfo {title}
  {{Computational Thermodynamics of Materials}}}}\ (\bibinfo  {publisher}
  {Cambridge University Press},\ \bibinfo {year} {2016})\BibitemShut {NoStop}%
\bibitem [{\citenamefont {Sundman}\ \emph {et~al.}(2018)\citenamefont
  {Sundman}, \citenamefont {Chen},\ and\ \citenamefont {Du}}]{106}%
  \BibitemOpen
  \bibfield  {author} {\bibinfo {author} {\bibfnamefont {B.}~\bibnamefont
  {Sundman}}, \bibinfo {author} {\bibfnamefont {Q.}~\bibnamefont {Chen}}, \
  and\ \bibinfo {author} {\bibfnamefont {Y.}~\bibnamefont {Du}},\ }\href
  {\doibase 10.1007/s11669-018-0671-y} {\bibfield  {journal} {\bibinfo
  {journal} {Journal of Phase Equilibria and Diffusion}\ }\textbf {\bibinfo
  {volume} {39}},\ \bibinfo {pages} {678} (\bibinfo {year} {2018})}\BibitemShut
  {NoStop}%
\bibitem [{\citenamefont {{van de Walle}}\ \emph {et~al.}(2017)\citenamefont
  {{van de Walle}}, \citenamefont {Sun}, \citenamefont {Hong},\ and\
  \citenamefont {Kadkhodaei}}]{107}%
  \BibitemOpen
  \bibfield  {author} {\bibinfo {author} {\bibfnamefont {A.}~\bibnamefont {{van
  de Walle}}}, \bibinfo {author} {\bibfnamefont {R.}~\bibnamefont {Sun}},
  \bibinfo {author} {\bibfnamefont {Q.-J.}\ \bibnamefont {Hong}}, \ and\
  \bibinfo {author} {\bibfnamefont {S.}~\bibnamefont {Kadkhodaei}},\ }\href
  {\doibase https://doi.org/10.1016/j.calphad.2017.05.005} {\bibfield
  {journal} {\bibinfo  {journal} {Calphad}\ }\textbf {\bibinfo {volume} {58}},\
  \bibinfo {pages} {70} (\bibinfo {year} {2017})}\BibitemShut {NoStop}%
\bibitem [{\citenamefont {Liu}(2009)}]{108}%
  \BibitemOpen
  \bibfield  {author} {\bibinfo {author} {\bibfnamefont {Z.-K.}\ \bibnamefont
  {Liu}},\ }\href {\doibase 10.1007/s11669-009-9570-6} {\bibfield  {journal}
  {\bibinfo  {journal} {Journal of Phase Equilibria and Diffusion}\ }\textbf
  {\bibinfo {volume} {30}},\ \bibinfo {pages} {517} (\bibinfo {year}
  {2009})}\BibitemShut {NoStop}%
\bibitem [{\citenamefont {Bajaj}\ \emph {et~al.}(2011)\citenamefont {Bajaj},
  \citenamefont {Landa}, \citenamefont {Söderlind}, \citenamefont {Turchi},\
  and\ \citenamefont {Arróyave}}]{109}%
  \BibitemOpen
  \bibfield  {author} {\bibinfo {author} {\bibfnamefont {S.}~\bibnamefont
  {Bajaj}}, \bibinfo {author} {\bibfnamefont {A.}~\bibnamefont {Landa}},
  \bibinfo {author} {\bibfnamefont {P.}~\bibnamefont {Söderlind}}, \bibinfo
  {author} {\bibfnamefont {P.~E.}\ \bibnamefont {Turchi}}, \ and\ \bibinfo
  {author} {\bibfnamefont {R.}~\bibnamefont {Arróyave}},\ }\href {\doibase
  https://doi.org/10.1016/j.jnucmat.2011.08.050} {\bibfield  {journal}
  {\bibinfo  {journal} {Journal of Nuclear Materials}\ }\textbf {\bibinfo
  {volume} {419}},\ \bibinfo {pages} {177} (\bibinfo {year}
  {2011})}\BibitemShut {NoStop}%
\bibitem [{\citenamefont {Li}\ \emph {et~al.}(2021)\citenamefont {Li},
  \citenamefont {Li}, \citenamefont {Chen}, \citenamefont {Ren}, \citenamefont
  {Wang}, \citenamefont {Liu}, \citenamefont {Zhang},\ and\ \citenamefont
  {Chen}}]{110}%
  \BibitemOpen
  \bibfield  {author} {\bibinfo {author} {\bibfnamefont {X.}~\bibnamefont
  {Li}}, \bibinfo {author} {\bibfnamefont {Z.}~\bibnamefont {Li}}, \bibinfo
  {author} {\bibfnamefont {C.}~\bibnamefont {Chen}}, \bibinfo {author}
  {\bibfnamefont {Z.}~\bibnamefont {Ren}}, \bibinfo {author} {\bibfnamefont
  {C.}~\bibnamefont {Wang}}, \bibinfo {author} {\bibfnamefont {X.}~\bibnamefont
  {Liu}}, \bibinfo {author} {\bibfnamefont {Q.}~\bibnamefont {Zhang}}, \ and\
  \bibinfo {author} {\bibfnamefont {S.}~\bibnamefont {Chen}},\ }\href {\doibase
  10.1039/D0TA12620A} {\bibfield  {journal} {\bibinfo  {journal} {J. Mater.
  Chem. A}\ }\textbf {\bibinfo {volume} {9}},\ \bibinfo {pages} {6634}
  (\bibinfo {year} {2021})}\BibitemShut {NoStop}%
\bibitem [{\citenamefont {van~de Walle}(2013)}]{111}%
  \BibitemOpen
  \bibfield  {author} {\bibinfo {author} {\bibfnamefont {A.}~\bibnamefont
  {van~de Walle}},\ }\href {\doibase 10.1007/s11837-013-0764-3} {\bibfield
  {journal} {\bibinfo  {journal} {JOM}\ }\textbf {\bibinfo {volume} {65}},\
  \bibinfo {pages} {1523} (\bibinfo {year} {2013})}\BibitemShut {NoStop}%
\bibitem [{\citenamefont {van~de Walle}\ \emph {et~al.}(2018)\citenamefont
  {van~de Walle}, \citenamefont {Nataraj},\ and\ \citenamefont {kui
  Liu}}]{112}%
  \BibitemOpen
  \bibfield  {author} {\bibinfo {author} {\bibfnamefont {A.}~\bibnamefont
  {van~de Walle}}, \bibinfo {author} {\bibfnamefont {C.}~\bibnamefont
  {Nataraj}}, \ and\ \bibinfo {author} {\bibfnamefont {Z.}~\bibnamefont {kui
  Liu}},\ }\href@noop {} {\bibfield  {journal} {\bibinfo  {journal} {Calphad}\
  } (\bibinfo {year} {2018})}\BibitemShut {NoStop}%
\bibitem [{\citenamefont {Hillert}\ and\ \citenamefont
  {Staffansson}(1970)}]{hillert1970regular}%
  \BibitemOpen
  \bibfield  {author} {\bibinfo {author} {\bibfnamefont {M.}~\bibnamefont
  {Hillert}}\ and\ \bibinfo {author} {\bibfnamefont {L.}~\bibnamefont
  {Staffansson}},\ }\href@noop {} {\bibfield  {journal} {\bibinfo  {journal}
  {Acta chem. scand.}\ }\textbf {\bibinfo {volume} {24}},\ \bibinfo {pages}
  {3618} (\bibinfo {year} {1970})}\BibitemShut {NoStop}%
\bibitem [{\citenamefont {Harvig}\ \emph {et~al.}(1971)\citenamefont {Harvig},
  \citenamefont {Kullberg}, \citenamefont {Roti}, \citenamefont {Okinaka},
  \citenamefont {Kosuge},\ and\ \citenamefont {Kachi}}]{Harvig1971AnEV}%
  \BibitemOpen
  \bibfield  {author} {\bibinfo {author} {\bibfnamefont {H.}~\bibnamefont
  {Harvig}}, \bibinfo {author} {\bibfnamefont {L.}~\bibnamefont {Kullberg}},
  \bibinfo {author} {\bibfnamefont {I.}~\bibnamefont {Roti}}, \bibinfo {author}
  {\bibfnamefont {H.}~\bibnamefont {Okinaka}}, \bibinfo {author} {\bibfnamefont
  {K.}~\bibnamefont {Kosuge}}, \ and\ \bibinfo {author} {\bibfnamefont
  {S.}~\bibnamefont {Kachi}},\ }\href
  {https://api.semanticscholar.org/CorpusID:97643340} {\bibfield  {journal}
  {\bibinfo  {journal} {Acta Chemica Scandinavica}\ }\textbf {\bibinfo {volume}
  {25}},\ \bibinfo {pages} {3199} (\bibinfo {year} {1971})}\BibitemShut
  {NoStop}%
\bibitem [{\citenamefont {Sundman}\ and\ \citenamefont
  {Ågren}(1981)}]{SUNDMAN1981297}%
  \BibitemOpen
  \bibfield  {author} {\bibinfo {author} {\bibfnamefont {B.}~\bibnamefont
  {Sundman}}\ and\ \bibinfo {author} {\bibfnamefont {J.}~\bibnamefont
  {Ågren}},\ }\href {\doibase https://doi.org/10.1016/0022-3697(81)90144-X}
  {\bibfield  {journal} {\bibinfo  {journal} {Journal of Physics and Chemistry
  of Solids}\ }\textbf {\bibinfo {volume} {42}},\ \bibinfo {pages} {297}
  (\bibinfo {year} {1981})}\BibitemShut {NoStop}%
\bibitem [{\citenamefont {Andersson}\ \emph {et~al.}(1986)\citenamefont
  {Andersson}, \citenamefont {Guillermet}, \citenamefont {Hillert},
  \citenamefont {Jansson},\ and\ \citenamefont
  {Sundman}}]{andersson1986compound}%
  \BibitemOpen
  \bibfield  {author} {\bibinfo {author} {\bibfnamefont {J.-O.}\ \bibnamefont
  {Andersson}}, \bibinfo {author} {\bibfnamefont {A.~F.}\ \bibnamefont
  {Guillermet}}, \bibinfo {author} {\bibfnamefont {M.}~\bibnamefont {Hillert}},
  \bibinfo {author} {\bibfnamefont {B.}~\bibnamefont {Jansson}}, \ and\
  \bibinfo {author} {\bibfnamefont {B.}~\bibnamefont {Sundman}},\ }\href@noop
  {} {\bibfield  {journal} {\bibinfo  {journal} {Acta metallurgica}\ }\textbf
  {\bibinfo {volume} {34}},\ \bibinfo {pages} {437} (\bibinfo {year}
  {1986})}\BibitemShut {NoStop}%
\bibitem [{\citenamefont {Hillert}(2001)}]{HILLERT2001161}%
  \BibitemOpen
  \bibfield  {author} {\bibinfo {author} {\bibfnamefont {M.}~\bibnamefont
  {Hillert}},\ }\href {\doibase https://doi.org/10.1016/S0925-8388(00)01481-X}
  {\bibfield  {journal} {\bibinfo  {journal} {Journal of Alloys and Compounds}\
  }\textbf {\bibinfo {volume} {320}},\ \bibinfo {pages} {161} (\bibinfo {year}
  {2001})}\BibitemShut {NoStop}%
\bibitem [{\citenamefont {Frisk}\ and\ \citenamefont
  {Selleby}(2001)}]{FRISK2001177}%
  \BibitemOpen
  \bibfield  {author} {\bibinfo {author} {\bibfnamefont {K.}~\bibnamefont
  {Frisk}}\ and\ \bibinfo {author} {\bibfnamefont {M.}~\bibnamefont
  {Selleby}},\ }\href {\doibase https://doi.org/10.1016/S0925-8388(00)01482-1}
  {\bibfield  {journal} {\bibinfo  {journal} {Journal of Alloys and Compounds}\
  }\textbf {\bibinfo {volume} {320}},\ \bibinfo {pages} {177} (\bibinfo {year}
  {2001})},\ \bibinfo {note} {materials Constitution and Thermochemistry.
  Examples of Methods, Measurements and Applications. In Memoriam Alan
  Prince}\BibitemShut {NoStop}%
\bibitem [{\citenamefont {Oates}\ \emph {et~al.}(1995)\citenamefont {Oates},
  \citenamefont {Eriksson},\ and\ \citenamefont {Wenzl}}]{Oates1995}%
  \BibitemOpen
  \bibfield  {author} {\bibinfo {author} {\bibfnamefont {W.}~\bibnamefont
  {Oates}}, \bibinfo {author} {\bibfnamefont {G.}~\bibnamefont {Eriksson}}, \
  and\ \bibinfo {author} {\bibfnamefont {H.}~\bibnamefont {Wenzl}},\ }\href
  {\doibase https://doi.org/10.1016/0925-8388(94)06018-5} {\bibfield  {journal}
  {\bibinfo  {journal} {Journal of Alloys and Compounds}\ }\textbf {\bibinfo
  {volume} {220}},\ \bibinfo {pages} {48} (\bibinfo {year} {1995})},\ \bibinfo
  {note} {proceedings of the 5th International Meeting on Thermodynamics of
  Alloys}\BibitemShut {NoStop}%
\bibitem [{\citenamefont {Chen}\ \emph {et~al.}(1998)\citenamefont {Chen},
  \citenamefont {Hillert}, \citenamefont {Sundman}, \citenamefont {Oates},
  \citenamefont {Fries},\ and\ \citenamefont {Schmid-Fetzer}}]{Chen1998}%
  \BibitemOpen
  \bibfield  {author} {\bibinfo {author} {\bibfnamefont {Q.}~\bibnamefont
  {Chen}}, \bibinfo {author} {\bibfnamefont {M.}~\bibnamefont {Hillert}},
  \bibinfo {author} {\bibfnamefont {B.}~\bibnamefont {Sundman}}, \bibinfo
  {author} {\bibfnamefont {W.~A.}\ \bibnamefont {Oates}}, \bibinfo {author}
  {\bibfnamefont {S.~G.}\ \bibnamefont {Fries}}, \ and\ \bibinfo {author}
  {\bibfnamefont {R.}~\bibnamefont {Schmid-Fetzer}},\ }\href {\doibase
  10.1007/s11664-998-0128-x} {\bibfield  {journal} {\bibinfo  {journal}
  {Journal of Electronic Materials}\ }\textbf {\bibinfo {volume} {27}},\
  \bibinfo {pages} {961} (\bibinfo {year} {1998})}\BibitemShut {NoStop}%
\bibitem [{\citenamefont {Li}\ and\ \citenamefont {Kerr}(2013)}]{LI20131213}%
  \BibitemOpen
  \bibfield  {author} {\bibinfo {author} {\bibfnamefont {J.}~\bibnamefont
  {Li}}\ and\ \bibinfo {author} {\bibfnamefont {L.~L.}\ \bibnamefont {Kerr}},\
  }\href {\doibase https://doi.org/10.1016/j.optmat.2013.01.014} {\bibfield
  {journal} {\bibinfo  {journal} {Optical Materials}\ }\textbf {\bibinfo
  {volume} {35}},\ \bibinfo {pages} {1213} (\bibinfo {year}
  {2013})}\BibitemShut {NoStop}%
\bibitem [{\citenamefont {Hu}\ \emph {et~al.}(2018)\citenamefont {Hu},
  \citenamefont {{Paz Soldan Palma}}, \citenamefont {Wang}, \citenamefont
  {Firdosy}, \citenamefont {Star}, \citenamefont {Fleurial}, \citenamefont
  {Ravi},\ and\ \citenamefont {Liu}}]{117}%
  \BibitemOpen
  \bibfield  {author} {\bibinfo {author} {\bibfnamefont {Y.}~\bibnamefont
  {Hu}}, \bibinfo {author} {\bibfnamefont {J.}~\bibnamefont {{Paz Soldan
  Palma}}}, \bibinfo {author} {\bibfnamefont {Y.}~\bibnamefont {Wang}},
  \bibinfo {author} {\bibfnamefont {S.}~\bibnamefont {Firdosy}}, \bibinfo
  {author} {\bibfnamefont {K.}~\bibnamefont {Star}}, \bibinfo {author}
  {\bibfnamefont {J.}~\bibnamefont {Fleurial}}, \bibinfo {author}
  {\bibfnamefont {V.}~\bibnamefont {Ravi}}, \ and\ \bibinfo {author}
  {\bibfnamefont {Z.}~\bibnamefont {Liu}},\ }\href {\doibase
  10.1016/j.calphad.2018.03.003} {\bibfield  {journal} {\bibinfo  {journal}
  {Calphad: Computer Coupling of Phase Diagrams and Thermochemistry}\ }\textbf
  {\bibinfo {volume} {61}},\ \bibinfo {pages} {227} (\bibinfo {year}
  {2018})}\BibitemShut {NoStop}%
\bibitem [{\citenamefont {Peters}\ \emph {et~al.}(2019)\citenamefont {Peters},
  \citenamefont {Doak}, \citenamefont {Saal}, \citenamefont {Olson},\ and\
  \citenamefont {Voorhees}}]{Peters2019}%
  \BibitemOpen
  \bibfield  {author} {\bibinfo {author} {\bibfnamefont {M.~C.}\ \bibnamefont
  {Peters}}, \bibinfo {author} {\bibfnamefont {J.~W.}\ \bibnamefont {Doak}},
  \bibinfo {author} {\bibfnamefont {J.~E.}\ \bibnamefont {Saal}}, \bibinfo
  {author} {\bibfnamefont {G.~B.}\ \bibnamefont {Olson}}, \ and\ \bibinfo
  {author} {\bibfnamefont {P.~W.}\ \bibnamefont {Voorhees}},\ }\href {\doibase
  10.1007/s11664-018-6819-z} {\bibfield  {journal} {\bibinfo  {journal}
  {Journal of Electronic Materials}\ }\textbf {\bibinfo {volume} {48}},\
  \bibinfo {pages} {1031} (\bibinfo {year} {2019})}\BibitemShut {NoStop}%
\bibitem [{\citenamefont {Peters}\ \emph {et~al.}(2017)\citenamefont {Peters},
  \citenamefont {Doak}, \citenamefont {Zhang}, \citenamefont {Saal},
  \citenamefont {Olson},\ and\ \citenamefont {Voorhees}}]{119}%
  \BibitemOpen
  \bibfield  {author} {\bibinfo {author} {\bibfnamefont {M.}~\bibnamefont
  {Peters}}, \bibinfo {author} {\bibfnamefont {J.}~\bibnamefont {Doak}},
  \bibinfo {author} {\bibfnamefont {W.-W.}\ \bibnamefont {Zhang}}, \bibinfo
  {author} {\bibfnamefont {J.}~\bibnamefont {Saal}}, \bibinfo {author}
  {\bibfnamefont {G.}~\bibnamefont {Olson}}, \ and\ \bibinfo {author}
  {\bibfnamefont {P.}~\bibnamefont {Voorhees}},\ }\href {\doibase
  https://doi.org/10.1016/j.calphad.2017.05.001} {\bibfield  {journal}
  {\bibinfo  {journal} {Calphad}\ }\textbf {\bibinfo {volume} {58}},\ \bibinfo
  {pages} {17} (\bibinfo {year} {2017})}\BibitemShut {NoStop}%
\bibitem [{\citenamefont {Bajaj}\ \emph {et~al.}(2015)\citenamefont {Bajaj},
  \citenamefont {Pomrehn}, \citenamefont {Doak}, \citenamefont {Gierlotka},
  \citenamefont {jay Wu}, \citenamefont {Chen}, \citenamefont {Wolverton},
  \citenamefont {Goddard},\ and\ \citenamefont {{Jeffrey Snyder}}}]{120}%
  \BibitemOpen
  \bibfield  {author} {\bibinfo {author} {\bibfnamefont {S.}~\bibnamefont
  {Bajaj}}, \bibinfo {author} {\bibfnamefont {G.~S.}\ \bibnamefont {Pomrehn}},
  \bibinfo {author} {\bibfnamefont {J.~W.}\ \bibnamefont {Doak}}, \bibinfo
  {author} {\bibfnamefont {W.}~\bibnamefont {Gierlotka}}, \bibinfo {author}
  {\bibfnamefont {H.}~\bibnamefont {jay Wu}}, \bibinfo {author} {\bibfnamefont
  {S.-W.}\ \bibnamefont {Chen}}, \bibinfo {author} {\bibfnamefont
  {C.}~\bibnamefont {Wolverton}}, \bibinfo {author} {\bibfnamefont {W.~A.}\
  \bibnamefont {Goddard}}, \ and\ \bibinfo {author} {\bibfnamefont
  {G.}~\bibnamefont {{Jeffrey Snyder}}},\ }\href {\doibase
  https://doi.org/10.1016/j.actamat.2015.03.034} {\bibfield  {journal}
  {\bibinfo  {journal} {Acta Materialia}\ }\textbf {\bibinfo {volume} {92}},\
  \bibinfo {pages} {72} (\bibinfo {year} {2015})}\BibitemShut {NoStop}%
\bibitem [{\citenamefont {Chen}\ and\ \citenamefont
  {Hillert}(1996)}]{Chen1996}%
  \BibitemOpen
  \bibfield  {author} {\bibinfo {author} {\bibfnamefont {Q.}~\bibnamefont
  {Chen}}\ and\ \bibinfo {author} {\bibfnamefont {M.}~\bibnamefont {Hillert}},\
  }\href {\doibase https://doi.org/10.1016/S0925-8388(96)02441-3} {\bibfield
  {journal} {\bibinfo  {journal} {Journal of Alloys and Compounds}\ }\textbf
  {\bibinfo {volume} {245}},\ \bibinfo {pages} {125} (\bibinfo {year}
  {1996})}\BibitemShut {NoStop}%
\bibitem [{\citenamefont {Sundman}\ \emph {et~al.}(2011)\citenamefont
  {Sundman}, \citenamefont {Guéneau},\ and\ \citenamefont {Dupin}}]{123}%
  \BibitemOpen
  \bibfield  {author} {\bibinfo {author} {\bibfnamefont {B.}~\bibnamefont
  {Sundman}}, \bibinfo {author} {\bibfnamefont {C.}~\bibnamefont {Guéneau}}, \
  and\ \bibinfo {author} {\bibfnamefont {N.}~\bibnamefont {Dupin}},\ }\href
  {\doibase https://doi.org/10.1016/j.actamat.2011.06.012} {\bibfield
  {journal} {\bibinfo  {journal} {Acta Materialia}\ }\textbf {\bibinfo {volume}
  {59}},\ \bibinfo {pages} {6039} (\bibinfo {year} {2011})}\BibitemShut
  {NoStop}%
\bibitem [{\citenamefont {Guan}\ \emph {et~al.}(2017)\citenamefont {Guan},
  \citenamefont {Shang}, \citenamefont {Lindwall}, \citenamefont {Anderson},\
  and\ \citenamefont {Liu}}]{124}%
  \BibitemOpen
  \bibfield  {author} {\bibinfo {author} {\bibfnamefont {P.-W.}\ \bibnamefont
  {Guan}}, \bibinfo {author} {\bibfnamefont {S.-L.}\ \bibnamefont {Shang}},
  \bibinfo {author} {\bibfnamefont {G.}~\bibnamefont {Lindwall}}, \bibinfo
  {author} {\bibfnamefont {T.}~\bibnamefont {Anderson}}, \ and\ \bibinfo
  {author} {\bibfnamefont {Z.-K.}\ \bibnamefont {Liu}},\ }\href {\doibase
  https://doi.org/10.1016/j.calphad.2017.10.006} {\bibfield  {journal}
  {\bibinfo  {journal} {Calphad}\ }\textbf {\bibinfo {volume} {59}},\ \bibinfo
  {pages} {171} (\bibinfo {year} {2017})}\BibitemShut {NoStop}%
\bibitem [{\citenamefont {Anand}\ \emph {et~al.}(2021)\citenamefont {Anand},
  \citenamefont {Male}, \citenamefont {Wolverton},\ and\ \citenamefont
  {Snyder}}]{Anand2021Visualizing}%
  \BibitemOpen
  \bibfield  {author} {\bibinfo {author} {\bibfnamefont {S.}~\bibnamefont
  {Anand}}, \bibinfo {author} {\bibfnamefont {J.~P.}\ \bibnamefont {Male}},
  \bibinfo {author} {\bibfnamefont {C.}~\bibnamefont {Wolverton}}, \ and\
  \bibinfo {author} {\bibfnamefont {G.~J.}\ \bibnamefont {Snyder}},\ }\href
  {\doibase 10.1039/D1MH00397F} {\bibfield  {journal} {\bibinfo  {journal}
  {Mater. Horiz.}\ }\textbf {\bibinfo {volume} {8}},\ \bibinfo {pages} {1966}
  (\bibinfo {year} {2021})}\BibitemShut {NoStop}%
\bibitem [{\citenamefont {Adekoya}\ and\ \citenamefont
  {Snyder}(2024{\natexlab{a}})}]{Adekoya2024}%
  \BibitemOpen
  \bibfield  {author} {\bibinfo {author} {\bibfnamefont {A.~H.}\ \bibnamefont
  {Adekoya}}\ and\ \bibinfo {author} {\bibfnamefont {G.~J.}\ \bibnamefont
  {Snyder}},\ }\href {\doibase https://doi.org/10.1002/adfm.202403926}
  {\bibfield  {journal} {\bibinfo  {journal} {Advanced Functional Materials}\
  ,\ \bibinfo {pages} {2403926}} (\bibinfo {year} {2024}{\natexlab{a}})},\
  \Eprint
  {http://arxiv.org/abs/https://onlinelibrary.wiley.com/doi/pdf/10.1002/adfm.202403926}
  {https://onlinelibrary.wiley.com/doi/pdf/10.1002/adfm.202403926} \BibitemShut
  {NoStop}%
\bibitem [{\citenamefont {Adekoya}\ and\ \citenamefont
  {Snyder}(2024{\natexlab{b}})}]{ADEKOYA2024100109}%
  \BibitemOpen
  \bibfield  {author} {\bibinfo {author} {\bibfnamefont {A.~H.}\ \bibnamefont
  {Adekoya}}\ and\ \bibinfo {author} {\bibfnamefont {G.~J.}\ \bibnamefont
  {Snyder}},\ }\href {\doibase https://doi.org/10.1016/j.mtelec.2024.100109}
  {\bibfield  {journal} {\bibinfo  {journal} {Materials Today Electronics}\
  }\textbf {\bibinfo {volume} {9}},\ \bibinfo {pages} {100109} (\bibinfo {year}
  {2024}{\natexlab{b}})}\BibitemShut {NoStop}%
\bibitem [{\citenamefont {Sze}\ and\ \citenamefont {Ng}(2006)}]{Sze2006}%
  \BibitemOpen
  \bibfield  {author} {\bibinfo {author} {\bibfnamefont {S.}~\bibnamefont
  {Sze}}\ and\ \bibinfo {author} {\bibfnamefont {K.~.}\ \bibnamefont {Ng}},\
  }\enquote {\bibinfo {title} {Physics and properties of semiconductors—a
  review},}\ in\ \href {\doibase https://doi.org/10.1002/9780470068328.ch1}
  {\emph {\bibinfo {booktitle} {Physics of Semiconductor Devices}}}\ (\bibinfo
  {publisher} {John Wiley \& Sons, Ltd},\ \bibinfo {year} {2006})\ pp.\
  \bibinfo {pages} {5--75},\ \Eprint
  {http://arxiv.org/abs/https://onlinelibrary.wiley.com/doi/pdf/10.1002/9780470068328.ch1}
  {https://onlinelibrary.wiley.com/doi/pdf/10.1002/9780470068328.ch1}
  \BibitemShut {NoStop}%
\bibitem [{See(2009)}]{Seebauer2009}%
  \BibitemOpen
  \enquote {\bibinfo {title} {Fundamentals of defect ionization and
  transport},}\ in\ \href {\doibase 10.1007/978-1-84882-059-3_2} {\emph
  {\bibinfo {booktitle} {Charged Semiconductor Defects: Structure,
  Thermodynamics and Diffusion}}}\ (\bibinfo  {publisher} {Springer London},\
  \bibinfo {address} {London},\ \bibinfo {year} {2009})\ pp.\ \bibinfo {pages}
  {5--37}\BibitemShut {NoStop}%
\bibitem [{\citenamefont {Freysoldt}\ \emph {et~al.}(2022)\citenamefont
  {Freysoldt}, \citenamefont {Neugebauer}, \citenamefont {Tan},\ and\
  \citenamefont {Hennig}}]{PhysRevB.105.014103}%
  \BibitemOpen
  \bibfield  {author} {\bibinfo {author} {\bibfnamefont {C.}~\bibnamefont
  {Freysoldt}}, \bibinfo {author} {\bibfnamefont {J.}~\bibnamefont
  {Neugebauer}}, \bibinfo {author} {\bibfnamefont {A.~M.~Z.}\ \bibnamefont
  {Tan}}, \ and\ \bibinfo {author} {\bibfnamefont {R.~G.}\ \bibnamefont
  {Hennig}},\ }\href {\doibase 10.1103/PhysRevB.105.014103} {\bibfield
  {journal} {\bibinfo  {journal} {Phys. Rev. B}\ }\textbf {\bibinfo {volume}
  {105}},\ \bibinfo {pages} {014103} (\bibinfo {year} {2022})}\BibitemShut
  {NoStop}%
\end{thebibliography}


%%%%%%%
%%%%%%%
\appendix 

\section{Notes on DEF End-member Gibbs energy}\label{sec:equivalence}
In a recent study, we derived the DEF end-member Gibbs energy directly from DFT supercell energy calculations \cite{Adekoya2024}.  In Ref.\cite{Adekoya2024}, end-members containing defects were defined by scaling the difference between the pristine supercell and the defective supercell to arrive at a stoichiometry containing the same composition as the end-member (see Eq. 12 in Ref.\cite{Adekoya2024}). In this section, we show that the general derivation presented in this paper can be reformatted to arrive at the same formulation as Ref.\cite{Adekoya2024}.

For a B-vacancy defect in \AB, represented by the (A)\ce{_p}(Va)\ce{_q} end-member, the general derivation presented in this paper gives the Gibbs energy as (see equation \ref{eq:DEFmapping2_pq})
$$ {}^0G_{\text{A:Vac}}  = {}^0G_{\text{A:B}} + q \Delta E_{\text{B-vac}}  =  p \mu_A^0 + q \mu_B^0 + (p+q) \hp + q E_{\text{def}} -q E_{\text{prinstine}}$$
Reformatting the supercell total energy by the formation energy defined in equation \ref{eq:formationenergies} results in $\frac{E_{\text{prinstine}}}{l}= (p+q) \hp + p \mu_A^0 + q \mu_B^0$. Substituting this equality gives the end-member Gibbs energy as
$$ {}^0G_{\text{A:Vac}}  = {}^0G_{\text{A:B}} + \Delta E_{\text{B-vac}}  =  \frac{E_{\text{prinstine}}}{l} + q E_{\text{def}} -q E_{\text{prinstine}} = qE_{\text{def}} - \frac{ql-1}{l}E_{\text{prinstine}}$$
Eq. 12 in Ref.\cite{Adekoya2024} presents the above equation for $q=1$ and $l=32$.

Similarly, for an A-interstitial defect represented by the (A)\ce{_p}(B)\ce{_q}(A$^i$)\ce{_m} end-member, we can reformat the end-member Gibbs energy of equation \ref{eq:DEFmapping3_pq} in terms of total supercell energies of defective and pristine compounds as the following 

$$ {}^0G_{\text{A:B:A}}  = {}^0G_{\text{A:B:Vac}} + m \Delta E_{\text{A-interstitial}}  =  \frac{E_{\text{prinstine}}}{l} + m E_{\text{def}} -m E_{\text{prinstine}} = mE_{\text{def}} - \frac{ml-1}{l}E_{\text{prinstine}}$$

\newpage
\section{(A,Vac)(B,Vac)(Vac\textsuperscript{i},B\textsuperscript{i}) Model}
\setcounter{figure}{0}  
\renewcommand{\thefigure}{A\arabic{figure}} % Change the figure numbering
\begin{figure*}[tph]
    \centering
    \includegraphics[width=0.75\textwidth]{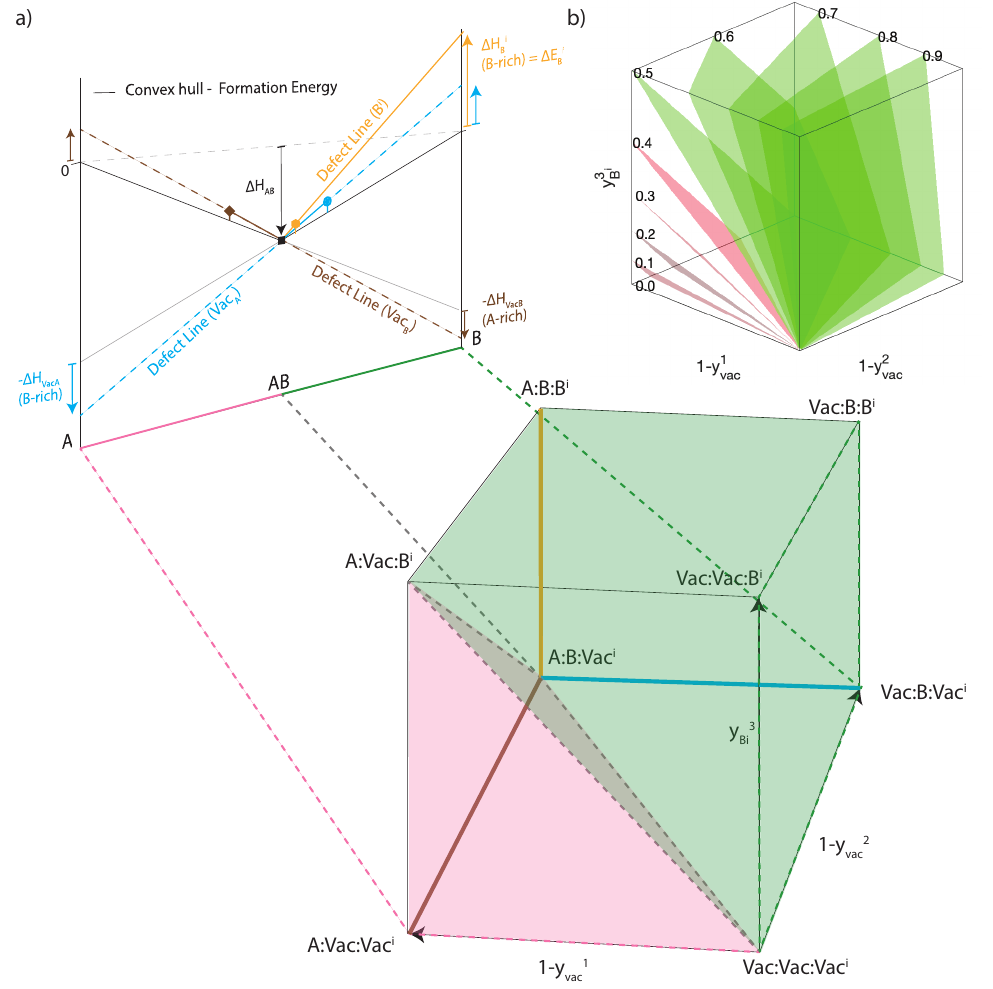}
    \caption{(a) Graphic representation of the mapping between $X_B$ composition on the convex hull and the constitutional cube of the three-sublattice model ($Y$-space cube). The $X_B$ composition on the AB-B and A-AB sides maps into the green and pink regions of the $Y$-space cube, respectively. Lines on the constitutional cube relevant to the defect lines on the convex hull are shown in the respective colors: blue for A-vacancy, brown for B-vacancy, and orange for B-interstitial. (b) Mapping of the constitutional cube for the (A,Vac)(B,Vac)(Vac\textsuperscript{i},B\textsuperscript{i}) DEF model, ($1-\yt$)-($1-\yu$)-$\yi$, into the composition $X_B$. Contour planes in the constitutional cube are labeled with corresponding $X_B$ values. Contour planes for $X_B < 0.5$ and $X_B \geq 0.5$ are colored red and green, respectively.}
    \label{fig:Bi}
\end{figure*}
\end{document}